\documentclass[10pt,oneside,final]{IEEEtran}
\usepackage{amssymb}
\usepackage{amsmath}

\usepackage{graphicx}
\usepackage{color}
\usepackage{multicol}

\usepackage{subfigure}
\usepackage{bm}
\usepackage{latexsym}
\usepackage{stfloats}
\usepackage{float}
\usepackage{exscale}
\usepackage{relsize}
\usepackage{cite}
\usepackage{cases}
\usepackage{algorithm}
\usepackage{algpseudocode}

\usepackage{epsfig}
\usepackage{epstopdf}
\usepackage{breqn}
\usepackage{enumerate}
\usepackage{multirow}
\usepackage[utf8]{inputenc}

\usepackage{mathtools}

\usepackage{algorithmicx}
\usepackage{setspace}
\usepackage{url}

\begin{document}

\title{Robust Transmission Design for RIS-assisted Secure Multiuser Communication Systems in the Presence of Hardware Impairments}
\author{
	Zhangjie~Peng,
	Ruisong~Weng,
	Cunhua Pan,~\IEEEmembership{Member,~IEEE},
	Gui Zhou,~\IEEEmembership{Graduate Student Member,~IEEE},
	\\
	Marco Di Renzo,~\IEEEmembership{Fellow,~IEEE},
	and~A. Lee Swindlehurst,~\IEEEmembership{Fellow,~IEEE}
\vspace{-0.3cm}
\thanks{Z. Peng, and R. Weng are with the College of Information, Mechanical and Electrical Engineering,
	Shanghai Normal University, Shanghai 200234, China (e-mails: pengzhangjie@shnu.edu.cn, 1000497102@smail.shnu.edu.cn).}
\thanks{C. Pan is with the National Mobile Communications Research Laboratory, Southeast University, China. (cunhuapan21@gmail.com).}
\thanks{G. Zhou is with the School of Electronic Engineering and Computer Science at Queen Mary University of London, London E1 4NS, U.K. (email: g.zhou@qmul.ac.uk).}
\thanks{M. Di Renzo is with Université Paris-Saclay, CNRS, CentraleSupélec, Laboratoire des Signaux et Systèmes, 91192 Gif-sur-Yvette, France (e-mail: marco.di-renzo@universite-paris-saclay.fr).}
\thanks{A. L. Swindlehurst is with the Department of Electrical Engineering and Computer Science,
	University of California at Irvine, Irvine, CA 92697 USA (e-mail: swindle@uci.edu).}
}
%
%\IEEEoverridecommandlockouts

%%%%%%%%%%%%%%%%%%wrs%%%%%%%%%%%%%%%%%%%%%%%%%%%%%%%%%%%%%%%
%
%   \author{
%            Author 1, Author 2, Author 3, and Author 4
%   \vspace{-0.9cm}
%   \vspace{-0.6cm}
%   \vspace{-0.3cm}
%
%   \thanks{Message of Author1}
%   \thanks{Message of Author2}
%}
%%%%%%%%%%%%%%%%%%%%%%%%%%%%%%%%20201126%%%%%%%%%%%%%%%%%%%%

\maketitle

\newtheorem{lemma}{Lemma}
\newtheorem{theorem}{Theorem}
\newtheorem{remark}{Remark}
\newtheorem{corollary}{Corollary}
\newtheorem{proposition}{Proposition}
\allowdisplaybreaks[4]

%%%%%%%%%%%%%%%%%%wrs%%%%%%%%%%%%%%%%%%%%%%%%%%%%%%%%%%%%%%%
%
%%%%%%%%%%%%%%%%%%%%%%%%%%%%%%%%20201126%%%%%%%%%%%%%%%%%%%%
\begin{abstract}
	
This paper investigates reconfigurable intelligent surface (RIS)-assisted secure multiuser communication systems in the presence of the hardware impairments (HIs) of the RIS and the transceiver.
We jointly optimize the beamforming vectors at the base station (BS) and the phase shifts of the reflecting elements at the RIS so as to maximize the weighted minimum secrecy rate (WMSR), subject to both transmission power constraints at the BS and unit-modulus constraints at the RIS. 
To address the formulated optimization problem, we first decouple it into two tractable subproblems and then use the block coordinate descent (BCD) method to alternately optimize the subproblems. 
Two different methods are proposed to solve the two obtained subproblems.
The first method transforms each subproblem into a second order cone programming (SOCP) problem by invoking the penalty convex–concave procedure (CCP) method and the closed-form fractional programming (FP) criterion, and then directly solves them by using CVX.
The second method leverages the Minorization-Maximization (MM) algorithm.  
Specifically, we first derive a concave approximation function, which is a lower bound of the original objective function, and then the two subproblems are transformed into two simple surrogate problems with closed-form solutions.
Simulation results verify the performance gains of the proposed robust transmission method over existing non-robust designs. 
In addition, the MM algorithm is shown to have much lower complexity than the SOCP-based algorithm.

\begin{IEEEkeywords}
%Keywords are supposed to be here
Intelligent reflecting surface (IRS),  reconfigurable intelligent surface (RIS), hardware impairments (HIs), physical layer security (PLS).
\end{IEEEkeywords}

\end{abstract}

\section{Introduction}
Thanks to the growing popularization of mobile devices, the global wireless network capacity is expected to increase 100-fold by 2030 \cite{itu2015vision}.
Furthermore, emerging applications, such as the industrial Internet of things, virtual reality (VR) and augmented reality (AR)\cite{9170653}, have high quality of service (QoS) requirements, such as ultra-low latency, ultra-high reliability and extremely high data rates \cite{9475160}.
Some potential techniques, such as massive  multiple-input multiple-output (m-MIMO) arrays, millimeter wave (mmWave) and terahertz (THz) communications\cite{8849960}, have been proposed to meet the above requirements.
However, these technologies usually result in increasing the cost of network deployment and the network power consumption\cite{8466374}.

%Meanwhile, some emerging applications such as the Industrial Internet of Things, virtual reality (VR) and augmented reality (AR) and so on \cite{9170653}, need ultra-low latency, ultra-high reliability and extremely high data rates \cite{9475160}.
%To satisfy these quality of service (QoS) requirements, some potential techniques have been proposed, such as massive  multiple-input multiple-output (m-MIMO) arrays\cite{9400853}, millimeter wave (mmWave) communications and THz communications, and so on\cite{9318531,8796365}.
%However, most of these techniques are difficult to avoid the negative effects of the uncontrollable wireless environment, such as the performance loss caused by uncontrolled scattering of electromagnetic waves \cite{8466374}.

Another emerging technology for fulfilling the high QoS requirements of future networks \cite{9318531, 9326394} is the use of reconfigurable intelligent surfaces (RISs). 
RIS is a thin metamaterial layer that is composed of an array of low cost reflecting elements integrated with low power and controllable electronics\cite{9140329}.
Due to the absence of power amplifiers, digital signal processing units, and multiple radio frequency chains, the main features of an RIS include a low implementation cost, a low power consumption, and an easy deployment, as well as the capability of reconfiguring the wireless environment\cite{9195133,9119122}.
Broadly speaking, an RIS is a dynamic metasurface whose electromagnetic characteristics can be dynamically adjusted through control signals. 
For example, the electromagnetic waves that impinge upon an RIS can be steered towards different directions, by simply optimizing the phase response of each of its constituent scattering elements\cite{direnzo2021communication}.
An RIS can be utilized to enhance the desired signal power, to mitigate the network interference, and to reduce the electromagnetic pollution since no additional signals are generated \cite{8811733}.
Compared with traditional active antenna arrays that are equipped with multiple active radio frequency transceivers, an RIS reradiates the incident signals by simply adjusting the amplitude and the phase shift of the reflecting elements, which can be realized by controlling the junction voltage of PIN diodes or varactors\cite{8910627}.
RISs can be deployed on, e.g., the facades of buildings, the interior walls of offices, and windows.

%It is foreseeable that the 5G and 6G applications will generate large amounts of sensitive and private data\cite{9140329,xu2019resource,2019arXiv190409573Y}.
RISs can be utilized for enhancing the security of wireless networks and have been recently amalgamated with physical layer security (PLS) \cite{2019arXiv190409573Y, 9384499}.
Traditional wireless security methods encrypt the data  at the network layer.
This usually requires a high overhead due to the frequent  distribution and management of secrecy keys \cite{xu2019resource, 6626661, 9198898}.
PLS is an alternative solution that makes use of the properties of the wireless communication medium and the transceiver hardware to enable critical aspects of secure communications.
However, conventional PLS techniques only focus on beamforming design at the transceivers, and may not provide good performance in some scenarios, e.g., when the legitimate user and the eavesdropper have highly correlated channels (e.g., when they are located in the same direction from the transmitter) \cite{8723525}.
%Especially, when the power of the eavesdropper is greater than the average power of the legal users, the system security performance will be deteriorated\cite{8723525}.
Thanks to the capability of reconfiguring the propagation environment in a desired manner, RIS can change the reflection direction of the incident signal to enhance the desired signal power at the legitimate users, while suppressing the signal received by eavesdroppers.
RISs have several applications in the context of PLS for improving the security of wireless communication systems \cite{9134962,9133130,9206080, 9528924}.
For example, the authors of \cite{9134962} studied the secrecy outage probability of an RIS-assisted single-antenna system where only one eavesdropper exist.
%The authors of \cite{9291402} investigated the ergodic secrecy rate of an RIS-assisted communication system in the presence of multiple eavesdroppers.
In \cite{9133130}, the authors proposed a robust algorithm to maximize the achievable secrecy rate of a multi-user multiple-input single-output (MISO) system.
%The authors of \cite{9201173} proposed artificial noise (AN) to achieve secure communications in RIS-assisted MIMO systems.
In \cite{9206080}, the authors proposed a deep reinforcement learning (DRL)-based scheme to improve the security performance of RIS-assisted MIMO systems.
The authors of \cite{9528924} analyzed the security performance gains when deploying an RIS in unmanned aerial vehicle (UAV)-assisted mmWave wireless communication networks.

The existing contributions on RIS-assisted PLS assume that the transceivers are constructed with ideal and perfect hardware components.
In practical communication systems, low-cost hardware is often preferred even though such hardware may be subject to hardware impairments (HIs) such as I/Q-imbalances, amplifier non-linearities, quantization errors, and phase noise\cite{6891254}.
If these hardware impairments are ignored at the design stage, the performance usually degrades \cite{9239335}.
Recently, the impact of HIs on the security performance of RIS-assisted single-user systems has been analyzed \cite{9374557,9376864}.
Specifically, the authors of \cite{9374557} proposed a robust algorithm to maximize the secrecy rate in the presence of HIs.
In \cite{9376864}, the authors derived an approximate closed-form expression for the secrecy outage probability and studied the impact of HIs on the system performance.

In this paper, we investigate the security performance of RIS-assisted multiuser MISO systems in the presence of HIs. 
Unlike the single-user scenarios considered in \cite{9374557} and \cite{9376864}, we assume a scenario with multiple legitimate users whose information security is threatened by an eavesdropper.
By deploying an RIS, we aim to improve the security performance under the premise of ensuring fairness among the users.
However, due to the considered complex scenario, the resulting optimization problem cannot be directly solved by using existing methods.
Thus, we propose tractable algorithms to tackle the formulated optimization problem. 
Specifically, the main contributions of this paper are summarized as follows:
\begin{enumerate}
\item
This work is the first to consider RIS-aided secure communications in multiuser MISO systems, where the base station (BS), the RIS and the legitimate users are subject to HIs.
By optimizing the BS precoding matrix and the RIS reflection coefficients,
we formulate a fairness-based joint optimization problem that maximizes the weighted minimum secrecy rate (WMSR), subject to both transmission power and unit modulus constraints.

\item
To efficiently solve the non-convex problem, we propose a benchmark algorithm based on the block coordinate descent (BCD) method.
Specifically, we first decouple the original problem into multiple tractable subproblems by invoking the penalty convex–concave procedure (CCP) and the closed-form fractional programming (FP) criterion.
The precoding and the reflection coefficient subproblems are transformed into second order cone programming (SOCP) problems.
Then, these two subproblems are alternately solved until convergence.

\item
Also, we propose a minorization-maximization (MM) algorithm to reduce the computational complexity.
In particular, we first derive a concave smooth function as a lower bound of the original non-differentiable objective function.
Then, we apply the MM algorithm to obtain a surrogate function which has a closed-form solution.

\item
Finally, we present simulation results to verify the effectiveness of the proposed schemes and the advantages of the proposed robust transmission design for secure communications.
We demonstrate that deploying an RIS can effectively improve the security performance of multiuser wireless communication systems in the presence of HIs.
The convergence and effectiveness of the proposed algorithm are verified as well.
\end{enumerate}

The rest of this paper is organized as follows.
Section \ref{SYSTEM MODEL AND PROBLEM FORMULATION} introduces the RIS-assisted wireless communication system model subject to HIs and formulates the WMSR problem.
Section \ref{BCD METHOD} decouples the original problem into multiple tractable sub-problems and proposes a benchmark optimization algorithm based on the BCD method.
In Section \ref{MM METHOD}, a low-complexity MM algorithm is introduced.
Simulation results are given in Section
\ref{Simulation Results} and Section \ref{conclusion} concludes this paper.

\emph{Notations}: Constants, column vectors and matrices are denoted by italics, boldface lowercase letters and boldface uppercase letters, respectively.
$\mathrm{Re}\left\{ b \right\}$, $\left| b \right|$ and $\angle \left( b \right) $ denote the real part, modulus and angle of the complex number $b$, respectively.
$\left\| \mathbf{b} \right\| _1$, $\left\| \mathbf{b} \right\| _2$ and $\left\| \mathbf{b} \right\| _F$ denote the 1-norm, 2-norm and Frobenius-norm of vector $\mathbf{b}$, respectively.
$\mathrm{diag}\left( \cdot \right) $ and $\mathrm{vec}\left( \cdot \right) $ represent the diagonalization and vectorization operators, respectively.
$\mathbf{B}^{\mathrm{T}}$, $\mathbf{B}^*$, $\mathbf{B}^{\mathrm{H}}$, $\mathrm{Tr}\left[ \mathbf{B} \right]$ and $\left\| \mathbf{B} \right\| _F$ denote the transpose, conjugate, Hermitian, trace and Frobenius norm of matrix $\mathbf{B}$, respectively.
The Hadamard product and Kronecker product of two matrices $\mathbf{B}$ and $\mathbf{C}$ are expressed as $\mathbf{B}\odot \mathbf{C}$ and $\mathbf{B}\otimes \mathbf{C}$, respectively.
$\mathbf{B}\succeq \mathbf{C}$ indicates that $\mathbf{B}-\mathbf{C}$ is a positive semidefinite matrix.
${\mathbb C}$ denotes the complex field and $j\triangleq \sqrt{-1}$ is the imaginary unit.

\section{System Model}\label{SYSTEM MODEL AND PROBLEM FORMULATION}

\begin{figure}
\centering
\includegraphics[scale=0.55]{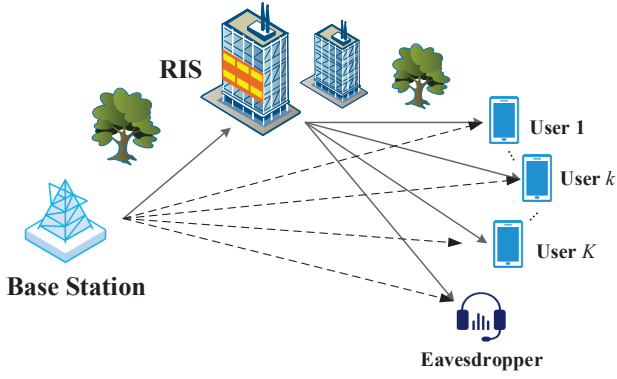}
\caption{An RIS-assisted MISO downlink system with an $N$-antenna BS, a single-antenna eavesdropper and $K$ single-antenna users.}
\label{systemModel}
\end{figure}

\subsection{Signal Transmission Model}
We consider an RIS-assisted MISO downlink system with a BS, an eavesdropper and $K$ legitimate users, as illustrated in Fig. \ref{systemModel}.
The BS is equipped with $N > 1$ transmit antennas to serve the legitimate users in the presence of the eavesdropper.
In this case, an RIS consisting of $M$ reflecting elements is deployed to ensure the secure transmission of data.
$\phi _m=e^{j\theta _m}$ is the reflection coefficient of the $m$-th reflecting element of the RIS, and the phase shift $\theta _m\in \left[ 0,2\pi \right] $.
The set of RIS reflection coefficients is collected in the diagonal  matrix $\mathbf{\Phi }=\mathrm{diag}\left( \boldsymbol{\phi } \right) $, where $\bm{\phi}  = {\left[ {{\phi _1}, \cdots ,{\phi _M}} \right]^{\rm T}}$ with ${\left| \phi _m \right|^2} = 1$, $\forall m \in \mathcal{M}$, $\mathcal{M}\triangleq \left\{ 1,2,...,M \right\}$.
$\mathbf{\Lambda }=\mathrm{diag}\left( \boldsymbol{\psi } \right)$ is the random phase noise matrix, wherein $\boldsymbol{\psi }=\left[ \psi _1,\cdots ,\psi _M \right] ^{\mathrm{T}}$ and $\psi _m=e^{j\vartheta _m}$.
$e^{j\vartheta _m}$ is the $m$-th RIS element's phase noise caused by RIS HI, and $\vartheta _m$ is uniformly distributed on $[-\pi/2,\pi/2]$ \cite{9390410}.
\footnote{The more complex and practical reflection model, such as with a phase-dependent amplitude \cite{9115725,9389801,9064547}, will be investigated in our future work.}
The direct channels from the BS to the legitimate user $k$ and from the BS to the eavesdropper, the indirect channel from the BS to the RIS, and the reflection channels from the RIS to the legitimate user $k$ and from the RIS to the eavesdropper, are denoted by ${\mathbf{h}_{{\rm BU},k}} \in {{\mathbb C}^{{N} \times 1}}$, ${\mathbf{h}_{\rm BE}} \in {{\mathbb C}^{{N} \times 1}}$, ${\mathbf{H}_{\rm BR}} \in {{\mathbb C}^{{M} \times {N}}}$, ${\mathbf{h}_{{\rm RU},k}} \in {{\mathbb C}^{{M} \times 1}}$ and ${\mathbf{h}_{\rm RE}} \in {{\mathbb C}^{{M} \times 1}}$, respectively.

%   1
The signal transmitted from the BS is modeled as
\setlength\abovedisplayskip{1pt}
\setlength\belowdisplayskip{1pt}
\begin{equation}
\mathbf{x}=\mathbf{\hat{x}}+\boldsymbol{\eta }_{\mathrm{t}},
\end{equation}
\begin{equation}
\mathbf{\hat{x}}\triangleq \sum_{k=1}^K{\mathbf{w}_ks_k},
\end{equation}
where $s_k$ is assumed to be an independent random Gaussian signal with zero mean and variance $\mathbb{E}\left[ \left| s_k \right|^2 \right] =1$.
In addition, $\mathbf{w}_k \in {{\mathbb C}^{{N} \times 1}}$ is the corresponding beamforming vector.
Hence, the precoding matrix of the BS can be defined as $\mathbf{W}\triangleq \left[ \mathbf{w}_1,\cdots ,\mathbf{w}_K \right] \in \mathbb{C}^{N\times K}$, which satisfies the constraint $\mathrm{Tr(}\mathbf{W}^{\mathrm{H}}\mathbf{W})\leqslant P$, where $P$ represents the maximum transmit power.

The additional distortion noise term $\boldsymbol{\eta }_{\mathrm{t}}$ describes the impact of HIs at the transmitter.
According to the model in \cite{7835110,9322510}, the distortion noise is assumed to be proportional to the signal power.
In particular, the entries of $\boldsymbol{\eta }_{\mathrm{t}}$ are independent zero-mean Gaussian random variables whose distribution is $\mathcal{C}\mathcal{N}(0,\mathbf{\Upsilon }_{\mathrm{t}})$, where $\mathbf{\Upsilon }_{\mathrm{t}} =\kappa _{\mathrm{t}}\mathrm{diag}\left( \mathbf{WW}^{\mathrm{H}} \right)$ and $\kappa _{\mathrm{t}}\geqslant 0 $ is the ratio between the transmit distorted noise power and the transmit signal power.

%  4
The signal received at user $k$ is given by
\begin{align}\label{signal_received_at_user}
y_{\mathrm{U},k}=\mathbf{h}_{\mathrm{U},k}^{\mathrm{H}}\mathbf{x}+\eta _{\mathrm{r},k}+n_{\mathrm{U},k}\triangleq \hat{y}_{\mathrm{U},k}+\eta _{\mathrm{r},k},
\end{align}
where $\mathbf{h}_{\mathrm{U},k}^{\mathrm{H}}\triangleq \mathbf{h}_{\mathrm{RU},k}^{\mathrm{H}}\mathbf{\Lambda \Phi H}_{\mathrm{BR}}+\mathbf{h}_{\mathrm{BU},k}^{\mathrm{H}}
$ and $n_{\mathrm{U},k}$ is the additive white Gaussian noise (AWGN) whose distribution is $\mathcal{C}\mathcal{N}(0,\delta _{\mathrm{U},k}^{2})$.
$\eta _{\mathrm{r},k}$ is an additional distortion noise term that is independent of $\hat{y}_{\mathrm{U},k}$ and whose distribution is ${\mathcal C}{\mathcal N}(0, \gamma _{\mathrm{r},k})$, with $\gamma _{\mathrm{r},k}$ being defined as $\gamma _{\mathrm{r},k}=\mathbb{E}\left\{ \kappa _{\mathrm{r},k}\left\|\hat{y}_{\mathrm{U},k}\right\|_2^2 \right\}$, where $\kappa _{\mathrm{r},k} \geqslant 0$ is the
ratio between the distorted noise power and the undistorted received signal
power \cite{9374557}.

We consider the worst-case assumption that the eavesdropper can eliminate most of the noise with the exception of the distortion noise of the transmitter hardware. 
Also, we assume that it can decode and cancel the interference from other users \cite{6851195}.
In addition, the eavesdropper is supposed to actively attack the communication system.
Specifically, by pretending to be a legitimate user sending pilot signals to the BS during the channel estimation procedure \cite{8335290}, the eavesdropper can mislead the BS to send signals to the eavesdropper. 
Furthermore, some low-complexity channel estimation methods \cite{9130088,9603291} can be adopted to estimate the RIS-user and RIS-eavesdropper channels.
Then, the signal received at the eavesdropper is given by
\begin{equation}\label{signal_eve}
y_{\mathrm{E}}=\mathbf{h}_{\mathrm{E}}^{\mathrm{H}}\mathbf{x}+n_{\mathrm{E}}
\end{equation}
where $\mathbf{h}_{\mathrm{E}}^{\mathrm{H}}\triangleq \mathbf{h}_{\mathrm{RE}}^{\mathrm{H}}\mathbf{\Lambda \Phi H}_{\mathrm{BI}}+\mathbf{h}_{\mathrm{BE}}^{\mathrm{H}}$ and $n_{\mathrm{E}}$ is AWGN whose distribution is $\mathcal{C}\mathcal{N}(0,\delta _{\mathrm{E}}^{2})$.

\subsection{HIs Model}
From \eqref{signal_received_at_user}, the signal-to-interference-plus-noise ratio (SINR) at the legitimate user $k$ can be expressed as
{\small\begin{align}
	&\gamma _k= \notag
	\\
	&\frac{\mathbf{w}_{k}^{\mathrm{H}}\mathbb{E}_{\boldsymbol{\psi }}\left\{ \mathbf{h}_{\mathrm{U},k}\mathbf{h}_{\mathrm{U},k}^{\mathrm{H}} \right\} \mathbf{w}_k}{\sum\limits_{\substack{i=1 \\ i\ne k}}^K{\mathbf{w}_{i}^{\mathrm{H}}\mathbb{E}_{\boldsymbol{\psi }}\left\{ \mathbf{h}_{\mathrm{U},k}\mathbf{h}_{\mathrm{U},k}^{\mathrm{H}} \right\} \mathbf{w}_i}\!\!+\!\!\mathrm{Tr}\left[ \mathbf{\Upsilon }_{\mathrm{t}}\mathbb{E}_{\boldsymbol{\psi }}\left\{ \mathbf{h}_{\mathrm{U},k}\mathbf{h}_{\mathrm{U},k}^{\mathrm{H}} \right\} \right] \!\!+\!\!\gamma _{\mathrm{r},k}\!\!+\!\!\delta _{\mathrm{U},k}^{2}},
\end{align}}
where
{\small\begin{align}\label{E_h_h_1}
	&\mathbb{E}_{\boldsymbol{\psi }}\left\{ \mathbf{h}_{\mathrm{U},k}\mathbf{h}_{\mathrm{U},k}^{\mathrm{H}} \right\}=2\mathrm{Re}\left\{ \mathbf{H}_{\mathrm{BR}}^{\mathrm{H}}\mathbf{\Phi }^{\mathrm{H}}\mathrm{diag}\left( \mathbf{h}_{\mathrm{RU},k} \right) \mathbb{E}_{\boldsymbol{\psi }}\left\{ \boldsymbol{\psi }^* \right\} \mathbf{h}_{\mathrm{BU},k}^{\mathrm{H}} \right\} \notag
	\\
	&\quad+\mathbf{H}_{\mathrm{BR}}^{\mathrm{H}}\mathbf{\Phi }^{\mathrm{H}}\mathrm{diag}\left( \mathbf{h}_{\mathrm{RU},k} \right)\mathbb{E}_{\boldsymbol{\psi }}\left\{ \boldsymbol{\psi }^*\boldsymbol{\psi }^{\mathrm{T}} \right\} \mathrm{diag}\left( \mathbf{h}_{\mathrm{RU},k}^{\mathrm{H}} \right) \mathbf{\Phi H}_{\mathrm{BR}}  \notag
	\\
	&\quad+\mathbf{h}_{\mathrm{BU},k}\mathbf{h}_{\mathrm{BU},k}^{\mathrm{H}}.
\end{align}}
Then, we will further calculate $\mathbb{E}_{\boldsymbol{\psi }}\left\{ \boldsymbol{\psi }^*\boldsymbol{\psi }^{\mathrm{T}} \right\}$ and $\mathbb{E}_{\boldsymbol{\psi }}\left\{ \boldsymbol{\psi }^* \right\}$.

Denote $\delta _{\vartheta} = \vartheta _i -\vartheta _j, \forall i,j \in \mathcal{M}$.
Note that $\vartheta _i$ and $\vartheta _j$ are uniformly distributed on $[-\pi/2,\pi/2]$, whose probability density function can be expressed as $f\left( \vartheta _i \right) =\frac{1}{\pi}$.
Hence, $\delta _{\vartheta}$ obeys triangular distribution on $[-\pi,\pi]$, whose probability density function can be expressed as \cite{9390410}
\begin{align}
f\left( \delta _{\vartheta} \right) =\begin{cases}
	\frac{1}{\pi ^2}\delta _{\vartheta}+\frac{1}{\pi},&		\delta _{\vartheta}\in \left[ -\pi ,0 \right],\\
	-\frac{1}{\pi ^2}\delta _{\vartheta}+\frac{1}{\pi},&		\delta _{\vartheta}\in \left[ 0,\pi \right].\\
\end{cases}
\end{align}
Hence, we have 
{\small\begin{align}
\mathbb{E}_{\delta _{\vartheta}}\left\{ e^{j\vartheta _i-j\vartheta _j} \right\} =\mathbb{E}_{\delta _{\vartheta}}\left\{ e^{j\delta _{\vartheta}} \right\} =\int_{-\pi}^{\pi}{f\left( \delta _{\vartheta} \right)}e^{j\delta _{\vartheta}}d\delta _{\vartheta}=\frac{4}{\pi ^2},
\end{align}}
and
$\mathbb{E}_{\boldsymbol{\psi }}\left\{ \boldsymbol{\psi }^*\boldsymbol{\psi }^{\mathrm{T}} \right\}$ can be given by
{\small	
	\begin{align} \label{E_psiHpsi}
		&\mathbb{E}_{\boldsymbol{\psi }}\left\{ \boldsymbol{\psi }^*\boldsymbol{\psi }^{\mathrm{T}} \right\} \notag
		\\
		&\!=\!\left( \begin{matrix}
			1&		\mathbb{E}_{\delta _{\vartheta}}\left\{ e^{j\vartheta _2-j\vartheta _1} \right\}&		\cdots&		\mathbb{E}_{\delta _{\vartheta}}\left\{ e^{j\vartheta _M-j\vartheta _1} \right\}\\
			\mathbb{E}_{\delta _{\vartheta}}\left\{ e^{j\vartheta _1-j\vartheta _2} \right\}&		1&		\cdots&		\mathbb{E}_{\delta _{\vartheta}}\left\{ e^{j\vartheta _M-j\vartheta _2} \right\}\\
			\vdots&		\vdots&		\ddots&		\vdots\\
			\mathbb{E}_{\delta _{\vartheta}}\left\{ e^{j\vartheta _1-j\vartheta _M} \right\}&		\mathbb{E}_{\delta _{\vartheta}}\left\{ e^{j\vartheta _2-j\vartheta _M} \right\}&		\cdots&		1\\
			\end{matrix} \right) \notag
		\\
		&=\left( \begin{matrix}
			1&		\frac{4}{\pi ^2}&		\cdots&		\frac{4}{\pi ^2}\\
			\frac{4}{\pi ^2}&		1&		\cdots&		\frac{4}{\pi ^2}\\
			\vdots&		\vdots&		\ddots&		\vdots\\
			\frac{4}{\pi ^2}&		\frac{4}{\pi ^2}&		\cdots&		1\\
		\end{matrix} \right)  =\mathbf{I}_M+\mathbf{G},
	\end{align}}
where
\begin{align}
\left[ \mathbf{G} \right] _{\left( i,j \right)}=\begin{cases}
	0,&		i=j,\\
	\frac{4}{\pi ^2},&		i\ne j.\\
\end{cases}
\end{align}
In addition, we can obtain
{\small\begin{align}
\mathbb{E}_{\vartheta _i}\left\{ e^{-j\vartheta _i} \right\} =\int_{-\frac{\pi}{2}}^{\frac{\pi}{2}}{f\left( \vartheta _i \right)}\left( \cos \vartheta _i-j\sin \vartheta _i \right) d\vartheta _i=\frac{2}{\pi}.
\end{align}}
Then, we have
\begin{align} \label{E_psi}
\mathbb{E}_{\boldsymbol{\psi }}\left\{ \boldsymbol{\psi }^* \right\} =\frac{2}{\pi}\mathbf{1},
\end{align}
where $\mathbf{1}$ represents the unit column vector with all elements of 1.

By substituting \eqref{E_psiHpsi} and \eqref{E_psi} into \eqref{E_h_h_1}, we have
{\small\begin{align}\label{E_h_U}
&\mathbb{E}_{\boldsymbol{\psi }}\left\{ \mathbf{h}_{\mathrm{U},k}\mathbf{h}_{\mathrm{U},k}^{\mathrm{H}} \right\} \notag
\\
&=\mathbf{H}_{\mathrm{BR}}^{\mathrm{H}}\mathbf{\Phi }^{\mathrm{H}}\mathrm{diag}\left( \mathbf{h}_{\mathrm{RU},k} \right) \left( \mathbf{I}_M+\mathbf{G} \right) \mathrm{diag}\left( \mathbf{h}_{\mathrm{RU},k}^{\mathrm{H}} \right) \mathbf{\Phi H}_{\mathrm{BR}} \notag
\\
&\quad+2\mathrm{Re}\left\{ \mathbf{H}_{\mathrm{BR}}^{\mathrm{H}}\mathbf{\Phi }^{\mathrm{H}}\mathrm{diag}\left( \mathbf{h}_{\mathrm{RU},k} \right) \frac{2}{{\pi }}\mathbf{1h}_{\mathrm{BU},k}^{\mathrm{H}} \right\} +\mathbf{h}_{\mathrm{BU},k}\mathbf{h}_{\mathrm{BU},k}^{\mathrm{H}} \notag
\\
&=\mathbf{H}_{\mathrm{BR}}^{\mathrm{H}}\mathbf{\Phi }^{\mathrm{H}}\mathrm{diag}\left( \mathbf{h}_{\mathrm{RU},k} \right) \mathbf{TT}^{\mathrm{T}} \mathrm{diag}\left( \mathbf{h}_{\mathrm{RU},k}^{\mathrm{H}} \right) \mathbf{\Phi H}_{\mathrm{BR}} \notag
\\
&\quad+\left( \frac{2}{{\pi }}\mathbf{H}_{\mathrm{BR}}^{\mathrm{H}}\mathbf{\Phi }^{\mathrm{H}}\mathbf{h}_{\mathrm{RU},k}+\mathbf{h}_{\mathrm{BU},k} \right) \left( \frac{2}{{\pi }}\mathbf{h}_{\mathrm{RU},k}^{\mathrm{H}}\mathbf{\Phi H}_{\mathrm{BR}}+\mathbf{h}_{\mathrm{BU},k}^{\mathrm{H}} \right) \notag
\\
&=\mathbf{\hat{h}}_{\mathrm{U},k}\mathbf{\hat{h}}_{\mathrm{U},k}^{\mathrm{H}}+\mathbf{\hat{H}}_{\mathrm{U},k}\mathbf{\hat{H}}_{\mathrm{U},k}^{\mathrm{H}}=\mathbf{\bar{H}}_{\mathrm{U},k}\mathbf{\bar{H}}_{\mathrm{U},k}^{\mathrm{H}},
\end{align}}
where
\begin{subequations}
\begin{align}
\mathbf{TT}^{\mathrm{T}}&\triangleq \mathrm{diag}\left( \left( 1-\frac{4}{\pi ^2} \right) \mathbf{I}_M \right), 
\\
\mathbf{\hat{h}}_{\mathrm{U},k}^{\mathrm{H}}&\triangleq \frac{2}{{\pi }}\mathbf{h}_{\mathrm{RU},k}^{\mathrm{H}}\mathbf{\Phi H}_{\mathrm{BR}}+\mathbf{h}_{\mathrm{BU},k}^{\mathrm{H}},
\\
\mathbf{\hat{H}}_{\mathrm{U},k}^{\mathrm{H}}&\triangleq \mathbf{T}^{\mathrm{T}}\mathrm{diag}\left( \mathbf{h}_{\mathrm{RU},k}^{\mathrm{H}} \right) \mathbf{\Phi H}_{\mathrm{BR}},
\\
\mathbf{\bar{H}}_{\mathrm{U},k}&\triangleq \left[ \begin{matrix}
	\mathbf{\hat{h}}_{\mathrm{U},k}&		\mathbf{\hat{H}}_{\mathrm{U},k}\\
\end{matrix} \right], 
\end{align}
\end{subequations}
and $\mathbf{I}_M$ denotes the $M \times M$ identity matrix.

Note that $\gamma _{\mathrm{r},k}=\mathbb{E}\left\{ \kappa _{\mathrm{r},k}\left\|\hat{y}_{\mathrm{U},k}\right\|_2^2 \right\}$.
Hence, $\gamma _{\mathrm{r},k}$ can be rewritten as
{\small\begin{align}\label{gamma_expression}
		\gamma _{\mathrm{r},k}=\mathrm{Tr}\left[ \kappa _{\mathrm{r},k}\left( \mathbf{WW}^{\mathrm{H}}+\kappa _{\mathrm{t}}\mathrm{diag}\left( \mathbf{WW}^{\mathrm{H}} \right) \right) \mathbf{\bar{H}}_{\mathrm{U},k}\mathbf{\bar{H}}_{\mathrm{U},k}^{\mathrm{H}} \right],
\end{align}}
and the achievable rate of user $k$ is given by
\begin{align}\label{user_speed}
	R_{\mathrm{U},k}\!&=\log \left( 1+\gamma _k \right), 
\end{align}
where
{\small\begin{align}
	\gamma _k&\triangleq\!\frac{\left\| \mathbf{\bar{H}}_{\mathrm{U},k}^{\mathrm{H}}\mathbf{w}_k \right\|_2 ^2}{\sum\limits_{\substack{i=1 \\ i\ne k}}^K{\left\| \mathbf{\bar{H}}_{\mathrm{U},k}^{\mathrm{H}}\mathbf{w}_i \right\|_2 ^2}\!+\mathrm{Tr}\left[ \mathbf{\Upsilon }_{\mathrm{t}}\mathbf{\bar{H}}_{\mathrm{U},k}\mathbf{\bar{H}}_{\mathrm{U},k}^{\mathrm{H}} \right] \!\!+\!\gamma _{\mathrm{r},k}\!+\!\delta _{\mathrm{U},k}^{2}}\!.
\end{align}}

Similarly, from \eqref{signal_eve}, the SINR at the eavesdropper associated with user $k$ can be expressed as
\begin{align}
	\gamma _{\mathrm{E},k}=\frac{\mathbf{w}_{k}^{\mathrm{H}}\mathbb{E}_{\boldsymbol{\psi }}\left\{ \mathbf{h}_{\mathrm{E}}\mathbf{h}_{\mathrm{E}}^{\mathrm{H}} \right\} \mathbf{w}_k}{\mathrm{Tr}\left[ \mathbf{\Upsilon }_{\mathrm{t}}\mathbb{E}_{\boldsymbol{\psi }}\left\{ \mathbf{h}_{\mathrm{E}}\mathbf{h}_{\mathrm{E}}^{\mathrm{H}} \right\} \right] +\delta _{\mathrm{E}}^{2}},
\end{align}
where
\begin{subequations}
\begin{align}
	\mathbb{E}_{\boldsymbol{\psi }}\left\{ \mathbf{h}_{\mathrm{E}}\mathbf{h}_{\mathrm{E}}^{\mathrm{H}} \right\} &=\mathbf{\bar{H}}_{\mathrm{E}}\mathbf{\bar{H}}_{\mathrm{E}}^{\mathrm{H}}
	\\
	\mathbf{\bar{H}}_{\mathrm{E}}&\triangleq \left[ \begin{matrix}
	\mathbf{\hat{h}}_{\mathrm{E}}&		\mathbf{\hat{H}}_{\mathrm{E}}\\
	\end{matrix} \right]. 
	\\
	\mathbf{\hat{h}}_{\mathrm{E}}^{\mathrm{H}}&\triangleq \frac{2}{\pi}\mathbf{h}_{\mathrm{RE}}^{\mathrm{H}}\mathbf{\Phi H}_{\mathrm{BR}}+\mathbf{h}_{\mathrm{BE}}^{\mathrm{H}},
	\\
	\mathbf{\hat{H}}_{\mathrm{E}}^{\mathrm{H}}&\triangleq \mathbf{T}^{\mathrm{T}}\mathrm{diag}\left( \mathbf{h}_{\mathrm{RE}}^{\mathrm{H}} \right) \mathbf{\Phi H}_{\mathrm{BR}}.
\end{align}
\end{subequations}
Then, the achievable rate of the eavesdropper associated with user $k$ is
\begin{equation}\label{eavesdropper_speed}
R_{\mathrm{E},k}=\log \left( 1+\frac{\left\| \mathbf{\bar{H}}_{\mathrm{E}}^{\mathrm{H}}\mathbf{w}_k \right\|_2 ^2}{\mathrm{Tr}\left[ \mathbf{\Upsilon }_{\mathrm{t}}\mathbf{\bar{H}}_{\mathrm{E}}\mathbf{\bar{H}}_{\mathrm{E}}^{\mathrm{H}} \right] +\delta _{\mathrm{E}}^{2}} \right).
\end{equation}
Accordingly, the secrecy rate $R_k$ of the legitimate user $k$ in nats/second/Hertz (nat/s/Hz) is given by
%  6
\begin{equation}\label{the_achievable_rate}
R_k\triangleq \left[ R_{\mathrm{U},k}-R_{\mathrm{E},k} \right] ^+,
\end{equation}
where $[a]^+\triangleq \max \left( a,0 \right)$.

% 11
\subsection{Problem Formulation}
To maximize the WMSR while ensuring fairness, we consider the joint optimization of the precoding matrix $\bf{W}$ and the reflection coefficient vector $\bm{\phi}$.
By denoting the weighting factor of user $k$ by $\omega_k$, the WMSR problem is formulated as
\begin{subequations}\label{WMSR_Problem}
\begin{align}
\max_{\mathbf{W},\boldsymbol{\phi }} \quad& \min_{k\in \mathcal{K}} 
 \quad \left\{ \omega _kR_k \right\}
\\
\mathrm{s}.\mathrm{t}.\quad& \mathrm{Tr(}\mathbf{W}^{\mathrm{H}}\mathbf{W})\leqslant P,
\\
& \boldsymbol{\phi }\in \mathcal{S},
\end{align}
\end{subequations}
where $\mathcal{K}\triangleq \left\{ 1,2,...,K \right\}$, $\mathcal{M}\triangleq \left\{ 1,2,...,M \right\}$ and the set $ \mathcal{S} \triangleq \left\{ {\bm \phi | \left| {{\phi _m}} \right| = 1,\forall m \in \mathcal{M}} \right\}$ imposes the unit-modulus constraint on $\bm \phi$.
Compared to a system model with no HIs, the objective function of the problem in \eqref{WMSR_Problem} is more complex.
The analysis of the secrecy rate instead of the information rate further complicates the objective to the point that a direct solution becomes intractable.
To circumvent these issues, we propose two efficient algorithms in the next sections.

% BCD
\section{BCD-SOCP Algorithm}\label{BCD METHOD}
In this section, we propose a BCD-SOCP algorithm to solve the WMSR problem in \eqref{WMSR_Problem}.
Specifically, we first decouple the problem in \eqref{WMSR_Problem} into two subproblems, each of which is converted into an SOCP problem that can be efficiently solved. The two subproblems are then alternately solved until convergence.  

\subsection{Problem Reformulation}
To reduce the complexity of the objective function in \eqref{WMSR_Problem}, we write $R_k$ as the sum of three parts, i.e.,
{\small\begin{align}	\label{divide_OF}
&R_k(\mathbf{W},\boldsymbol{\phi })=R_{\mathrm{U},k}(\mathbf{W},\boldsymbol{\phi })\!-\!R_{\mathrm{E},k}(\mathbf{W},\boldsymbol{\phi }) \notag
\\
&\quad=R_{\mathrm{U},k}(\mathbf{W},\boldsymbol{\phi })\!-\!\log \left( \frac{\left\| \mathbf{\bar{H}}_{\mathrm{E}}^{\mathrm{H}}\mathbf{w}_k \right\|_2 ^2+\mathrm{Tr}\left[ \mathbf{\Upsilon }_{\mathrm{t}}\mathbf{\bar{H}}_{\mathrm{E}}\mathbf{\bar{H}}_{\mathrm{E}}^{\mathrm{H}} \right] +\delta _{\mathrm{E}}^{2}}{\mathrm{Tr}\left[ \mathbf{\Upsilon }_{\mathrm{t}}\mathbf{\bar{H}}_{\mathrm{E}}\mathbf{\bar{H}}_{\mathrm{E}}^{\mathrm{H}} \right] +\delta _{\mathrm{E}}^{2}} \right) \notag
\\
&\quad=f_{1,k}(\mathbf{W},\boldsymbol{\phi })+f_{2,k}(\mathbf{W},\boldsymbol{\phi })+f_3(\mathbf{W},\boldsymbol{\phi }),
\end{align}
where
{\small\begin{align}
f_{1,k}(\mathbf{W},\boldsymbol{\phi })&\triangleq R_{\mathrm{U},k}(\mathbf{W},\boldsymbol{\phi })
\\
f_{2,k}(\mathbf{W},\boldsymbol{\phi })&\triangleq -\log \left( 1+\frac{\left\| \mathbf{\bar{H}}_{\mathrm{E}}^{\mathrm{H}}\mathbf{w}_k \right\|_2 ^2+\mathrm{Tr}\left[ \mathbf{\Upsilon }_{\mathrm{t}}\mathbf{\bar{H}}_{\mathrm{E}}\mathbf{\bar{H}}_{\mathrm{E}}^{\mathrm{H}} \right]}{\delta _{\mathrm{E}}^{2}} \right) 
\\
f_3(\mathbf{W},\boldsymbol{\phi })&\triangleq \log \left( 1+\frac{\mathrm{Tr}\left[ \mathbf{\Upsilon }_{\mathrm{t}}\mathbf{\bar{H}}_{\mathrm{E}}\mathbf{\bar{H}}_{\mathrm{E}}^{\mathrm{H}} \right]}{\delta _{\mathrm{E}}^{2}} \right).  \label{f_3}
\end{align}}
In the following, we derive lower bounds for $f_{1,k}$, $f_{2,k}$ and $f_{3}$.

As far as $f_{1,k}$ is concerned, we derive a lower bound by applying the closed-form FP approach \cite{8314727}.
First of all, $f_{1,k}$ can be tackled based on the following lemma.

\emph{Lemma 1}: Consider the function $f\left( \bar{y} \right) =\log \left( 1+\bar{y} \right) -\bar{y}+\frac{\left( 1+\bar{y} \right) \bar{x}}{1+\bar{x}}$ for any $\bar{x} > 0$.
Then, we have
\begin{equation}
\log \left( 1+\bar{x} \right) =\max_{\bar{y}\geqslant 0}\,\,f\left( \bar{y} \right),
\end{equation}
and the optimal solution is $\bar{y}=\bar{x}$.
\hfill $\blacksquare$

The lemma provides a lower bound for $\log \left( 1+\bar{x} \right)$, which is tight when $\bar{y}=\bar{x}$.
Hence, by introducing a set of auxiliary variables  $\mathcal{V}=\left\{ v_k\geqslant 0,\forall k\in \mathcal{K} \right\} $, we have
\begin{align}\label{f_1_1}
f_{1,k}(\mathbf{W},\boldsymbol{\phi },\mathcal{V}) \geqslant \log \left( 1+v_k \right) -v_k+\frac{\left( 1+v_k \right) \gamma _k}{1+\gamma _k}.
\end{align}

Due to the fact that variables $\mathbf{W},\boldsymbol{\phi },\mathcal{V}$ are coupled together in the $\frac{\left( 1+v_k \right) \gamma _k}{1+\gamma _k}$.
Then, by introducing  a set of auxiliary variables $\mathcal{U}=\left\{ \mathbf{u}_k\in \mathbb{C}^{\left( M+1 \right) \times 1},\forall k\in \mathcal{K} \right\}$ and adopting the quadratic transform, a lower bound for $f_{1,k}$ can be expressed as \eqref{MSE_user_speed} in the next page.

\begin{align}\label{MSE_user_speed}
&\tilde{f}_{1,k}(\mathbf{W},\boldsymbol{\phi },\mathcal{U},\mathcal{V})=\log \left( 1+v_k \right) -v_k-\!\delta _{\mathrm{U},k}^{2}\mathbf{u}_{k}^{\mathrm{H}}\mathbf{u}_k \notag
\\
&\quad-\left( 1+\kappa _{\mathrm{r},k} \right) \mathbf{u}_{k}^{\mathrm{H}}\mathbf{u}_k\mathrm{Tr}\left[ \left( \mathbf{WW}^{\mathrm{H}}\!+\!\kappa _{\mathrm{t}}\mathrm{diag}\left( \mathbf{WW}^{\mathrm{H}} \right) \right)  \right. \notag
\\
&\quad\left. \mathbf{\bar{H}}_{\mathrm{U},k} \mathbf{\bar{H}}_{\mathrm{U},k}^{\mathrm{H}} \right] +2\sqrt{\left( 1+v_k \right)}\mathrm{Re}\left\{ \mathbf{u}_{k}^{\mathrm{H}}\mathbf{\bar{H}}_{\mathrm{U},k}^{\mathrm{H}}\mathbf{w}_k \right\}.
\end{align}
The relationship between $f_{1,k}$ and $\tilde{f}_{1,k}$ is
\begin{align}
	f_{1,k}(\mathbf{W},\boldsymbol{\phi }) =\max_{\mathcal{U}, \mathcal{V}} \,  \tilde{f}_{1,k}(\mathbf{W}, \boldsymbol{\phi }, \mathcal{U}, \mathcal{V}),
\end{align}
where the optimal $\mathbf{u}_{k}^{\mathrm{opt}}$ and $v_k^{\mathrm{opt}}$ can be obtained as
{\small\begin{align}\label{optimal_U}
&\mathbf{u}_{k}^{\mathrm{opt}} \notag
\\
&=\frac{\sqrt{\left( 1+v_k \right)}\mathbf{\bar{H}}_{\mathrm{U},k}^{\mathrm{H}}\mathbf{w}_k}{\mathrm{Tr}\left[ \left( 1+\kappa _{\mathrm{r},k} \right) \left( \mathbf{WW}^{\mathrm{H}}\!+\!\kappa _{\mathrm{t}}\mathrm{diag}\left( \mathbf{WW}^{\mathrm{H}} \right) \right) \mathbf{\bar{H}}_{\mathrm{U},k}\mathbf{\bar{H}}_{\mathrm{U},k}^{\mathrm{H}} \right] +\!\delta _{\mathrm{U},k}^{2}},
\end{align}}
and
\begin{align}\label{optimal_V}
v_{k}^{\mathrm{opt}}=\gamma _k.
\end{align}

%The optimal $u_{\mathrm{U},k}$ can be obtained by setting the first-order derivative of $\tilde{f}_{1,k}$ with respect to $u_{\mathrm{U},k}$ equal to 0, which yields

As far as $f_{2,k}$ is concerned, we introduce the following lemma to obtain a lower bound.

\emph{Lemma 2 } \cite{8972400}: Consider the function $f \left( \bar{y} \right) =-\bar{y}\bar{x}+\log \bar{y}+1$ for any $\bar{x} > 0$.
Then, we have
\begin{equation}
-\log \bar{x}=\mathop{\max}_{\bar{y}>0}\,f \left( \bar{y} \right),
\end{equation}
and the optimal solution is $\bar{y}=\frac{1}{\bar{x}}$.
\hfill $\blacksquare$

The lemma shows that $f \left( \bar{y} \right)$ is a lower bound of $-\log \bar{x}$, and this bound is tight when $\bar{y}=\frac{1}{\bar{x}}$.
Let us denote $\mathcal{D}=\left\{ d_k\geqslant 0,k\in \mathcal{K} \right\}$ and define $\bar{x}=1+\frac{\left\| \mathbf{\bar{H}}_{\mathrm{E}}^{\mathrm{H}}\mathbf{w}_k \right\|_2 ^2+\mathrm{Tr}\left[ \mathbf{\Upsilon }_{\mathrm{t}}\mathbf{\bar{H}}_{\mathrm{E}}\mathbf{\bar{H}}_{\mathrm{E}}^{\mathrm{H}} \right]}{\delta _{\mathrm{E}}^{2}}$, $\bar{y}=d_k$.
Then, a lower bound for $f_{2,k}$ is given by $\tilde{f}_{2,k}$, which is defined as
\begin{align}\label{f_wave_2}
f_{2,k}(\mathbf{W},\boldsymbol{\phi })=\max_{\mathcal{D}}\,\,  \tilde{f}_{2,k}(\mathbf{W},\boldsymbol{\phi },\mathcal{D}), 
\end{align}
where
{\small\begin{align}\label{a_wave_Ek}
\tilde{f}_{2,k}(\mathbf{W},\boldsymbol{\phi },\mathcal{D})&=-d_k\left( 1+\frac{\left\| \mathbf{\bar{H}}_{\mathrm{E}}^{\mathrm{H}}\mathbf{w}_k \right\|_2 ^2+\mathrm{Tr}\left[ \mathbf{\Upsilon }_{\mathrm{t}}\mathbf{\bar{H}}_{\mathrm{E}}\mathbf{\bar{H}}_{\mathrm{E}}^{\mathrm{H}} \right]}{\delta _{\mathrm{E}}^{2}} \right) \notag
\\
 &\quad+\log d_k+1,
\end{align}}
and the optimal solution for $d_k$ is
\begin{align}\label{optimal_d}
d_{k}^{\mathrm{opt}}=\left( 1+\frac{\left\| \mathbf{\bar{H}}_{\mathrm{E}}^{\mathrm{H}}\mathbf{w}_k \right\|_2 ^2+\mathrm{Tr}\left[ \mathbf{\Upsilon }_{\mathrm{t}}\mathbf{\bar{H}}_{\mathrm{E}}\mathbf{\bar{H}}_{\mathrm{E}}^{\mathrm{H}} \right]}{\delta _{\mathrm{E}}^{2}} \right) ^{-1}.
\end{align}

Finally, to find a lower bound for $f_{3}$ that is given in a tractable analytical form, we utilize the following  lemma.

\textit{Lemma 3} \cite{zhou2021user}: Given the complex vector $\mathbf{\bar{y}}$, the function $f\left( \mathbf{\bar{y}},\mathbf{\bar{x}} \right) =\left( \left\| \mathbf{\bar{x}} \right\|_2^2+\delta ^2 \right) \left\| \mathbf{\bar{y}} \right\|_2^2-2\mathrm{Re}\left\{ \mathbf{\bar{y}}^{\mathrm{H}}\mathbf{\bar{x}} \right\} +1$ satisfies
\begin{equation}
\frac{\delta ^2}{\left\| \mathbf{\bar{x}} \right\|_2^2+\delta ^2}=\mathop {\min}_{\mathbf{\bar{y}}}f\left( \mathbf{\bar{y}},\mathbf{\bar{x}} \right),
\end{equation}
and the optimal solution is $\mathbf{\bar{y}}=\frac{\mathbf{\bar{x}}}{\left\| \mathbf{\bar{x}}\right\|_2^2+\delta ^2}$.
\hfill $\blacksquare$

The lemma provides an upper bound for $\frac{\delta ^2}{\left\| \mathbf{\bar{x}} \right\|_2^2+\delta ^2}$, which is tight when $\mathbf{\bar{y}}=\frac{\mathbf{\bar{x}}}{\left\| \mathbf{\bar{x}}\right\|_2^2+\delta ^2}$.
Then, let us introduce a new variable $\mathbf{\tilde{w}}=\mathrm{vec(}\mathbf{W})$.
Due to the complexity of $f_3$, we derive the corresponding lower bounds for the following two cases: 1) Case A: Given the other variables, $\mathbf{\tilde{w}}$ is the only variable to be optimized; 
2) Case B: Given the other variables, $\boldsymbol{\phi }$ is the only variable to be optimized.

\textit{1) Case A: Given the other variables, $\mathbf{\tilde{w}}$ is the only variable to be optimized.}
Based on Lemma 2 and Lemma 3, a lower bound for $f_3(\mathbf{\tilde{w}})$ can be obtained as stated in the following lemma.

\textit{Lemma 4}: Let us introduce the auxiliary variables $p_{\mathrm{w}}$ and $\mathbf{q}_{\mathrm{w}}$. 
A lower bound for $f_{3}(\mathbf{\tilde{w}})$ is given by
\begin{align}\label{f_3_w}
\tilde{f}_{3,\mathbf{\tilde{w}}}\left( \mathbf{\tilde{w}} \right) =-\mathbf{\tilde{w}}^{\mathrm{H}}\mathbf{\tilde{C}}_{3,\mathrm{w}}\mathbf{\tilde{w}}+2\mathrm{Re}\left\{ \mathbf{\tilde{b}}_{3,\mathrm{w}}^{\mathrm{H}}\mathbf{\tilde{w}} \right\} +\tilde{c}_{3,\mathrm{w}},
\end{align}
where
\begin{subequations}\label{f_3_w_paramter}
\begin{align}
\mathbf{\tilde{C}}_{3,\mathrm{w}}&\triangleq p_{\mathrm{w}}\left\| \mathbf{q}_{\mathrm{w}} \right\|_2 ^2\mathbf{LL}^{\mathrm{T}},
\\
\mathbf{\tilde{b}}_{3,\mathrm{w}}&\triangleq p_{\mathrm{w}}\mathbf{Lq}_{\mathrm{w}},
\\
\tilde{c}_{3,\mathrm{w}}&\triangleq -p_{\mathrm{w}}\left\| \mathbf{q}_{\mathrm{w}} \right\|_2 ^2\delta _{\mathrm{E}}^{2}-p_{\mathrm{w}}+\log p_{\mathrm{w}}+1,
\\
\mathbf{LL}^{\mathrm{T}}&\triangleq \left( \mathbf{I}_K\otimes \mathrm{diag}\left( \mathbf{\bar{H}}_{\mathrm{E}}\mathbf{\bar{H}}_{\mathrm{E}}^{\mathrm{H}} \right) \right) ,
\end{align}
\end{subequations}
and $\mathbf{I}_K$ denotes the $K \times K$ identity matrix.
Additionally, the optimal solutions for $p_{\mathrm{w}}$ and $\mathbf{q}_{\mathrm{w}}$ are given by
\begin{align}
\label{optimal_p_w}p_{\mathrm{w}}^{\mathrm{opt}}&=\left( 1+\frac{\mathrm{Tr}\left[ \mathbf{\Upsilon }_{\mathrm{t}}\mathbf{\bar{H}}_{\mathrm{E}}\mathbf{\bar{H}}_{\mathrm{E}}^{\mathrm{H}} \right]}{\delta _{\mathrm{E}}^{2}} \right) ,
\\
\label{optimal_g_w}\mathbf{q}_{\mathrm{w}}^{\mathrm{opt}}&=\frac{\mathbf{L}^{\mathrm{T}}\mathbf{\tilde{w}}}{\left\| \mathbf{L}^{\mathrm{T}}\mathbf{\tilde{w}} \right\|_2 ^2+\delta _{\mathrm{E}}^{2}}.
\end{align}

\emph{Proof:} See Appendix \ref{appendixA}.
\hfill $\blacksquare$

\textit{2) Case B: Given the other variables, $\boldsymbol{\phi }$ is the only variable to be optimized.}
Based on Lemma 1 and Lemma 2, a lower bound for $f_{3}\left( \boldsymbol{\phi } \right)$ can be obtained as stated in the following lemma.

\textit{Lemma 5}: Let us introduce the auxiliary variables $p_{\phi}$ and $\mathbf{Q}_{\phi}$, and denote $\mathbf{\hat{q}}_{\phi}$ and $\mathbf{\hat{Q}}_{\phi}$ as $\mathbf{Q}_{\phi}=\left[ \begin{matrix}
	\mathbf{\hat{q}}_{\phi}&		\mathbf{\hat{Q}}_{\phi}\\
\end{matrix} \right]$.
A lower bound for $f_{3}(\boldsymbol{\phi })$ is given by
\begin{align}\label{f_3_phi}
\tilde{f}_{3,\boldsymbol{\phi }}\left( \boldsymbol{\phi } \right) =-\boldsymbol{\phi }^{\mathrm{H}}\mathbf{\tilde{C}}_{3,\phi}\boldsymbol{\phi }+2\mathrm{Re}\left\{ \mathbf{\tilde{b}}_{3,\phi}^{\mathrm{H}}\boldsymbol{\phi } \right\} +\tilde{c}_{3,\phi},
\end{align}
where
{\begin{subequations}\label{f_3_phi_paramter_1}
\begin{align}
\mathbf{\tilde{C}}_{3,\phi}&\triangleq p_{\phi}\left\| \mathbf{Q}_{\phi} \right\| _{F}^{2}\mathbf{C}_{3,\phi}
\\
\mathbf{\tilde{b}}_{3,\phi}&\triangleq p_{\phi}\mathbf{a}_{3,\phi}^{*}
-p_{\phi}\left\| \mathbf{Q}_{\phi} \right\| _{F}^{2}\mathbf{b}_{3,\phi}
\\
\tilde{c}_{3,\phi}&\triangleq -p_{\phi}\left\| \mathbf{Q}_{\phi} \right\| _{F}^{2}\mathbf{h}_{\mathrm{BE}}^{\mathrm{H}}\mathbf{JJ}^{\mathrm{T}}\mathbf{h}_{\mathrm{BE}}-p_{\phi}\left\| \mathbf{Q}_{\phi} \right\| _{F}^{2}\delta _{\mathrm{E}}^{2} \notag
\\
&\quad+2p_{\phi}\mathrm{Re}\left\{ \mathrm{Tr}\left[ \mathbf{J\hat{q}}_{\phi}\mathbf{h}_{\mathrm{BE}}^{\mathrm{H}} \right] \right\} -p_{\phi}+\log p_{\phi}+1
\\
\mathbf{JJ}^{\mathrm{T}}&\triangleq \mathbf{\Upsilon }_{\mathrm{t}},
\end{align}
\end{subequations}}
and
{\small
\begin{subequations}\label{f_3_phi_paramter_2}
\begin{align}
	&\mathbf{C}_{3,\phi}\triangleq \left( \left( \frac{4}{\pi ^2}\mathbf{h}_{\mathrm{RE}}\mathbf{h}_{\mathrm{RE}}^{\mathrm{H}} \right) \odot \left( \mathbf{H}_{\mathrm{BR}}\mathbf{JJ}^{\mathrm{T}}\mathbf{H}_{\mathrm{BR}}^{\mathrm{H}} \right) ^{\mathrm{T}} \right) \notag
	\\
	&+\left( \left( \mathrm{diag}\left( \mathbf{h}_{\mathrm{RE}} \right) \mathbf{TT}^{\mathrm{T}}\mathrm{diag}\left( \mathbf{h}_{\mathrm{RE}}^{\mathrm{H}} \right) \right) \odot \left( \mathbf{H}_{\mathrm{BR}}\mathbf{JJ}^{\mathrm{T}}\mathbf{H}_{\mathrm{BR}}^{\mathrm{H}} \right) ^{\mathrm{T}} \right) 
	\\
	&\mathbf{b}_{3,\phi}^{\mathrm{H}}\triangleq \mathbf{h}_{\mathrm{RE}}^{\mathrm{H}}\mathrm{diag}\left( \mathbf{H}_{\mathrm{BR}}\mathbf{JJ}^{\mathrm{T}}\mathbf{h}_{\mathrm{BE}} \right) 
	\\
	&\mathbf{a}_{3,\phi}\!\triangleq \!\!\left[ \!\left[ \frac{2}{\pi }\mathbf{H}_{\mathrm{BR}}\mathbf{J\hat{q}}_{\phi}\mathbf{h}_{\mathrm{RE}}^{\mathrm{H}}+\mathbf{H}_{\mathrm{BR}}\mathbf{J\hat{Q}}_{\phi}\mathbf{T}^{\mathrm{T}}\mathrm{diag}\left( \mathbf{h}_{\mathrm{RE}}^{\mathrm{H}} \right) \! \right] _{\!1,1},...,\right. \notag
	\\
	&\left.\!\left[ \frac{2}{\pi }\mathbf{H}_{\mathrm{BR}}\mathbf{J\hat{q}}_{\phi}\mathbf{h}_{\mathrm{RE}}^{\mathrm{H}}+\mathbf{H}_{\mathrm{BR}}\mathbf{J\hat{Q}}_{\phi}\mathbf{T}^{\mathrm{T}}\mathrm{diag}\left( \mathbf{h}_{\mathrm{RE}}^{\mathrm{H}} \right) \! \right] _{M,M}\! \right] \!^{\mathrm{T}}\!.
\end{align}
\end{subequations}}

Additionally, the optimal solutions for $p_{\phi}$ and $\mathbf{Q}_{\phi}$ are given by
\begin{align}
\label{optimal_p_phi}p_{\phi}^{\mathrm{opt}}&=\left( 1+\frac{\mathrm{Tr}\left[ \mathbf{\Upsilon }_{\mathrm{t}}\mathbf{\bar{H}}_{\mathrm{E}}\mathbf{\bar{H}}_{\mathrm{E}}^{\mathrm{H}} \right]}{\delta _{\mathrm{E}}^{2}} \right) ,
\\
\label{optimal_g_phi}\mathbf{Q}_{\phi}^{\mathrm{opt}}&=\frac{\mathbf{J}^{\mathrm{T}}\mathbf{\bar{H}}_{\mathrm{E}}}{\left\| \mathbf{J}^{\mathrm{T}}\mathbf{\bar{H}}_{\mathrm{E}} \right\| _{F}^{2}+\delta _{\mathrm{E}}^{2}}
\end{align}

\emph{Proof:} See Appendix \ref{appendixB}.
\hfill $\blacksquare$ 

Thus, by denoting $\mathcal{P}=\left\{ p_{\mathrm{w}},p_{\phi} \right\}$, $\mathcal{Q}=\left\{ \mathbf{q}_{\mathrm{w}},\mathbf{Q}_{\phi} \right\} 
$, a lower bound for $f_3$ is expressed  as
\begin{align}\label{f_wave_3}
\tilde{f}_3(\mathbf{\tilde{w}},\boldsymbol{\phi },\mathcal{P},\mathcal{Q})\triangleq\begin{cases}
	\tilde{f}_{3,\mathbf{\tilde{w}}}(\mathbf{\tilde{w}},\boldsymbol{\phi },\mathcal{P},\mathcal{Q}),\,\,&\text{Case A}\,\,\\
	\tilde{f}_{3,\boldsymbol{\phi }}(\mathbf{\tilde{w}},\boldsymbol{\phi },\mathcal{P},\mathcal{Q}),\,\,&\text{Case B}\\
\end{cases}.
\end{align}

Finally, from \eqref{divide_OF}, \eqref{MSE_user_speed}, \eqref{f_wave_2} and \eqref{f_wave_3}, a lower bound for ${R}_k$ can be expressed as
{\small\begin{align}\label{new_achievable_rate}
&\tilde{R}_k\!\!=\!\!\left[ \tilde{f}_{1,k}(\mathbf{\tilde{w}},\boldsymbol{\phi },\mathcal{U},\mathcal{V}) \!+\!\tilde{f}_{2,k}(\mathbf{\tilde{w}},\boldsymbol{\phi },\mathcal{D}) \!+\!\tilde{f}_3(\mathbf{\tilde{w}},\boldsymbol{\phi },\mathcal{P},\mathcal{Q}) \right]^+,
\end{align}}
where $[a]^+\triangleq \max \left( a,0 \right)$.

The problem in \eqref{WMSR_Problem} can be reformulated as
\begin{subequations}\label{WMMSE_Problem}
\begin{align}
&\max_{\mathbf{\tilde{w}},\boldsymbol{\phi },\mathcal{U},\mathcal{V},\mathcal{D},\mathcal{P},\mathcal{Q}}
 \quad \min_{k\in \mathcal{K}}\quad \left\{ \omega _k\tilde{R}_k \right\}
\\
&\quad\quad\mathrm{s}.\mathrm{t}.\quad \mathbf{\tilde{w}}^{\mathrm{H}}\mathbf{\tilde{w}}\leqslant P,
\\
\label{19-c}&\quad\qquad\quad \boldsymbol{\phi }\in \mathcal{S}.
\end{align}
\end{subequations}

To solve the problem in \eqref{WMMSE_Problem}, we use the BCD method to alternately optimize each variable in the objective function, while keeping the other variables fixed.
The optimal solutions for $\mathcal{U}$, $\mathcal{V}$, $\mathcal{D}$,
$\mathcal{P}$ and $\mathcal{Q}$ are given in \eqref{optimal_U}, \eqref{optimal_V}, \eqref{optimal_d}, \eqref{optimal_p_w}, \eqref{optimal_p_phi}, \eqref{optimal_g_w}, and \eqref{optimal_g_phi}, respectively.
On the other hand, the optimization of the precoding vector $\mathbf{\tilde{w}}$ and the reflection coefficient vector $\boldsymbol{\phi }$ are addressed in the following sections.

\subsection{Optimization of the Precoding Vector $\mathbf{\tilde{w}}$}
In this subsection, $\mathbf{\tilde{w}}$ is optimized under the assumption that all the other variables are kept fixed.
Since the lower bound $\tilde{f}_{3,\mathbf{\tilde{w}}}$ in \eqref{f_3_w} is a quadratic function in the optimization variable, we rewrite $\tilde{f}_{1,k}\left( \mathbf{\tilde{w}} \right)$ and $\tilde{f}_{2,k}\left( \mathbf{\tilde{w}} \right)$ as quadratic functions as well.

\textit{1) Mathematical Derivation of $\tilde{f}_{1,k}\left( \mathbf{\tilde{w}} \right)$.}
Denote $\mathbf{t}_k$ as a vector whose single non-zero element is “1” at the $k$-th position.
$\tilde{f}_{1,k}\left( \mathbf{\tilde{w}} \right)$ in \eqref{MSE_user_speed} can be reformulated as
{\small\begin{align} 
&\tilde{f}_{1,k}\left( \mathbf{\tilde{w}} \right) =2\sqrt{\left( 1+v_k \right)}\mathrm{Re}\left\{ \mathrm{Tr}\left[ \mathbf{t}_k\mathbf{u}_{k}^{\mathrm{H}}\mathbf{\bar{H}}_{\mathrm{U},k}^{\mathrm{H}}\mathbf{W} \right] \right\} +\tilde{c}_{1,\mathrm{w},k} \notag
\\
&-\!\left( 1\!+\!\kappa _{\mathrm{r},k} \right) \mathbf{u}_{k}^{\mathrm{H}}\mathbf{u}_k\mathrm{Tr}\left[ \mathbf{W}^{\mathrm{H}}\left( \mathbf{\bar{H}}_{\mathrm{U},k}\mathbf{\bar{H}}_{\mathrm{U},k}^{\mathrm{H}}\!+\mathrm{diag}\left( \mathbf{\bar{H}}_{\mathrm{U},k}\mathbf{\bar{H}}_{\mathrm{U},k}^{\mathrm{H}} \right) \right) \mathbf{W} \right] \notag
\\
&=2\mathrm{Re}\left\{ \mathrm{Tr}\left[ \mathbf{B}_{1,\mathrm{w},k}\mathbf{W} \right] \right\} -\mathrm{Tr}\left[ \mathbf{W}^{\mathrm{H}}\mathbf{C}_{1,\mathrm{w},k}\mathbf{W} \right] +\tilde{c}_{1,\mathrm{w},k},
\end{align}}
where
{\small\begin{subequations}
\begin{align}
&\mathbf{B}_{1,\mathrm{w},k}\!\triangleq\! \sqrt{\left( 1+v_k \right)}\mathbf{t}_k\mathbf{u}_{k}^{\mathrm{H}}\mathbf{\bar{H}}_{\mathrm{U},k}^{\mathrm{H}}
\\
&\mathbf{C}_{1,\mathrm{w},k}\!\triangleq\! \left( 1+\kappa _{\mathrm{r},k} \right) \mathbf{u}_{k}^{\mathrm{H}}\mathbf{u}_k( \mathbf{\bar{H}}_{\mathrm{U},k}\mathbf{\bar{H}}_{\mathrm{U},k}^{\mathrm{H}}+\mathrm{diag}( \mathbf{\bar{H}}_{\mathrm{U},k}\mathbf{\bar{H}}_{\mathrm{U},k}^{\mathrm{H}} ) ) 
\\
&\tilde{c}_{1,\mathrm{w},k}\!\triangleq\! \log \left( 1+v_k \right) -\eta _k-\!\delta _{\mathrm{U},k}^{2}\mathbf{u}_{k}^{\mathrm{H}}\mathbf{u}_k.
\end{align}
\end{subequations}}
Then, by using the identity $\mathrm{Tr}\left[ \mathbf{A}^{\mathrm{T}}\mathbf{D} \right] =\left( \mathrm{vec}\left( \mathbf{A} \right) \right) ^{\mathrm{T}}\mathrm{vec}\left( \mathbf{D} \right) 
$ \cite{Zhang2017Matrix} and $\mathrm{Tr}\left[ \mathbf{ABC} \right] =\left( \mathrm{vec}\left( \mathbf{A}^{\mathrm{T}} \right) \right) ^{\mathrm{T}}\left( \mathbf{I} \otimes \mathbf{B} \right) \mathrm{vec}\left( \mathbf{C} \right)$, we have 
\begin{align}\label{R_U_w_wave}
\tilde{f}_{1,k}\left( \mathbf{\tilde{w}} \right) =2\mathrm{Re}\left\{ \mathbf{\tilde{b}}_{1,\mathrm{w},k}^{\mathrm{H}}\mathbf{\tilde{w}} \right\} -\mathbf{\tilde{w}}^{\mathrm{H}}\mathbf{\tilde{C}}_{1,\mathrm{w},k}\mathbf{\tilde{w}}+\tilde{c}_{1,\mathrm{w},k},
\end{align}
where
\begin{subequations}
\begin{align}
	\mathbf{\tilde{b}}_{1,\mathrm{w},k}&\triangleq \mathrm{vec}\left( \mathbf{B}_{1,\mathrm{w},k}^{\mathrm{H}} \right),
	\\
	\mathbf{\tilde{C}}_{1,\mathrm{w},k}&\triangleq \mathbf{I}_K\otimes \mathbf{C}_{1,\mathrm{w},k}.
\end{align}
\end{subequations}

\textit{2) Mathematical Derivation of $\tilde{f}_{2,k}\left( \mathbf{\tilde{w}} \right)$.}
By using the identity $\mathrm{Tr}\left[ \mathbf{ABCD} \right] =\left( \mathrm{vec}\left( \mathbf{D}^{\mathrm{T}} \right) \right) ^{\mathrm{T}}\left( \mathbf{C}^{\mathrm{T}}\otimes \mathbf{A} \right) \mathrm{vec}\left( \mathbf{B} \right)$ \cite{Zhang2017Matrix}, $\tilde{f}_{2,k}\left( \mathbf{\tilde{w}} \right)$ in \eqref{a_wave_Ek} can be reformulated as
{\small\begin{align}
\tilde{f}_{2,k}\left( \mathbf{\tilde{w}} \right) &=-\frac{d_k}{\delta _{\mathrm{E},k}^{2}}\left( \mathrm{Tr}\left[ \mathbf{\bar{H}}_{\mathrm{E}}\mathbf{\bar{H}}_{\mathrm{E}}^{\mathrm{H}}\mathbf{Wt}_k\mathbf{t}_{k}^{\mathrm{H}}\mathbf{W}^{\mathrm{H}} \right] \right. \notag
\\
&\quad \left.+\kappa _{\mathrm{t}}\mathrm{Tr}\left[ \mathbf{W}^{\mathrm{H}}\mathrm{diag}\left( \mathbf{\bar{H}}_{\mathrm{E}}\mathbf{\bar{H}}_{\mathrm{E}}^{\mathrm{H}}
 \right) \mathbf{W} \right] \right) +\log d_k+1-d_k \notag
\\
&=-\frac{d_k}{\delta _{\mathrm{E},k}^{2}}\left( \mathbf{\tilde{w}}^{\mathrm{H}}\left( \left( \mathbf{t}_k\mathbf{t}_{k}^{\mathrm{H}} \right) ^{\mathrm{T}}\otimes \left( \mathbf{\bar{H}}_{\mathrm{E}}\mathbf{\bar{H}}_{\mathrm{E}}^{\mathrm{H}} \right) \right) \mathbf{\tilde{w}} \right. \notag
\\
&\quad \left.+\kappa _{\mathrm{t}}\mathbf{\tilde{w}}^{\mathrm{H}}\left( \mathbf{I}_K\otimes \mathrm{diag}\left( \mathbf{\bar{H}}_{\mathrm{E}}\mathbf{\bar{H}}_{\mathrm{E}}^{\mathrm{H}} \right) \right) \mathbf{\tilde{w}} \right) +\log d_k+1-d_k \notag
\\
&=-\mathbf{\tilde{w}}^{\mathrm{H}}\mathbf{\tilde{C}}_{2,\mathrm{w},k}\mathbf{\tilde{w}}+\tilde{c}_{2,\mathrm{w},k}, \label{a_wave_Ek_w_wave}
\end{align}}
where
{\small
\begin{subequations}
\begin{align}
&\mathbf{\tilde{C}}_{2,\mathrm{w},k}\!\triangleq\! \frac{d_k}{\delta _{\mathrm{E}}^{2}}( ( \mathbf{t}_k\mathbf{t}_{k}^{\mathrm{H}} )^{\mathrm{T}}\! \otimes \!( \mathbf{\bar{H}}_{\mathrm{E}}\mathbf{\bar{H}}_{\mathrm{E}}^{\mathrm{H}} ) )\!+\!\kappa _{\mathrm{t}}( \mathbf{I}_{K}\otimes \mathrm{diag}( \mathbf{\bar{H}}_{\mathrm{E}}\mathbf{\bar{H}}_{\mathrm{E}}^{\mathrm{H}} ) ),
\\
&\tilde{c}_{2,\mathrm{w},k}\!\triangleq\! \log d_k+1-d_k.
\end{align}
\end{subequations}}

Substituting \eqref{f_3_w}, \eqref{R_U_w_wave} and \eqref{a_wave_Ek_w_wave} into \eqref{WMMSE_Problem}, the subproblem for $\mathbf{\tilde{w}}$ can be transformed into the following equivalent problem
\begin{subequations}\label{WMMSE_Problem_W}
\begin{align}
\max_{\mathbf{\tilde{w}}} \quad& \min_{k\in \mathcal{K}} 
 \quad \left\{ \tilde{r}_{\mathrm{w},k}\left( \mathbf{\tilde{w}} \right) \right\}
\\
\mathrm{s}.\mathrm{t}.\quad& \mathbf{\tilde{w}}^{\mathrm{H}}\mathbf{\tilde{w}}\leqslant P,
\end{align}
\end{subequations}
where
\begin{align}
\tilde{r}_{\mathrm{w},k}\left( \mathbf{\tilde{w}} \right) =-\mathbf{\tilde{w}}^{\mathrm{H}}\mathbf{\tilde{C}}_{\mathrm{w},k}\mathbf{\tilde{w}}+2\mathrm{Re}\left\{ \mathbf{\tilde{b}}_{\mathrm{w},k}^{\mathrm{H}}\mathbf{\tilde{w}} \right\} +\tilde{c}_{\mathrm{w},k},
\end{align}
and $\mathbf{\tilde{C}}_{\mathrm{w},k}$, $\mathbf{\tilde{b}}_{\mathrm{w},k}$ and $\tilde{c}_{\mathrm{w},k}$ are defined, respectively, as follows
{\small\begin{subequations}
\begin{align}
\mathbf{\tilde{C}}_{\mathrm{w},k}&\triangleq \omega _k\left( \mathbf{\tilde{C}}_{1,\mathrm{w},k}+\mathbf{\tilde{C}}_{2,\mathrm{w},k}+\mathbf{\tilde{C}}_{3,\mathrm{w}} \right) ,
\\
\mathbf{\tilde{b}}_{\mathrm{w},k}&\triangleq \omega _k\left( \mathbf{\tilde{b}}_{1,\mathrm{w},k}+\mathbf{\tilde{b}}_{3,\mathrm{w}} \right) ,
\\
\tilde{c}_{\mathrm{w},k}&\triangleq \omega _k\left( \tilde{c}_{1,\mathrm{w},k}+\tilde{c}_{2,\mathrm{w},k}+\tilde{c}_{3,\mathrm{w}} \right) .
\end{align}
\end{subequations}}
Finally, by introducing the auxiliary variable $\delta_{\mathrm{w}}$, the optimization problem in \eqref{WMMSE_Problem_W} can be reformulated as
\begin{subequations}\label{r_k_w_wave_SOCP}
\begin{align}
\max_{\mathbf{\tilde{w}},\delta_{\mathrm{w}}} \quad &\delta_{\mathrm{w}}
\\
\mathrm{s}.\mathrm{t}. \quad &\tilde{r}_{\mathrm{w},k}\left( \mathbf{\tilde{w}} \right) \geqslant \delta_{\mathrm{w}},\forall k\in \mathcal{K},
\\
 \quad &\mathbf{\tilde{w}}^{\mathrm{H}}\mathbf{\tilde{w}}\leqslant P.
\end{align}
\end{subequations}
The obtained reformulation in \eqref{r_k_w_wave_SOCP} is an SOCP problem whose globally optimum solution $\mathbf{\tilde{w}}$ can be obtained by using standard numerical optimization methods, such as CVX.

\vspace{-0.4cm}
\subsection{Optimization of the Reflection Coefficient Vector $\boldsymbol{\phi }$}
In this subsection, $\boldsymbol{\phi }$ is optimized under the assumption that all the other variables are kept fixed.

The lower bound $\tilde{f}_{3,\boldsymbol{\phi }}$  in \eqref{f_3_phi} is a quadratic function in the optimization variable.
Therefore, we rewrite $\tilde{f}_{1,k}\left( \boldsymbol{\phi } \right)$ and $\tilde{f}_{2,k}\left( \boldsymbol{\phi } \right)$ as quadratic functions.

\textit{1) Mathematical Derivation of $\tilde{f}_{1,k}\left( \boldsymbol{\phi } \right)$.}
First of all, $\mathbf{u}_{k}^{\mathrm{H}}\mathbf{\bar{H}}_{\mathrm{U},k}^{\mathrm{H}}\mathbf{w}_k$ can be rewritten as
\begin{align}
&\mathbf{u}_{k}^{\mathrm{H}}\mathbf{\bar{H}}_{\mathrm{U},k}^{\mathrm{H}}\mathbf{w}_k=\left[ \begin{matrix}
	u_{\phi ,k}^{*}&		\mathbf{u}_{\phi ,k}^{\mathrm{H}}\\
\end{matrix} \right] \left[ \begin{array}{c}
	\mathbf{\hat{h}}_{\mathrm{U},k}^{\mathrm{H}}\\
	\mathbf{\hat{H}}_{\mathrm{U},k}^{\mathrm{H}}\\
\end{array} \right] \mathbf{w}_k \notag
\\
&=u_{\phi ,k}^{*}\frac{2}{\pi}\mathbf{h}_{\mathrm{RU},k}^{\mathrm{H}}\mathbf{\Phi H}_{\mathrm{BR}}\mathbf{w}_k+u_{\phi ,k}^{*}\mathbf{h}_{\mathrm{BU},k}^{\mathrm{H}}\mathbf{w}_k \notag
\\
&\quad+\mathbf{u}_{\phi ,k}^{\mathrm{H}}\mathbf{T}^{\mathrm{T}}\mathrm{diag}\left( \mathbf{h}_{\mathrm{RU},k}^{\mathrm{H}} \right) \mathbf{\Phi H}_{\mathrm{BR}}\mathbf{w}_k \notag
\\
&=( u_{\phi ,k}^{*}\frac{2}{\pi}\mathbf{h}_{\mathrm{RU},k}^{\mathrm{H}}\mathrm{diag}\left( \mathbf{H}_{\mathrm{BR}}\mathbf{w}_k \right) +\mathbf{u}_{\phi ,k}^{\mathrm{H}}\mathbf{T}^{\mathrm{T}}\mathrm{diag}\left( \mathbf{h}_{\mathrm{RU},k}^{\mathrm{H}} \right)  \notag 
\\
&\quad \mathrm{diag}\left( \mathbf{H}_{\mathrm{BR}}\mathbf{w}_k \right) ) \boldsymbol{\phi }+u_{\phi ,k}^{*}\mathbf{h}_{\mathrm{BU},k}^{\mathrm{H}}\mathbf{w}_k \notag
\\
&=\mathbf{a}_{1,\phi ,k}^{\mathrm{H}}
\boldsymbol{\phi }+u_{1,\phi ,k}^{*}\mathbf{h}_{\mathrm{BU},k}^{\mathrm{H}}\mathbf{w}_k,
\end{align}
where
{\small\begin{align}
\mathbf{a}_{1,\phi ,k}^{\mathrm{H}}
&\triangleq u_{\phi ,k}^{*}\frac{2}{\pi}\mathbf{h}_{\mathrm{RU},k}^{\mathrm{H}}\mathrm{diag}\left( \mathbf{H}_{\mathrm{BR}}\mathbf{w}_k \right) \notag
\\
&\quad+\mathbf{u}_{\phi ,k}^{\mathrm{H}}\mathbf{T}^{\mathrm{T}}\mathrm{diag}\left( \mathbf{h}_{\mathrm{RU},k}^{\mathrm{H}} \right) \mathrm{diag}\left( \mathbf{H}_{\mathrm{BR}}\mathbf{w}_k \right). 
\end{align}}
Denote {\small$\mathbf{A}_{1,\phi ,k}\triangleq \left( 1+\kappa _{\mathrm{r},k} \right) \mathbf{u}_{k}^{\mathrm{H}}\mathbf{u}_k\left( \mathbf{WW}^{\mathrm{H}}\!+\!\kappa _{\mathrm{t}}\mathrm{diag}\left( \mathbf{WW}^{\mathrm{H}} \right) \right) $}.
By using the matrix identity in \cite[Eq. (1.10.6)]{Zhang2017Matrix}, we have 
{\small\begin{align}
&\mathrm{Tr}\left[ \mathbf{\bar{H}}_{\mathrm{U},k}^{\mathrm{H}}\mathbf{A}_{1,\phi ,k}\mathbf{\bar{H}}_{\mathrm{U},k} \right] =\mathbf{A}_{1,\phi ,k}\mathbf{\hat{h}}_{\mathrm{U},k}\mathbf{\hat{h}}_{\mathrm{U},k}^{\mathrm{H}}+\mathbf{A}_{1,\phi ,k}\mathbf{\hat{H}}_{\mathrm{U},k}\mathbf{\hat{H}}_{\mathrm{U},k}^{\mathrm{H}} \notag
\\
&=\frac{4}{\pi ^2}\mathrm{Tr}\left[ \mathbf{\Phi }^{\mathrm{H}}\mathbf{h}_{\mathrm{RU},k}\mathbf{h}_{\mathrm{RU},k}^{\mathrm{H}}\mathbf{\Phi H}_{\mathrm{BR}}\mathbf{A}_{1,\phi ,k}\mathbf{H}_{\mathrm{BR}}^{\mathrm{H}} \right] \notag 
\\
&\quad+\mathrm{Tr}\left[ \mathbf{\Phi }^{\mathrm{H}}\mathrm{diag}\left( \mathbf{h}_{\mathrm{RU},k} \right) \mathbf{TT}^{\mathrm{T}}\mathrm{diag}\left( \mathbf{h}_{\mathrm{RU},k}^{\mathrm{H}} \right) \mathbf{\Phi H}_{\mathrm{BR}}\mathbf{A}_{1,\phi ,k}\mathbf{H}_{\mathrm{BR}}^{\mathrm{H}} \right] \notag 
\\
&\quad+2\mathrm{Re}\left\{ \frac{2}{\pi}\mathbf{h}_{\mathrm{RU},k}^{\mathrm{H}}\mathbf{\Phi H}_{\mathrm{BR}}\mathbf{A}_{1,\phi ,k}\mathbf{h}_{\mathrm{BU},k} \right\} +\mathbf{h}_{\mathrm{BU},k}^{\mathrm{H}}\mathbf{A}_{1,\phi ,k}\mathbf{h}_{\mathrm{BU},k} \notag
\\
&=\boldsymbol{\phi }^{\mathrm{H}}\mathbf{\tilde{C}}_{1,\phi ,k}\boldsymbol{\phi }+2\mathrm{Re}\left\{ \mathbf{b}_{1,\phi ,k}^{\mathrm{H}}\boldsymbol{\phi } \right\} +\mathbf{h}_{\mathrm{BU},k}^{\mathrm{H}}\mathbf{A}_{1,\phi ,k}\mathbf{h}_{\mathrm{BU},k},
\end{align}}
where
{\small
\begin{subequations}
\begin{align}
&\mathbf{\tilde{C}}_{1,\phi ,k}\triangleq ( ( \frac{4}{\pi ^2}\mathbf{h}_{\mathrm{RU},k}\mathbf{h}_{\mathrm{RU},k}^{\mathrm{H}} ) \odot ( \mathbf{H}_{\mathrm{BR}}\mathbf{A}_{1,\phi ,k}\mathbf{H}_{\mathrm{BR}}^{\mathrm{H}} ) ^{\mathrm{T}} ) 
\\
&+\!((\mathrm{diag}\left( \mathbf{h}_{\mathrm{RU},k} \right) \mathbf{TT}^{\mathrm{T}}\mathrm{diag}( \mathbf{h}_{\mathrm{RU},k}^{\mathrm{H}} ) ) \!\odot \!( \mathbf{H}_{\mathrm{BR}}\mathbf{A}_{1,\phi ,k}\mathbf{H}_{\mathrm{BR}}^{\mathrm{H}} ) ^{\mathrm{T}}\! ), \notag
\\
&\mathbf{b}_{1,\phi ,k}^{\mathrm{H}}\triangleq \frac{2}{\pi}\mathbf{h}_{\mathrm{RU},k}^{\mathrm{H}}\mathrm{diag}\left( \mathbf{H}_{\mathrm{BR}}\mathbf{A}_{1,\phi ,k}\mathbf{h}_{\mathrm{BU},k} \right). 
\end{align}
\end{subequations}}
Then, $\tilde{f}_{1,k}\left( \boldsymbol{\phi } \right)$ in \eqref{MSE_user_speed} can be reformulated as
{\small\begin{align}\label{R_U_phi}
\tilde{f}_{1,k}\left( \boldsymbol{\phi } \right) &=2\sqrt{\left( 1+\eta _k \right)}\mathrm{Re}\left\{ \mathbf{u}_{k}^{\mathrm{H}}\mathbf{\bar{H}}_{\mathrm{U},k}^{\mathrm{H}}\mathbf{w}_k \right\} \!-\!\mathrm{Tr}\left[ \mathbf{\bar{H}}_{\mathrm{U},k}^{\mathrm{H}}\mathbf{A}_{1,\phi ,k}\mathbf{\bar{H}}_{\mathrm{U},k} \right] \notag
\\
&\quad+\log \left( 1+\eta _k \right) -\eta _k-\!\delta _{\mathrm{U},k}^{2}\mathbf{u}_{k}^{\mathrm{H}}\mathbf{u}_k \notag
\\
&=2\mathrm{Re}\left\{ \mathbf{\tilde{b}}_{1,\phi ,k}^{\mathrm{H}}\boldsymbol{\phi } \right\} -\boldsymbol{\phi }^{\mathrm{H}}\mathbf{\tilde{C}}_{1,\phi ,k}\boldsymbol{\phi }+\tilde{c}_{1,\phi,k}
,
\end{align}}
where
\begin{subequations}
\begin{align}
&\mathbf{\tilde{b}}_{1,\phi ,k}\triangleq \sqrt{\left( 1+\eta _k \right)}\mathbf{a}_{1,\phi ,k}-\mathbf{b}_{1,\phi ,k},
\\
&\tilde{c}_{1,\phi ,k}\triangleq \log \left( 1+\eta _k \right) -\eta _k-\!\delta _{\mathrm{U},k}^{2}\mathbf{u}_{k}^{\mathrm{H}}\mathbf{u}_k-\mathbf{h}_{\mathrm{BU},k}^{\mathrm{H}}\mathbf{A}_{1,\phi ,k}\mathbf{h}_{\mathrm{BU},k} \notag
\\
&\quad+2\sqrt{\left( 1+\eta _k \right)}\mathrm{Re}\left\{ u_{\phi ,k}^{*}\mathbf{h}_{\mathrm{BU},k}^{\mathrm{H}}\mathbf{w}_k \right\}.
\end{align}
\end{subequations}

\textit{2) Mathematical Derivation of $\tilde{f}_{2,k}\left( \boldsymbol{\phi } \right)$.}
Similarly, denote $\mathbf{A}_{2,\phi ,k}\triangleq \mathbf{w}_k\mathbf{w}_{k}^{\mathrm{H}}+\kappa _{\mathrm{t}}\mathrm{diag}( \mathbf{WW}^{\mathrm{H}} )$, $\mathrm{Tr}\left[ \mathbf{\bar{H}}_{\mathrm{E}}^{\mathrm{H}}\mathbf{A}_{2,\phi ,k}\mathbf{\bar{H}}_{\mathrm{E}} \right] $ can be rewritten as
{\small\begin{align}
&\mathrm{Tr}\left[ \mathbf{\bar{H}}_{\mathrm{E}}^{\mathrm{H}}\mathbf{A}_{2,\phi ,k}\mathbf{\bar{H}}_{\mathrm{E}} \right] =\mathbf{A}_{2,\phi ,k}\mathbf{\hat{h}}_{\mathrm{E}}\mathbf{\hat{h}}_{\mathrm{E}}^{\mathrm{H}}+\mathbf{A}_{2,\phi ,k}\mathbf{\hat{H}}_{\mathrm{E}}\mathbf{\hat{H}}_{\mathrm{E}}^{\mathrm{H}} \notag
\\
&=\frac{4}{\pi ^2}\mathrm{Tr}\left[ \mathbf{\Phi }^{\mathrm{H}}\mathbf{h}_{\mathrm{RE}}\mathbf{h}_{\mathrm{RE}}^{\mathrm{H}}\mathbf{\Phi H}_{\mathrm{BR}}\mathbf{A}_{2,\phi ,k}\mathbf{H}_{\mathrm{BR}}^{\mathrm{H}} \right] \notag
\\
&\quad+\mathrm{Tr}\left[ \mathbf{\Phi }^{\mathrm{H}}\mathrm{diag}\left( \mathbf{h}_{\mathrm{RE}} \right) \mathbf{TT}^{\mathrm{T}}\mathrm{diag}\left( \mathbf{h}_{\mathrm{RE}}^{\mathrm{H}} \right) \mathbf{\Phi H}_{\mathrm{BR}}\mathbf{A}_{2,\phi ,k}\mathbf{H}_{\mathrm{BR}}^{\mathrm{H}} \right] \notag
\\
&\quad+2\mathrm{Re}\left\{ \frac{2}{\pi}\mathbf{h}_{\mathrm{RE}}^{\mathrm{H}}\mathbf{\Phi H}_{\mathrm{BR}}\mathbf{A}_{2,\phi ,k}\mathbf{h}_{\mathrm{BE}} \right\} +\mathbf{h}_{\mathrm{BE}}^{\mathrm{H}}\mathbf{A}_{2,\phi ,k}\mathbf{h}_{\mathrm{BE}} \notag
\\
&=\boldsymbol{\phi }^{\mathrm{H}}\mathbf{C}_{2,\phi ,k}\boldsymbol{\phi }+2\mathrm{Re}\left\{ \mathbf{b}_{2,\phi ,k}^{\mathrm{H}}\boldsymbol{\phi } \right\} +\mathbf{h}_{\mathrm{BE}}^{\mathrm{H}}\mathbf{A}_{2,\phi ,k}\mathbf{h}_{\mathrm{BE}},
\end{align}}
where
{\small\begin{subequations}
\begin{align}
&\mathbf{C}_{2,\phi ,k}\triangleq ( ( \frac{4}{\pi ^2}\mathbf{h}_{\mathrm{RE}}\mathbf{h}_{\mathrm{RE}}^{\mathrm{H}} ) \odot ( \mathbf{H}_{\mathrm{BR}}\mathbf{A}_{2,\phi ,k}\mathbf{H}_{\mathrm{BR}}^{\mathrm{H}} ) ^{\mathrm{T}} ) 
\\
&\quad+( ( \mathrm{diag}( \mathbf{h}_{\mathrm{RE}} ) \mathbf{TT}^{\mathrm{T}}\mathrm{diag}( \mathbf{h}_{\mathrm{RE}}^{\mathrm{H}} ) ) \odot ( \mathbf{H}_{\mathrm{BR}}\mathbf{A}_{2,\phi ,k}\mathbf{H}_{\mathrm{BR}}^{\mathrm{H}} ) ^{\mathrm{T}} ), \notag
\\
&\mathbf{b}_{2,\phi ,k}^{\mathrm{H}}\triangleq \frac{2}{\pi}\mathbf{h}_{\mathrm{RE}}^{\mathrm{H}}\mathrm{diag}\left( \mathbf{H}_{\mathrm{BR}}\mathbf{A}_{2,\phi ,k}\mathbf{h}_{\mathrm{BE}} \right). 
\end{align}
\end{subequations}}
Then, $\tilde{f}_{2,k}\left( \boldsymbol{\phi } \right)$ in \eqref{a_wave_Ek} can be reformulated as
{\small\begin{align}\label{a_wave_Ek_phi}
\tilde{f}_{2,k}(\boldsymbol{\phi })&=-\frac{d_k}{\delta _{\mathrm{E}}^{2}}\left( \mathrm{Tr}\left[ \mathbf{\bar{H}}_{\mathrm{E}}^{\mathrm{H}}\mathbf{A}_{2,\phi ,k}\mathbf{\bar{H}}_{\mathrm{E}} \right] \right) +\log d_k+1-d_k \notag
\\
&=-\boldsymbol{\phi }^{\mathrm{H}}\mathbf{\tilde{C}}_{2,\phi ,k}\boldsymbol{\phi }-2\mathrm{Re}\left\{ \mathbf{\tilde{b}}_{2,\phi ,k}^{\mathrm{H}}\boldsymbol{\phi } \right\} +\tilde{c}_{2,\phi ,k}, 
\end{align}}
where
{\small\begin{subequations}
\begin{align}
\mathbf{\tilde{C}}_{2,\phi ,k}&\triangleq \frac{d_k}{\delta _{\mathrm{E}}^{2}}\mathbf{C}_{2,\phi ,k}
\\
\mathbf{\tilde{b}}_{2,\phi ,k}&\triangleq \frac{d_k}{\delta _{\mathrm{E}}^{2}}\mathbf{b}_{2,\phi ,k}
\\
\tilde{c}_{2,\phi ,k}&\triangleq \log d_k+1-d_k-\frac{d_k}{\delta _{\mathrm{E}}^{2}}\mathbf{h}_{\mathrm{BE}}^{\mathrm{H}}\mathbf{A}_{2,\phi ,k}\mathbf{h}_{\mathrm{BE}}
\end{align}
\end{subequations}}
By substituting \eqref{f_3_phi}, \eqref{R_U_phi} and \eqref{a_wave_Ek_phi} into \eqref{WMMSE_Problem}, the optimization subproblem for $\boldsymbol{\phi }$ is equivalent to
\begin{subequations}\label{WMMSE_Problem_phi}
\begin{align}
\max_{\boldsymbol{\phi }} \quad& \min_{k\in \mathcal{K}} 
 \quad \left\{ \tilde{r}_{\phi,k}\left( \boldsymbol{\phi } \right) \right\}
\\
\mathrm{s}.\mathrm{t}.\quad& \boldsymbol{\phi }\in \mathcal{S},
\end{align}
\end{subequations}
where
\begin{align}
\tilde{r}_{\phi,k}\left( \boldsymbol{\phi } \right) =-\boldsymbol{\phi }^{\mathrm{H}}\mathbf{\tilde{C}}_{\phi ,k}\boldsymbol{\phi }+2\mathrm{Re}\left\{ \mathbf{\tilde{b}}_{\phi ,k}^{\mathrm{H}}\boldsymbol{\phi } \right\} +\tilde{c}_{\phi ,k},
\end{align}
and $\mathbf{\tilde{C}}_{\boldsymbol{\phi },k}$, $\mathbf{\tilde{b}}_{\boldsymbol{\phi },k}$ and $\tilde{c}_{\boldsymbol{\phi },k}$ are, respectively, given by
\begin{subequations}
\begin{align}
\mathbf{\tilde{C}}_{\boldsymbol{\phi },k}&\triangleq \omega _k( \mathbf{\tilde{C}}_{1,\boldsymbol{\phi },k}+\mathbf{\tilde{C}}_{2,\boldsymbol{\phi },k}+\mathbf{\tilde{C}}_{3,\boldsymbol{\phi }} ) ,
\\
\mathbf{\tilde{b}}_{\boldsymbol{\phi },k}&\triangleq \omega _k( \mathbf{\tilde{b}}_{1,\boldsymbol{\phi },k}-\mathbf{\tilde{b}}_{2,\boldsymbol{\phi },k}+\mathbf{\tilde{b}}_{3,\boldsymbol{\phi }} ) ,
\\
\tilde{c}_{\boldsymbol{\phi },k}&\triangleq \omega _k\left( \tilde{c}_{1,\boldsymbol{\phi },k}+\tilde{c}_{2,\boldsymbol{\phi },k}+\tilde{c}_{3,\boldsymbol{\phi }} \right).
\end{align}
\end{subequations}

By introducing the auxiliary variable $\delta _{\phi}$, the problem in \eqref{WMMSE_Problem_phi} can be rewritten as
\begin{subequations}\label{r_k_phi}
\begin{align}
\max_{\boldsymbol{\phi },\delta _{\phi}} \quad&\delta _{\phi}
\\
\mathrm{s}.\mathrm{t}. \quad&\tilde{r}_{\phi,k}\left( \boldsymbol{\phi } \right) \geqslant \delta _{\phi}, \forall k\in \mathcal{K},
\\
\label{30-b}&\boldsymbol{\phi }\in \mathcal{S}.
\end{align}
\end{subequations}
Due to the non-convex unit-modulus constraints in \eqref{30-b}, the problem in \eqref{r_k_phi} is still  non-convex.
Furthermore, if the semidefinite relaxation (SDR) is used to relax the problem with rank-1 constraint, it would be difficult to obtain a good solution for a phase-only beamforming problem by Gaussian randomization \cite{8334295}.
Hence, the penalty CCP \cite{9110587} is used to tackle this issue.
First, the constraints \eqref{r_k_phi} can be equivalently rewritten as $1\leqslant \left| \phi _m \right|^2\leqslant 1, \forall m \in \mathcal{M}$.
And then, the non-convex parts can be linearized by $\left| \phi _{m}^{[t]} \right|^2-2\mathrm{Re}\left( \phi _{m}^{*}\phi _{m}^{[t]} \right) \leqslant -1$ at fixed $\phi _{m}^{[t]}$ in the $t$-th iteration. 
By introducing a set of slack variables $\mathbf{b}=\left[ b_1,\dots ,b_{2M} \right] ^{\mathrm{T}}$ and a penalty multiplier $\lambda$, Problem \eqref{r_k_phi} can be reformulated as
\begin{subequations}\label{r_k_phi_SOCP}
\begin{align}
\max_{\boldsymbol{\phi },\delta _{\phi},\mathbf{b}} \quad& \delta _{\phi}-\lambda \sum_{m=1}^{2M}{b_m}
\\
\mathrm{s}.\mathrm{t}.\quad& \tilde{r}_k\left( \boldsymbol{\phi } \right) \geqslant \delta _{\phi},\forall k\in \mathcal{K},
\\
&\left| \phi _{m}^{[t]} \right|^2-2\mathrm{Re}\left( \phi _{m}^{*}\phi _{m}^{[t]} \right) \leqslant b_m-1,\forall m \in \mathcal{M},
\\
&\left| \phi _m \right|^2\leqslant 1+b_{M+m}
\\
&\mathbf{b}\geqslant 0.
\end{align}
\end{subequations}
The problem in \eqref{r_k_phi_SOCP} is an SOCP optimization problem, which can be solved by using conventional numerical optimization tools, such as CVX.
The detail of the proposed penalty CCP algorithm for solving Problem \eqref{r_k_phi_SOCP} is summarized in Algorithm 1.
More specifically, $\left\| \boldsymbol{\phi }^{[t]}-\boldsymbol{\phi }^{ [t-1] } \right\| _1 \leqslant \varepsilon _1$ controls the convergence of Algorithm 1, and $\left\| \mathbf{b} \right\| _1 \leqslant \varepsilon _2$ guarantees the unit-modulus constraints in Problem \eqref{r_k_phi} when $\varepsilon _2$ is sufficiently small.
Additionally, the maximum value $\lambda _{\max}$ is introduced to avoid the numerical difficulties caused by a large $\lambda$.

\begin{algorithm}[t] %算法开始
	\caption{Penalty CCP-Based Optimization for RIS Reflection Phase Shifts} %算法的题目
	\label{CCP Algorithm} %算法的标签	
	\textbf{Initialize}: Initialize  $\boldsymbol{\phi }^{[0]}=\boldsymbol{\phi }^{(n)}$, $\gamma > 1$, and set $t = 0$
	\begin{algorithmic}[1]
		\While{$\left\| \boldsymbol{\phi }^{ [t]}-\boldsymbol{\phi }^{[t-1]} \right\| _1>\varepsilon _1$ or $\left\| \mathbf{b} \right\| _1>\varepsilon _2$}
		\State Update $\boldsymbol{\phi }^{[t+1]}$ from Problem \eqref{r_k_phi_SOCP};
		\State $\lambda ^{[t+1]}=\min \left\{ \gamma \lambda ^{[t]},\lambda _{\max} \right\}$;
		\State Set $t \gets t+1$
		\EndWhile	
		\State Output $\boldsymbol{\phi }^{\left( n+1 \right)}$ = $\boldsymbol{\phi }^{[t]}$
	\end{algorithmic}
\end{algorithm}

\begin{algorithm}[t] %算法开始
\caption{BCD-SOCP Algorithm} %算法的题目
\label{BCD-SOCP Algorithm} %算法的标签	
\textbf{Initialize}: Initialize  $\mathbf{\tilde{w}}^{(0)}$, $\boldsymbol{\phi }^{(0)}$ to feasible values and set $n$=0
\begin{algorithmic}[1]
	\While{The value of the objective fuction in \eqref{WMMSE_Problem} has not converged}
	\State Given $\mathbf{\tilde{w}}^{(n)}$ and $\boldsymbol{\phi }^{(n)}$, calculate $\mathcal{U}^{(n+1)}$, $\mathcal{V}^{(n+1)}$, $\mathcal{D}^{(n+1)}$, $\mathcal{P}^{(n+1)}$ and $\mathcal{Q}^{(n+1)}$  by using \eqref{optimal_U}, \eqref{optimal_V}, \eqref{optimal_d}, \eqref{optimal_p_w}, \eqref{optimal_p_phi}, \eqref{optimal_g_w} and \eqref{optimal_g_phi};
	\State Calculate $\mathbf{\tilde{w}}^{(n+1)}$  as the solution of the problem in \eqref{r_k_w_wave_SOCP} while $\boldsymbol{\phi }^{(n)}$, $\mathcal{U}^{(n+1)}$, $\mathcal{V}^{(n+1)}$, $\mathcal{D}^{(n+1)}$, $\mathcal{P}^{(n+1)}$ and $\mathcal{Q}^{(n+1)}$ are kept fixed;
	\State Calculate $\boldsymbol{\phi }^{(n+1)}$ via Algorithm 1 while $\mathbf{\tilde{w}}^{(n+1)}$, $\mathcal{U}^{(n+1)}$, $\mathcal{V}^{(n+1)}$, $\mathcal{D}^{(n+1)}$, $\mathcal{P}^{(n+1)}$ and $\mathcal{Q}^{(n+1)}$ are kept fixed;
	\State Set $n \gets n+1$
	\EndWhile		
\end{algorithmic}
\end{algorithm}

\subsection{Algorithm Development}\label{AD}
\subsubsection{BCD-SOCP Algorithm}
In Algorithm 2, we present the complete BCD-SOCP algorithm.
Specifically, we maximize the WMSR by alternately optimizing the variables $\mathcal{U}$, $\mathcal{V}$, $\mathcal{D}$, $\mathcal{P}$, $\mathcal{Q}$, $\mathbf{\tilde{w}}$ and $\boldsymbol{\phi }$.
Note that the globally optimal solution of Problem \eqref{r_k_w_wave_SOCP} can be obtained at each iteration.
Hence, the convergence of Algorithm 2 can be guaranteed.

\subsubsection{Complexity Analysis}
The complexity of optimizing the auxiliary variables $\mathcal{U}$, $\mathcal{V}$, $\mathcal{D}$, $\mathcal{P}$ and $\mathcal{Q}$ is discussed first.
The complexity order for computing each $u_{k}$ in \eqref{optimal_U}, $v_k$ in \eqref{optimal_V}, $d_k$ in \eqref{optimal_d}, and $\mathcal{P}$ in \eqref{optimal_p_w} and \eqref{optimal_p_phi}, is given by $\mathcal{O}\left( K\left( M^2+MN \right) \right)$.
Hence, the computation of $\mathcal{U}$, $\mathcal{V}$, $\mathcal{D}$, and $\mathcal{P}$ has the same complexity order equal to $\mathcal{O}\left( K^2\left( M^2+MN \right) \right)$.
Since the Cholesky decomposition and the Kronecker product applied to compute $\mathbf{q}_{\mathrm{w}}$ in \eqref{optimal_g_w} and $\mathbf{Q}_{\phi}$ in \eqref{optimal_g_phi}, the overall complexity for computing $\mathcal{Q}$ is $\mathcal{O}\left( N^3K^3/ 3+N^2K^2+M^2+MN \right)$.
Thus, the total computational complexity for obtaining $\mathcal{U}$, $\mathcal{V}$, $\mathcal{D}$, $\mathcal{P}$ and $\mathcal{Q}$ is $\mathcal{O}\left( N^3K^3/ 3+N^2K^2+M^2K^2+MNK^2 \right) $.

The computational complexity of calculating the main optimization variables corresponds to the complexity of solving the SOCP problems formulated in \eqref{r_k_w_wave_SOCP} and \eqref{r_k_phi_SOCP}.
According to \cite{ben2001lectures}, since the problem in \eqref{r_k_w_wave_SOCP} includes a power constraint and $K$ rate constraints whose dimension is $NK$, the corresponding complexity is $\mathcal{O}\left( N^3K^{5.5}\right)$.
Similarly, the relaxed version of the problem in \eqref{r_k_phi_SOCP} includes 2$K$ rate constraints of dimension $M$ and $M$ constant modulus constraints of dimension one. 
Denote $t_{\max}$ as the maximum number that allows Algorithm 1 to converge.
Thus, the corresponding complexity is $\mathcal{O}\left( t_{\max}(M^{3.5}+M^3K^{2.5}+N^3K^{5.5})\right)$.

In summary, the computational complexity of each iteration of Algorithm 2 is $\mathcal{O}\left( t_{\max}(M^{3.5}+M^3K^{2.5}+N^3K^{5.5})\right)$.
}

% BCD-MM算法
\section{BCD-MM Algorithm}\label{MM METHOD}
In Algorithm 2, the use of CVX to solve the SOCP problems results in a large computational complexity, since high complexity optimization algorithms, such as the interior point method, are utilized.
To reduce the computational complexity, we introduce, a BCD-MM algorithm.
Specifically, since the objective functions in \eqref{WMMSE_Problem_W} and \eqref{WMMSE_Problem_phi} are non-differentiable, we first derive smooth lower bound functions, and then apply the MM algorithm by introducing surrogate objective functions for the obtained lower bounds. 
We show that this approach results in a simple closed-form solution.

\subsection{Approximate Functions}
Based on \cite{Xu2001Smoothing}, we approximate the objective functions in problems \eqref{WMMSE_Problem_W} and \eqref{WMMSE_Problem_phi} as
{\small\begin{align}
\label{f_W}\min_{k\in \mathcal{K}} \left\{ \tilde{r}_{\mathrm{w},k}\left( \mathbf{\tilde{w}} \right) \right\} &\approx f\left( \mathbf{\tilde{w}} \right) \!=\!-\frac{1}{\zeta}\log ( \sum_{k=1}^K{\exp \left\{ -\zeta \tilde{r}_{\mathrm{w},k}\left( \mathbf{\tilde{w}} \right) \right\}} ),
\\
\label{f_phi}\min_{k\in \mathcal{K}} \left\{ \tilde{r}_{\phi,k}\left( \boldsymbol{\phi } \right) \right\} &\approx f\left( \boldsymbol{\phi } \right) \!=\!-\frac{1}{\zeta}\log ( \sum_{k=1}^K{\exp \left\{ -\zeta \tilde{r}_{\phi,k}\left( \boldsymbol{\phi } \right) \right\}} ),
\end{align}}
where $f\left( \mathbf{\tilde{w}} \right)$ and $f\left( \boldsymbol{\phi } \right)$ are lower bounds for the objective functions in \eqref{WMMSE_Problem_W} and \eqref{WMMSE_Problem_phi}, respectively, and $\zeta > 0$ is a smoothing parameter that satisfies the conditions:
\begin{align}
f\left( \mathbf{\tilde{w}} \right) +\frac{1}{\zeta}\log \left( K \right) &\geqslant \min_{k\in \mathcal{K}} \left\{ \tilde{r}_{\mathrm{w},k}\left( \mathbf{\tilde{w}} \right) \right\} \geqslant f\left( \mathbf{\tilde{w}} \right)
\\
f\left( \boldsymbol{\phi } \right) +\frac{1}{\zeta}\log \left( K \right) &\geqslant \min_{k\in \mathcal{K}} \left\{ \tilde{r}_{\phi,k}\left( \boldsymbol{\phi } \right) \right\} \geqslant f\left( \boldsymbol{\phi } \right).
\end{align}
In \cite{zhou2019intelligent}, the authors proved that $- \frac{1}{\mu }\log ( {\sum\limits_{x \in {\cal X}} {\exp \left\{ { - \mu x } \right\}} } )$ is a concave function of $x$ and is monotonically increasing.
Additionally, $\tilde{r}_{\mathrm{w},k}\left( \mathbf{\tilde{w}} \right)$ is a quadratic concave function of $\mathbf{\tilde{w}}$, and hence $f\left( \mathbf{\tilde{w}} \right)$ is a concave function of $\mathbf{\tilde{w}}$.
Similarly, $f\left( \boldsymbol{\phi } \right)$ is a concave function of $\boldsymbol{\phi }$.
The smoothing parameter $\zeta$ is optimized as described in \cite{9318531}. Specifically, we set $\zeta$ equal to a small initial value, and then gradually increases it, to improve the approximation accuracy, until it reaches an upper limit $\zeta _{\max}$.
The advantage of this strategy is that it avoids local minima in the early stages of operation and avoids the loss of accuracy caused by the use of a large smoothing factor, which can degrade the performance of the MM algorithm.

\vspace{-0.4cm}
\subsection{Majorization-Minimization Method}
Armed with the approximated functions in \eqref{f_W} and \eqref{f_phi}, we adopt the MM algorithm \cite{7547360}.
The MM algorithm does not directly optimize the functions in \eqref{f_W} and \eqref{f_phi}, but it operates on surrogate functions that are easier to optimize.
Specifically, let us consider the maximization of the complex function $f\left( \mathbf{x} \right)$ where $\mathbf{x}$ belongs to a set $\mathcal{S}_{\mathbf{x}}$.
Let us consider the surrogate function $\tilde{f}\left( \mathbf{x}|\mathbf{x}^{(n)} \right)$ with given $\mathbf{x}^{(n)}$, were $\mathbf{x}^{(n)}$ is  the optimal solution that corresponds to the surrogate function at the ($n-1$)-th iteration.
The surrogate function $\tilde{f}\left( \mathbf{x}|\mathbf{x}^{(n)} \right)$ is said to minorize $f\left( \mathbf{x} \right)$ at the given point $\mathbf{x}^{(n)}$ if the following conditions are satisfied \cite{7547360}

\noindent(A1) $\tilde{f}( \mathbf{x}|\mathbf{x}^{(n)} )$
is continuous in $\mathbf{x}$ and $\mathbf{x}^{(n)}$; 

\noindent(A2) $\tilde{f}( \mathbf{x}^{(n)}|\mathbf{x}^{(n)} ) =f( \mathbf{x}^{(n)}) ,\forall \mathbf{x}^{(n)}\in \mathcal{{S}}_{\mathrm{x}}$;

\noindent(A3) $\tilde{f}( \mathbf{x}|\mathbf{x}^{(n)} ) \leqslant f( \mathbf{x} ) ,\forall \mathbf{x},\mathbf{x}^{(n)}\in \mathcal{{S}}_{\mathrm{x}}$;

\noindent(A4) $ \tilde{f}^{\prime}( \mathbf{x}^{(n)}|\mathbf{x}^{(n)};\eta ) |_{\mathbf{x}=\mathbf{x}^{(n)}}=f^{\prime}( \mathbf{x}^{(n)};\eta ) ,\forall \eta \,$with$\,\mathbf{x}^{(n)}+\eta \in \mathcal{{S}}_{\mathrm{x}}$, where $f^{\prime}( \mathbf{x}^{(n)};\eta )$ is the  directional derivative of $f\left( \mathbf{x}^{(n)} \right)$, which is defined as
\begin{align}
f^{\prime}( \mathbf{x}^{(n)};\eta ) =\lim_{\lambda \rightarrow 0}\frac{f\left( \mathbf{x}^{(n)}+\lambda \eta \right) -f\left( \mathbf{x}^{(n)} \right)}{\lambda}.
\end{align}
A drawback of the MM algorithm is that it may need many iterations to converge.
To circumvent this issue, the SQUAREM method \cite{varadhan2008simple} is used to accelerate the convergence of the MM algorithm and hence to reduce the computational overhead.

\vspace{-0.2cm}
\subsection{Optimization of the Precoding Vector $\mathbf{\tilde{w}}$}
With $f(\mathbf{\tilde{w}})$ defined in \eqref{f_W}, the subproblem in \eqref{WMMSE_Problem_W} can be transformed to the following problem
\begin{subequations}\label{MM_W_problem_temp}
\begin{align}
\max_{\mathbf{\tilde{w}}} &\quad f\left( \mathbf{\tilde{w}} \right)
\\
\mathrm{s}.\mathrm{t}. &\quad   \mathbf{\tilde{w}}^{\mathrm{H}}\mathbf{\tilde{w}}\leqslant P.
\end{align}
\end{subequations}
A surrogate function for $f\left( \mathbf{\tilde{w}} \right)$is given in the following lemma.

\textit{Lemma 5: }
Let $\mathbf{\tilde{w}}^{(n)}$ be the solution at the $(n-1)$-th iteration.
For any feasible $\mathbf{\tilde{w}}$, $f(\mathbf{\tilde{w}})$ is minorized by the following quadratic function
\begin{align}\label{MM_W}
\bar{f}\left( \mathbf{\tilde{w}}|\mathbf{\tilde{w}}^{(n)} \right) =\bar{c}_{\mathrm{w}}+2\mathrm{Re}\left\{ \mathbf{\bar{v}}_{\mathrm{w}}^{\mathrm{H}}\mathbf{\tilde{w}} \right\} +\bar{\alpha} \mathbf{\tilde{w}}^{\mathrm{H}}\mathbf{\tilde{w}},
\end{align}
where 
{\small\begin{subequations}
\begin{align}
&\label{v_MM_w}\mathbf{\bar{v}}_{\mathrm{w}} \triangleq \sum_{k=1}^K{h_{\mathrm{w},k}\left( \mathbf{\tilde{w}}^{(n)} \right) \left( \boldsymbol{\tilde{b}}_{\mathrm{w},k}-\mathbf{\tilde{C}}_{\mathrm{w},k}^{\mathrm{H}}\mathbf{\tilde{w}}^{(n)} \right)}-\bar{\alpha}\mathbf{\tilde{w}}^{(n)},
\\
&\label{cons_MM_w}\bar{c}_{\mathrm{w}}\triangleq f\left( \mathbf{\tilde{w}}^{(n)} \right) +\bar{\alpha}\mathbf{\tilde{w}}^{(n),\mathrm{H}}\mathbf{\tilde{w}}^{(n)} \notag
\\
&-2\mathrm{Re}\left\{ \sum_{k=1}^K{h_{\mathrm{w},k}\left( \mathbf{\tilde{w}}^{(n)} \right) \left( \boldsymbol{\tilde{b}}_{\mathrm{w},k}^{\mathrm{H}}-\mathbf{\tilde{w}}^{(n),\mathrm{H}}\mathbf{\tilde{C}}_{\mathrm{w},k} \right) \mathbf{\tilde{w}}^{(n)}} \right\}, 
\\
&\label{bar_alpha}\bar{\alpha} \triangleq -\max_k \left\{ \mathrm{Tr}\left[ \mathbf{\tilde{C}}_{\mathrm{w},k} \right] \right\} -2\zeta \max_k\left\{ \bar{o}_{\mathrm{w},k} \right\},
\end{align}
\end{subequations}}
and $h_{\mathrm{w},k}\left( \mathbf{\tilde{w}}^{(n)} \right)$ and $\bar{o}_{\mathrm{w},k}$ are, respectively, given by
{\small\begin{subequations}
\begin{align}
&h_{\mathrm{w},k}\left( \mathbf{\tilde{w}}^{(n)} \right) \triangleq \frac{\exp \left\{ -\zeta \tilde{r}_{\mathrm{w},k}\left( \mathbf{\tilde{w}}^{(n)} \right) \right\}}{\sum_{k=1}^K{\exp \left\{ -\zeta \tilde{r}_{\mathrm{w},k}\left( \mathbf{\tilde{w}}^{(n)} \right) \right\}}}, \label{lemma4_h}
\\
&\bar{o}_{\mathrm{w},k}\triangleq P \mathrm{Tr}\left[ \mathbf{\tilde{C}}_{\mathrm{w},k}\mathbf{\tilde{C}}_{\mathrm{w},k}^{\mathrm{H}} \right] \!+\!\left\| \boldsymbol{\tilde{b}}_{\mathrm{w},k} \right\| _{2}^{2} +2\sqrt{P}\left\| \mathbf{\tilde{C}}_{\mathrm{w},k}\boldsymbol{\tilde{b}}_{\mathrm{w},k} \right\| _2.
\end{align}
\end{subequations}}

\emph{Proof:} See Appendix \ref{appendixC}.
\hfill $\blacksquare$

Therefore, the problem in \eqref{MM_W_problem_temp} can be approximated as
\begin{subequations}\label{MM_W_replace}
\begin{align}
\max_{\mathbf{\tilde{w}}} &\quad \bar{f}\left( \mathbf{\tilde{w}}|\mathbf{\tilde{w}}^{(n)} \right), 
\\
\mathrm{s}.\mathrm{t}.&\quad \mathbf{\tilde{w}}^{\mathrm{H}}\mathbf{\tilde{w}}\leqslant P.
\end{align}
\end{subequations}
The optimization problem in \eqref{MM_W_replace} can be solved by using the method of Lagrangian multipliers.
Specifically, the Lagrangian function is given by
\begin{align}\label{MM_w_surrogate}
L\left( \mathbf{\tilde{w}},\varepsilon \right) =\bar{f}\left( \mathbf{\tilde{w}}|\mathbf{\tilde{w}}^{(n)} \right)-\varepsilon \left( \mathbf{\tilde{w}}^{\mathrm{H}}\mathbf{\tilde{w}}-P \right),
\end{align}
where $\varepsilon$ is the Lagrange multiplier.
Therefore, the optimal solution $\mathbf{\tilde{w}}$ of the surrogate optimization problem in \eqref{MM_w_surrogate} at the $n$-th iteration is
\begin{align}\label{MM_w_rules}
\mathbf{\tilde{w}}^{(n+1)}=-\sqrt{\frac{P}{\mathbf{\bar{v}}_{\mathrm{w}}^{\mathrm{H}}\mathbf{\bar{v}}_{\mathrm{w}}}}\mathbf{\bar{v}}_{\mathrm{w}}.
\end{align}

\subsection{Optimization of the Reflection Coefficient Vector $\boldsymbol{\phi }$}
With $f(\boldsymbol{\phi})$ defined in \eqref{f_phi}, the subproblem in \eqref{WMMSE_Problem_phi} can be transformed to the following problem
\begin{subequations}\label{MM_phi_temp}
\begin{align}
\max_{\phi} &\quad f\left( \boldsymbol{\phi } \right)
\\
\mathrm{s}.\mathrm{t}.&\quad \boldsymbol{\phi }\in \mathcal{S}.
\end{align}
\end{subequations}
A surrogate function for $f\left( \boldsymbol{\phi } \right)$ is given in the following lemma.

\textit{Lemma 6}:
Let $\boldsymbol{\phi }^{(n)}$ be the solution at the $(n-1)$-th iteration.
For any feasible $\boldsymbol{\phi }$, $f(\boldsymbol{\phi })$ is minorized by the following function 
\begin{align}\label{MM_phi}
\bar{f}\left( \boldsymbol{\phi }|\boldsymbol{\phi }^{(n)} \right) =\!\bar{c}_{\phi}+2\mathrm{Re}\left\{ \mathbf{\bar{v}}_{\phi}^{\mathrm{H}}\boldsymbol{\phi } \right\},
\end{align}
where $n$ is the iteration number, and
{\small\begin{subequations}
\begin{align}
\label{v_MM_phi}\mathbf{\bar{v}}_{\phi}&\triangleq \sum_{k=1}^K{h_{\phi ,k}\left( \boldsymbol{\phi }^{(n)} \right) \left( \boldsymbol{\tilde{b}}_{\phi ,k}-\mathbf{\tilde{C}}_{\phi ,k}^{\mathrm{H}}\boldsymbol{\phi }^{(n)} \right)}-\bar{\beta}\boldsymbol{\phi }^{(n)},
\\
\label{const_MM_phi}\!\bar{c}_{\phi}&\triangleq \bar{f}\left( \boldsymbol{\phi }^{(n)} \right) +2M\bar{\beta} \notag
\\
&-2\mathrm{Re}\left\{ \sum_{k=1}^K{h_{\phi ,k}\left( \boldsymbol{\phi }^{(n)} \right) \left( \boldsymbol{\tilde{b}}_{\phi ,k}^{\mathrm{H}}-\boldsymbol{\phi }^{n,\mathrm{H}}\mathbf{\tilde{C}}_{\phi ,k} \right)}\boldsymbol{\phi }^{(n)} \right\},
\end{align}
\end{subequations}}
with
{\small\begin{subequations}
\begin{align}
&\label{h_phi}h_{\phi ,k}\left( \boldsymbol{\phi }^{(n)} \right) \triangleq \frac{\exp \left\{ -{\zeta}\tilde{r}_{\phi,k}\left( \boldsymbol{\phi }^{(n)} \right) \right\}}{\sum_{k=1}^K{\exp \left\{ -{\zeta}\tilde{r}_{\phi,k}\left( \boldsymbol{\phi }^{(n)} \right) \right\}}},
\\
&\label{beta_phi}\bar{\beta}\triangleq -\max_k\left\{ \lambda _{\max}\left( \mathbf{\tilde{C}}_{\phi ,k} \right) \right\}
\\
&-2\zeta \max_k \!\left\{\! \left\| \boldsymbol{\tilde{b}}_{\phi ,k} \right\|_2 ^{2}\!+\!M\lambda _{\max}\left( \mathbf{\tilde{C}}_{\phi ,k}\mathbf{\tilde{C}}_{\phi ,k}^{\mathrm{H}} \right) \!+\!2\left\| \mathbf{\tilde{C}}_{\phi ,k}\boldsymbol{\tilde{b}}_{\phi ,k} \right\| _1 \!\right\}\!. \notag
\end{align}
\end{subequations}}

\emph{Proof:} See Appendix \ref{appendixD}.
\hfill $\blacksquare$

Therefore, the problem in \eqref{MM_phi_temp} can be approximated as
\begin{subequations}\label{MM_phi_replace}
\begin{align}
\max_{\phi} &\quad \bar{f}( \boldsymbol{\phi }|\boldsymbol{\phi }^{(n)} )
\\
\mathrm{s}.\mathrm{t}.&\quad \boldsymbol{\phi }\in \mathcal{S}.
\end{align}
\end{subequations}
The optimal solution $\boldsymbol{\phi }^{(n+1)}$ at the $n$-th iteration is given by
\begin{align}\label{MM_phi_rules}
\boldsymbol{\phi }^{(n+1)}=\exp \left\{ j\angle \mathbf{\bar{v}}_{\phi} \right\},
\end{align}
where $\exp \left( \cdot \right) $ and $\angle \left( \cdot \right) $ are intended as element-wise functions. 

\begin{algorithm}[t]
\caption{BCD-MM algorithm}
\label{BCD-MM}
\textbf{Initialize}: Initialize feasible $\mathbf{\tilde{w}}^0$, $\boldsymbol{\phi }^0$. Set $n=0$, the smoothing factor $\zeta$, the maximum value of the smoothing factor $\zeta _{\max}$, the adjustment factor $\iota$, the maximum number of iterations $n_{\max}$ and the error tolerance $\varepsilon$.
\begin{algorithmic}[1]
	\While{$\varepsilon \leqslant \left| \mathcal{R}( \mathbf{\tilde{w}}^{(n+1)},\boldsymbol{\phi }^{(n+1)}) -\mathcal{R}\left( \mathbf{\tilde{w}}^{(n)},\boldsymbol{\phi }^{(n)} \right) \right|$ $/\left|\mathcal{R}\left( \mathbf{\tilde{w}}^{(n)},\boldsymbol{\phi }^{(n)} \right) \right|$ and $n\leqslant n_{\max}$}
	\State Given $\mathbf{\tilde{w}}^{(n)}$ and $\boldsymbol{\phi }^{(n)}$, calculate $\mathcal{U}^{(n+1)}$, $\mathcal{V}^{(n+1)}$, $\mathcal{D}^{(n+1)}$, $\mathcal{P}^{(n+1)}$ and $\mathcal{Q}^{(n+1)}$  by using \eqref{optimal_U}, \eqref{optimal_V}, \eqref{optimal_d}, \eqref{optimal_p_w}, \eqref{optimal_p_phi}, \eqref{optimal_g_w} and \eqref{optimal_g_phi};
	\State Calculate $\mathbf{\tilde{w}}_1=\mathcal{F}_{\mathbf{W}}\left( \mathbf{\tilde{w}}^{(n)} \right)$ and $\mathbf{\tilde{w}}_2=\mathcal{F}_{\mathbf{W}}\left( \mathbf{\tilde{w}}_1 \right)$;
	\State Calculate $\mathbf{j}_1=\mathbf{\tilde{w}}_1-\mathbf{\tilde{w}}^{(n)}$ and $\mathbf{j}_2=\mathbf{\tilde{w}}_2-\mathbf{\tilde{w}}_1-\mathbf{j}_1$;
	\State Calculate the step factor $\alpha  =  - \frac{{{{\left\| {\mathbf{j}_1} \right\|}_2}}}{{{{\left\| {\mathbf{j}_2} \right\|}_2}}}$;
	\State Calculate $\mathbf{\tilde{w}}^{(n+1)} =\mathbf{\tilde{w}}^{(n)}-2\alpha \mathbf{j}_1+\alpha ^2\mathbf{j}_2$;
	\State If ${{{\left\| \mathbf{\tilde{w}}^{(n+1)} \right\|}_2}} \!>\! \sqrt{P}$, set $\mathbf{\tilde{w}}^{(n+1)}\gets \frac{\sqrt{P}}{\left\| \mathbf{\tilde{w}}^{(n+1)} \right\| _2}\mathbf{\tilde{w}}^{(n+1)}$;
	\State If $f\left( \mathbf{\tilde{w}}^{(n+1)} \right) < f\left( \mathbf{\tilde{w}}_2 \right)$, set $\alpha \gets \frac{\left( \alpha -1 \right)}{2}$, back to step 4;
	\State Calculate ${\bm \phi}_1=\mathcal{F}_{\phi} \left({\bm \phi}^{(n)}\right)$ and ${\bm \phi}_2=\mathcal{F}_{\phi} \left({\bm \phi}_1\right)$;\label{iter_map_phi}\label{update_phi_BCDMM_begin}
	\State Calculate ${\bf k}_1={\bm \phi}_1-{\bm \phi}^{(n)}$ and ${\bf k}_2={\bm \phi}_2-{\bm \phi}_1-{\bf k}_1$;
	\State Calculate the step factor $\beta  =  - \frac{{{{\left\| {{{\bf{k}}_1}} \right\|}_2}}}{{{{\left\| {{{\bf{k}}_2}} \right\|}_2}}}$;
	\State Calculate $\boldsymbol{\phi }^{(n+1)} = \exp \left\{ {\angle \left( {{{\bm \phi}^{(n)}} - 2\beta {{\bf{k}}_1} + {\beta ^2}{{\bf{k}}_2}} \right)} \right\}$;\label{update_phi_BCDMM}
	\State If $f\left( \boldsymbol{\phi }^{(n+1)} \right) < f\left( {\bm \phi}_2 \right)$, set $\beta \gets \frac{\left( \beta -1 \right)}{2}$, back to step 10;
	\State Set $\zeta \gets \min \left( \zeta ^{\iota},\zeta _{\max} \right)$ and $n \gets n+1$;
	\EndWhile
\end{algorithmic}
\end{algorithm}

\subsection{Algorithm Development}
\subsubsection{BCD-MM Algorithm}
The accelerated version of the BCD-MM algorithm is summarized in  Algorithm 3.
Specifically, the optimization problems in \eqref{WMMSE_Problem_W} and \eqref{WMMSE_Problem_phi}
are transformed into the optimization problems in \eqref{MM_W_replace} and \eqref{MM_phi_replace}, whose approximate optimal solutions are given in \eqref{MM_w_rules} and \eqref{MM_phi_rules}, respectively.
In Algorithm 3, the following notation is used: $\mathcal{R}\left( \cdot \right) $ is the objective function of the problem in \eqref{WMSR_Problem}; $\mathcal{F}_{\mathbf{W}}\left( \cdot \right)$ and $\mathcal{F}_{\phi}\left( \cdot \right)$ 
denote the nonlinear fixed-point iteration map
of the MM algorithm in \eqref{MM_w_rules} and \eqref{MM_phi_rules}, respectively.
Specifically, steps 6 and 11 describe the gradient method proposed by the SQUAREM method.
Step 7 and 12 describe the projection operation to force wayward points to satisfy their nonlinear constraints.
In addition, step 8 and 13 are to ensure the ascent
property of Algorithm 3.
In step 14, the adjustment factor $\iota$ is used to successively increase the smoothness factor $\zeta$ from its initial value to $\zeta _{\max}$.
\subsubsection{Convergence analysis}
Note that the surrogate function $\bar{f}\left( \mathbf{\tilde{w}}|\mathbf{\tilde{w}}^{(n)} \right)$ satisfies the conditions $\bar{f}\left( \mathbf{\tilde{w}}|\mathbf{\tilde{w}}^{(n)}  \right) \leqslant f\left( \mathbf{\tilde{w}} \right)$, and $\bar{f}\left( \mathbf{\tilde{w}}^{(n)}|\mathbf{\tilde{w}}^{(n)} \right) =f\left( \mathbf{\tilde{w}}^{(n)} \right)$.
Hence, with given $\boldsymbol{\phi }^{(n)}$, we have
\begin{align}
f( \mathbf{\tilde{w}}^{(n)},\boldsymbol{\phi }^{(n)} ) =\bar{f}( \mathbf{\tilde{w}}^{(n)}|\mathbf{\tilde{w}}^{(n)} ) &\leqslant \bar{f}( \mathbf{\tilde{w}}^{(n+1)}|\mathbf{\tilde{w}}^{(n)} ) \notag
\\
&\leqslant f( \mathbf{\tilde{w}}^{(n+1)},\boldsymbol{\phi }^{(n)} ).
\end{align}
Similarly, with given $\mathbf{\tilde{w}}^{(n)}$, we have
\begin{align}
f( \boldsymbol{\phi }^{(n)},\mathbf{\tilde{w}}^{(n+1)} ) =\bar{f}( \boldsymbol{\phi }^{(n)}|\boldsymbol{\phi }^{(n)} ) &\leqslant \bar{f}( \boldsymbol{\phi }^{(n+1)}|\boldsymbol{\phi }^{(n)} )
\\
&\leqslant f( \boldsymbol{\phi }^{(n+1)},\mathbf{\tilde{w}}^{(n+1)}). \notag
\end{align}
Then, the objective function values generated by BCD-MM algorithm are monotonically increasing.
In addition, subject to the the maximum transmit power constraints, the value of the objective function in Problem \eqref{WMSR_Problem} has an upper bound.
Hence, the BCD-MM algorithm is guaranteed to converge.

\subsubsection{Complexity Analysis}
The computational complexity of optimizing the variables $\mathcal{U}$, $\mathcal{V}$, $\mathcal{D}$,  $\mathcal{P}$ and $\mathcal{Q}$ is the same as in Section \ref{AD}, which is $\mathcal{O}\left( N^3K^3/ 3+N^2K^2+M^2K^2+MNK^2 \right)$.
Next, we analyze the computational complexity of the two remaining optimization variables.
Note that $r_{\mathrm{w},k}\left( \mathbf{\tilde{w}}^{(n)} \right)$ and $r_{\phi,k}\left( \mathbf{\tilde{w}}^{(n)} \right)$ can be reused when calculating $h_{\mathrm{w},k}\left( \mathbf{\tilde{w}}^{(n)} \right)$ and $h_{\phi,k}\left( \mathbf{\tilde{w}}^{(n)} \right)$, respectively. 
First, we note that the complexity required to calculate $h_{\mathrm{w},k}\left( \mathbf{\tilde{w}}^{(n)} \right)$ and $h_{\phi,k}\left( \mathbf{\tilde{w}}^{(n)} \right)$ is $\mathcal{O}\left( K\left( N^3K^3/3+N^2K^2+M^2+MN \right) \right) $ and $\mathcal{O}\left( KM^3 \right) $, respectively.

As far as the optimization of $\mathbf{\tilde{w}}$ is concerned, the complexity of computing $o_{\mathrm{w},k}$ and $\bar{\alpha}$ are $\mathcal{O}\left( N^2K^2 \right)$ and $\mathcal{O}\left( K\left( N^2K^2 \right) \right) $, respectively.
The complexity of calculating $\mathbf{\bar{v}}_{\mathrm{w}}$ mainly depends on $h_{\mathrm{w},k}\left( \mathbf{\tilde{w}}^{(n)} \right)$.
Hence, the complexity of computing $\mathbf{\tilde{w}}^{(n+1)}$ is $\mathcal{O}\left( K\left( N^3K^3/3+N^2K^2+M^2+MN \right) \right)$.
As far as the computational complexity of the subproblems corresponding to $\boldsymbol{\phi }$ is concerned, the complexity of computing $\lambda _{\max}\left( \mathbf{\tilde{C}}_{\phi ,k}\mathbf{\tilde{C}}_{\phi ,k}^{\mathrm{H}} \right)$ is $\mathcal{O}\left( M^3 \right) $ and the computational complexity required to find $\bar{\beta}$ is  $\mathcal{O}\left( KM^3 \right) $.
Hence, the  complexity of calculating $\boldsymbol{\phi }^{(n+1)}$ is $\mathcal{O}\left( KM^3 \right) $.

Finally, the computational complexity of each iteration of Algorithm 2 is $\mathcal{O}\left( M^2K+MNK+N^3K^4/3+N^2K^3 \right) +\mathcal{O}\left( M^3K \right) $.
Therefore, the complexity of Algorithm 3 is lower than that of Algorithm 2.

\vspace{-0.3cm}
\section{Simulation Results}\label{Simulation Results}
\subsection{Simulation Setup}
\begin{figure}
\centering
\includegraphics[scale=0.14]{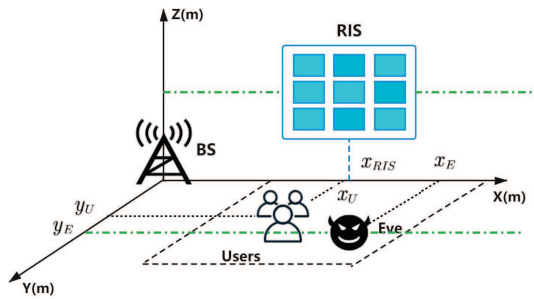}
\caption{The simulated RIS-assisted MISO communication scenario.}
\label{simulateModel}
\end{figure}

In this section, simulation results are illustrated to evaluate the performance of the proposed BCD-SOCP and BCD-MM algorithms.
Figure \ref{simulateModel} depicts the considered simulation setup, wherein the BS and the RIS are  located at (0 m, 0 m, 30 m) and ($x_{\mathrm{RIS}}$, 0 m, 10 m), respectively.
Unless stated otherwise, $x_{\mathrm{RIS}}$ = 50 m.
Three legitimate users are randomly located in a 10 m $ \times$ 10 m area, whose center is ($x_{\mathrm{U}}$, $y_{\mathrm{U}}$, 1.5 m), and the eavesdropper is located at ($x_{\mathrm{E}}$, $y_{\mathrm{E}}$, 1.5 m).
We assume that $x_{\mathrm{U}} = x_{\mathrm{E}} = 300$ m and $y_{\mathrm{U}} = y_{\mathrm{E}} = 10$ m.
In addition, unless stated otherwise, the number of BS transmit antennas and RIS reflecting elements is $N=4$ and $M=16$, respectively.

The large-scale path loss is defined as
\begin{align}\label{path_loss_model}
\mathrm{PL}=-30-10\alpha \log _{10}d,
\end{align}
where $\alpha$ is the path loss exponent and $d$ is the link distance in meters.
The path loss exponents of the BS-RIS channel, RIS-user channel, RIS-eavesdropper channel, BS-user channel and BS-eavesdropper channel are equal to $\alpha_{\mathrm{BR}}=\alpha_{\mathrm{RU}}=\alpha_{\mathrm{RE}}=2$ and $\alpha_{\mathrm{BU}}=\alpha_{\mathrm{BE}}=4$, respectively.

Due to rich scatters, the small scale fading of the BS-user channel and BS-eavesdropper channel is assumed to be Rayleigh fading.
In addition, the small scale fading of the RIS-related channels is assumed to obey a Rician distribution, and, therefore, the channel is
\begin{align}
\mathbf{\tilde{H}}=\sqrt{\frac{\kappa}{\kappa +1}}\mathbf{\tilde{H}}^{\mathrm{LoS}}+\sqrt{\frac{1}{\kappa +1}}\mathbf{\tilde{H}}^{\mathrm{NLoS}},
\end{align}
where $\kappa$ is the Rician factor, $\mathbf{\tilde{H}}^{\mathrm{LoS}}$ and $\mathbf{\tilde{H}}^{\mathrm{NLoS}}$ denote the line-of-sight (LoS) and the non-line-of-sight (NLoS) components, respectively.
$\mathbf{\tilde{H}}^{\mathrm{LoS}}$ is defined as the product of the steering vectors of the transmitter and receiver, while $\mathbf{\tilde{H}}^{\mathrm{NLoS}}$ is randomly generated according to a Rayleigh distribution with unit power. Unless stated otherwise, we set $\kappa=10$.

The MOSEK solver \cite{Themosekoptimizationtoolbox} in the CVX toolbox is used to solve the SOCP problem in Algorithm 2.
The final results are obtained  by averaging over 200 independent channels.
Unless stated otherwise, the simulation parameters are set as follows:
the HI factors are $\kappa_{\mathrm{t}} = \kappa_{\mathrm{r}, k} = 0.01$, the BS transmit power is $P$ = 1 W, the channel bandwidth is 10 MHz, the weighting factors are $\omega_k = 1$, $\forall k$, the noise power density is -174 dBm/Hz, the initial smoothing parameter is $\zeta = 1.25$, the adjustment factor is $\iota = 1.02$, the upper limit of the smoothing parameter is $\zeta _{\max} = 500$, and the error tolerance is $\varepsilon = 10^{-5}$.

\begin{figure}
\centering
\subfigure[Achievable WMSR versus the number of iterations]{
\label{iterationFigure} %% 第一幅图的标签
\includegraphics[scale=0.6]{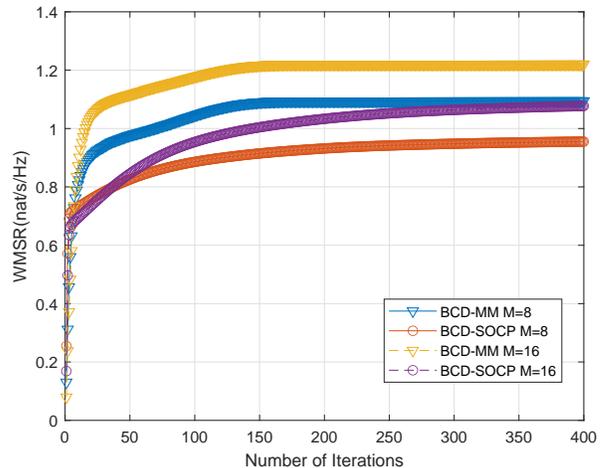}}
\subfigure[Achievable WMSR versus the CPU time]{
\label{timeFigure} %% 第二幅图的标签
\includegraphics[scale=0.6]{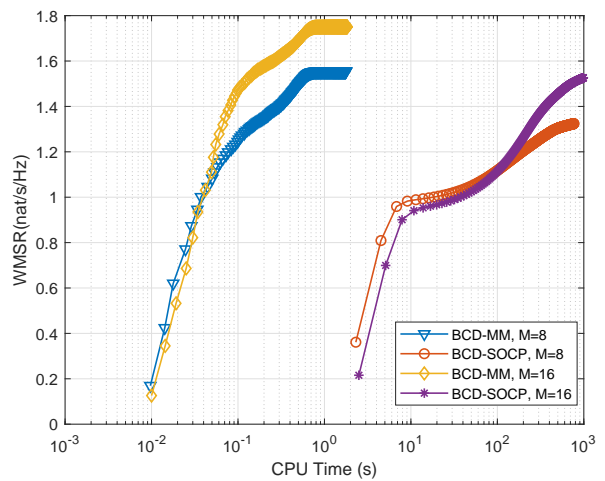}}
\caption{Convergence behavior of the proposed algorithms for $M=\left[8,16\right]$}
\label{ConvergenceFigure} %% label for entire figure
\end{figure}

\subsection{Baseline Schemes}
We compare the performance of the proposed algorithms with the following baseline schemes.
\begin{enumerate}
	\item
	To verify the effectiveness of the proposed robust design, we implement a \textbf{Non-Robust} version of the proposed approach that ignores the HIs at both the RIS and the transceiver.
	
	\item 
	To analyze the benefits of deploying RIS on the security communication system, we consider a scenario without RIS and only optimize the precoding vector $\mathbf{\tilde{w}}$ by applying the \textbf{BCD-MM} algorithm.
	The corresponding algorithm is referred to as \textbf{BCD-MM-No-RIS}.
	
	\item
	To study the advantages of jointly optimizing the precoding at the BS and the phase shifts at the RIS, we consider a scheme that only $\mathbf{\tilde{w}}$ is optimized and the reflection coefficient vector $\boldsymbol{\phi}$ is selected randomly.
	%	The first algorithm optimizes only the precoding vector $\mathbf{\tilde{w}}$ and randomly selects $\boldsymbol{\phi}$.
	The corresponding algorithm is referred to as \textbf{BCD-MM-Rand}.
	
	\item 
	To verify the effectiveness of the proposed MM algorithm for solving the subproblem of  $\boldsymbol{\phi}$, we consider a \textbf{BCD-MM-SDR} version of the proposed approach that uses the SDR \cite{8930608} method to optimize $\boldsymbol{\phi}$.
%	The second algorithm uses SDR \ci.te{8930608} method to optimize $\boldsymbol{\phi}$ and optimizes $\mathbf{\tilde{w}}$.
%	The corresponding algorithm is referred to as \textbf{BCD-MM-SDR}.

\item
In practice, it may be difficult and expensive to implement RISs that can adjust the phase shifts to any arbitrary continuous value.
Therefore, we study the performance of Algorithm 2 when the phase shifts of the RIS are quantized with two bits, i.e., only four phase shifts can be realized. 
The corresponding scheme is referred to as \textbf{BCD-MM-2bit}.
Specifically, let $\boldsymbol{\phi }_{m}^{con}$ be the optimal phase shift of the $m$-th element of the RIS, which obtained by applying the \textbf{BCD-MM} algorithm. Then, the corresponding 2-bit quantized phase shift is
\begin{align}
\boldsymbol{\phi }_{m}^{dis}=\exp \left\{ \,\,\mathrm{arg}  \min_{\theta}\left| \angle \boldsymbol{\phi }_{m}^{con}-\theta \right| \right\} ,
\end{align}
where $\theta \in \left\{ 0,\frac{\pi}{2},\pi ,\frac{3\pi}{2} \right\} $.

\end{enumerate}

\subsection{Convergence Behavior of the Proposed Algorithms}
Figure \ref{ConvergenceFigure} illustrates the convergence behavior of the two proposed algorithms as a function of the number of RIS elements $M$.
We see that the \textbf{BCD-MM} algorithm converges within 150 iterations, while the \textbf{BCD-SOCP} algorithm converges within 350 iterations.
Compared with the \textbf{BCD-SOCP} algorithm, the \textbf{BCD-MM} algorithm converges to a larger value of the WMSR, but it requires less CPU time, which confirms the superiority of the \textbf{BCD-MM} algorithm.
In addition, the obtained results show that the \textbf{BCD-MM} algorithm converges in almost the same number of iterations and CPU time for different values of $M$.
This is mainly because the convergence speed of the MM algorithm is closely related to the approximation accuracy of the surrogate function, which is affected by the strategy for updating the smoothing factor.

\begin{figure}
\centering
\includegraphics[scale=0.6]{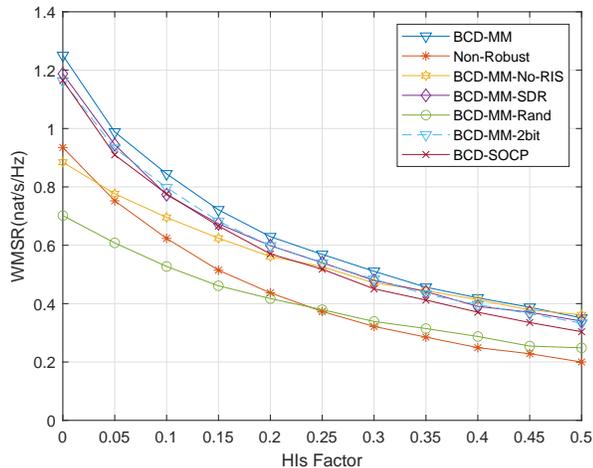}
\caption{Achievable WMSR versus the HIs factor}
\label{HIFigure}
\end{figure}

\begin{figure}
\centering
\includegraphics[scale=0.6]{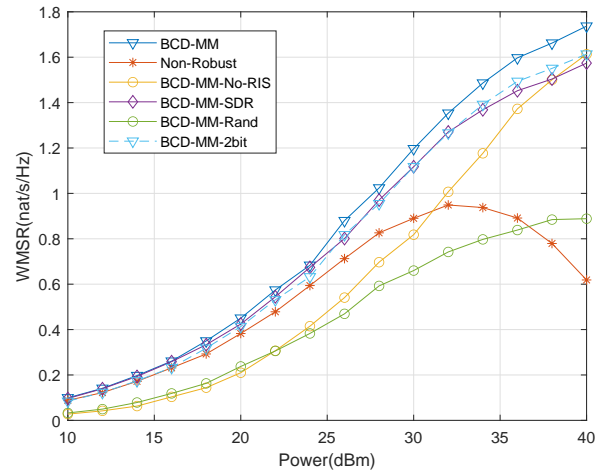}
\caption{Achievable WMSR versus the maximum transmit power}
\label{PowerFigure}
\end{figure}

\begin{figure}
\centering
\includegraphics[scale=0.6]{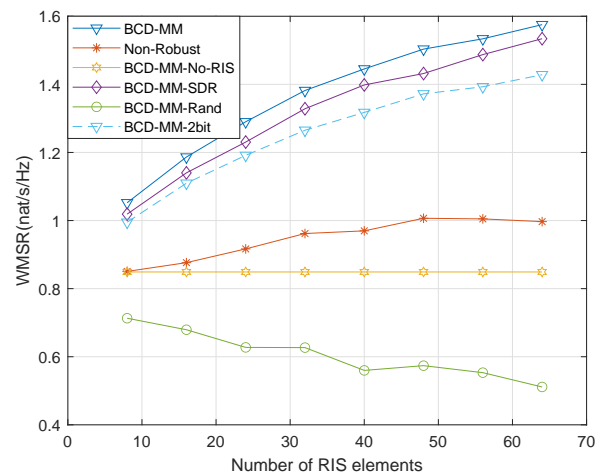}
\caption{Achievable WMSR versus the number of RIS elements $M$}
\label{MFigure}
\end{figure}

\subsection{Impact of the HIs Factor}
The impact of the HIs factor is shown in Figure \ref{HIFigure}.
We see that the security performance of the \textbf{Non-Robust} and \textbf{BCD-MM} algorithms degrades as the HIs factor increases.
However, as the HIs factor increases, the WMSR of the \textbf{BCD-MM} algorithm always outperforms the \textbf{Non-Robust} algorithm, which demonstrates the strength of the proposed robust transmission design.
Since the \textbf{BCD-MM-Rand} algorithm does not attempt to optimize the phase shifts of the RIS, it offers the worst security performance, which highlights the superiority of the joint optimization strategy.
In addition, we see that the security performance of the \textbf{BCD-SOCP} algorithm is always worse than that of the \textbf{BCD-MM} algorithm, which  further corroborates the superiority of the \textbf{BCD-MM} algorithm over the \textbf{BCD-SOCP} algorithm.

\begin{figure*}[hb]
	\hrulefill
	\begin{equation} \label{N_w_wave}
		\mathbf{\bar{N}}_{\mathrm{w}}=-\sum_{k=1}^K{\hat{g}_{\mathrm{w},k}\left( \eta \right) \left( \left[ \begin{matrix}
				\mathbf{\tilde{C}}_{\mathrm{w},k}&		0\\
				0&		 \mathbf{\tilde{C}}_{\mathrm{w},k}^{\mathrm{T}}\\
			\end{matrix} \right] \!+\!\zeta \left[ \begin{array}{c}
				\mathbf{e}_k\\
				\mathbf{e}_{k}^{*}\\
			\end{array} \right] \left[ \begin{array}{c}
				\mathbf{e}_k\\
				\mathbf{e}_{k}^{*}\\
			\end{array} \right] ^{\mathrm{H}} \right)}\!+\!\zeta\! \left[\! \begin{array}{c}
			\sum_{k=1}^K{\hat{g}_{\mathrm{w},k}\left( \eta \right) \mathbf{e}_k}\\
			\sum_{k=1}^K{\hat{g}_{\mathrm{w},k}\left( \eta \right) \mathbf{e}_{k}^{*}}\\
		\end{array}\! \right]\! \left[\! \begin{array}{c}
			\sum_{k=1}^K{\hat{g}_{\mathrm{w},k}\left( \eta \right) \mathbf{e}_k}\\
			\sum_{k=1}^K{\hat{g}_{\mathrm{w},k}\left( \eta \right) \mathbf{e}_{k}^{*}}\\
		\end{array} \!\right] ^{\mathrm{H}}. \tag{117}
	\end{equation}
\end{figure*}

\begin{figure*}[hb]
	\hrulefill
	\begin{align} \label{a_MM}
		\bar{\alpha}&=\lambda _{\min}\left( \mathbf{\bar{N}}_{\mathrm{w}} \right) \overset{\left( a1 \right)}{\geqslant}-\sum_{k=1}^K{\tilde{g}_{\mathrm{w},k}\left( \eta \right) \left( \lambda _{\max}\left( \left[ \begin{matrix}
				\mathbf{\tilde{C}}_{\mathrm{w},k}&		0\\
				0&		 \mathbf{\tilde{C}}_{\mathrm{w},k}^{\mathrm{T}}\\
			\end{matrix} \right] \right) +\zeta \lambda _{\max}\left( \left[ \begin{array}{c}
				\mathbf{e}_k\\
				\mathbf{e}_{k}^{*}\\
			\end{array} \right] \left[ \begin{array}{c}
				\mathbf{e}_k\\
				\mathbf{e}_{k}^{*}\\
			\end{array} \right] ^{\mathrm{H}} \right) \right)} \notag
		\\
		&+\zeta \lambda _{\min}\left( \left[ \begin{array}{c}
			\sum_{k=1}^K{\tilde{g}_{\mathrm{w},k}\left( \eta \right) \mathbf{e}_k}\\
			\sum_{k=1}^K{\tilde{g}_{\mathrm{w},k}\left( \eta \right) \mathbf{e}_{k}^{*}}\\
		\end{array} \right] \left[ \begin{array}{c}
			\sum_{k=1}^K{\tilde{g}_{\mathrm{w},k}\left( \eta \right) \mathbf{e}_k}\\
			\sum_{k=1}^K{\tilde{g}_{\mathrm{w},k}\left( \eta \right) \mathbf{e}_{k}^{*}}\\
		\end{array} \right] ^{\mathrm{H}} \right)  \notag
		\\
		&\overset{\left( a2 \right)}{=}-\sum_{k=1}^K{\tilde{g}_{\mathrm{w},k}\left( \eta \right) \left( \lambda _{\max}\left( \mathbf{\tilde{C}}_{\mathrm{w},k} \right) +2\zeta \mathbf{e}_{k}^{\mathrm{H}}\mathbf{e}_k \right)} \overset{\left( a3 \right)}{\geqslant}-\max_k\left\{ \lambda _{\max}\left( \mathbf{\tilde{C}}_{\mathrm{w},k} \right) \right\} -2\zeta \max_k\left\{ \left\| \mathbf{e}_k \right\| _{2}^{2} \right\}. \tag{121}
	\end{align}
\end{figure*}

\begin{figure*}[hb]
	\hrulefill
	\begin{align} \label{e_MM}
		\left\| \mathbf{e}_k \right\| _{2}^{2}&=\left\| \boldsymbol{\tilde{b}}_{\mathrm{w},k}-\mathbf{\tilde{C}}_{\mathrm{w},k}^{\mathrm{H}}\left( \mathbf{\tilde{w}}^{(n)}+\eta \left( \mathbf{\tilde{w}}^{(m)}-\mathbf{\tilde{w}}^{(n)} \right) \right) \right\| _{2}^{2} \notag
		\\
		&=\left\| \boldsymbol{\tilde{b}}_{\mathrm{w},k} \right\| _{2}^{2}+\left\| \mathbf{\tilde{C}}_{\mathrm{w},k}^{\mathrm{H}}\left( \mathbf{\tilde{w}}^{(n)}+\eta \left( \mathbf{\tilde{w}}^{(m)}-\mathbf{\tilde{w}}^{(n)} \right) \right) \right\| _{2}^{2}-2\mathrm{Re}\left\{ \boldsymbol{\tilde{b}}_{\mathrm{w},k}^{\mathrm{H}}\mathbf{\tilde{C}}_{\mathrm{w},k}^{\mathrm{H}}\left( \mathbf{\tilde{w}}^{(n)}+\eta \left( \mathbf{\tilde{w}}^{(m)}-\mathbf{\tilde{w}}^{(n)} \right) \right) \right\} \notag
		\\
		&\overset{\left( a4 \right)}{\leqslant} \lambda _{\max}\left( \mathbf{\tilde{C}}_{\mathrm{w},k}\mathbf{\tilde{C}}_{\mathrm{w},k}^{\mathrm{H}} \right) \left\| \mathbf{\tilde{w}}^{(n)}+\eta \left( \mathbf{\tilde{w}}^{(m)}-\mathbf{\tilde{w}}^{(n)} \right) \right\| _{2}^{2}+\left\| \boldsymbol{\tilde{b}}_{\mathrm{w},k} \right\| _{2}^{2}-2\mathrm{Re}\left\{ \boldsymbol{\tilde{b}}_{\mathrm{w},k}^{\mathrm{H}}\mathbf{\tilde{C}}_{\mathrm{w},k}^{\mathrm{H}}\left( \mathbf{\tilde{w}}^{(n)}+\eta \left( \mathbf{\tilde{w}}^{(m)}-\mathbf{\tilde{w}}^{(n)} \right) \right) \right\} \notag
		\\
		&\overset{\left( a5 \right)}{\leqslant} P\lambda _{\max}\left( \mathbf{\tilde{C}}_{\mathrm{w},k}\mathbf{\tilde{C}}_{\mathrm{w},k}^{\mathrm{H}} \right) +\left\| \boldsymbol{\tilde{b}}_{\mathrm{w},k} \right\| _{2}^{2}+2\sqrt{P}\left\| \mathbf{\tilde{C}}_{\mathrm{w},k}\boldsymbol{\tilde{b}}_{\mathrm{w},k} \right\| _2 \tag{122}
	\end{align}
\end{figure*}

\begin{figure*}[hb]
	\hrulefill
	\begin{equation} \label{N_phi}
		\mathbf{\bar{N}}_{\phi}=-\sum_{k=1}^K{\tilde{g}_{\phi ,k}\left( \eta \right) \left( \left[ \begin{matrix}
				\mathbf{\tilde{C}}_{\phi ,k}&		0\\
				0&		\mathbf{\tilde{C}}_{\phi ,k}^{\mathrm{T}}\\
			\end{matrix} \right] \!+\zeta \left[ \begin{array}{c}
				\mathbf{o}_k\\
				\mathbf{o}_{k}^{*}\\
			\end{array} \right] \left[ \begin{array}{c}
				\mathbf{o}_k\\
				\mathbf{o}_{k}^{*}\\
			\end{array} \right] ^{\mathrm{H}} \right)}+\zeta \left[ \!\begin{array}{c}
			\sum_{k=1}^K{\hat{g}_{\phi ,k}\left( \eta \right) \mathbf{o}_k}\\
			\sum_{k=1}^K{\hat{g}_{\phi ,k}\left( \eta \right) \mathbf{o}_{k}^{*}}\\
		\end{array} \right] \left[ \begin{array}{c}
			\sum_{k=1}^K{\hat{g}_{\phi ,k}\left( \eta \right) \mathbf{o}_k}\\
			\sum_{k=1}^K{\hat{g}_{\phi ,k}\left( \eta \right) \mathbf{o}_{k}^{*}}\\
		\end{array} \right] ^{\mathrm{H}}
		\tag{131}
	\end{equation}
\end{figure*}

\begin{figure*}[hb]
	\hrulefill
	\begin{align} \label{b_MM}
		\bar{\beta}&=\lambda _{\min}\left( \mathbf{\bar{N}}_{\phi} \right) \overset{\left( a1 \right)}{\geqslant}-\sum_{k=1}^K{\hat{g}_{\phi ,k}\left( \eta \right) \left( \lambda _{\max}\left( \left[ \begin{matrix}
				\mathbf{\tilde{C}}_{\phi ,k}&		0\\
				0&		\mathbf{\tilde{C}}_{\phi ,k}^{\mathrm{T}}\\
			\end{matrix} \right] \right) \!+\zeta \lambda _{\max}\left( \left[ \begin{array}{c}
				\mathbf{o}_k\\
				\mathbf{o}_{k}^{*}\\
			\end{array} \right] \left[ \begin{array}{c}
				\mathbf{o}_k\\
				\mathbf{o}_{k}^{*}\\
			\end{array} \right] ^{\mathrm{H}} \right) \right)} \notag
		\\
		&\quad +\zeta \lambda _{\min}\left( \left[ \!\begin{array}{c}
			\sum_{k=1}^K{\hat{g}_{\phi ,k}\left( \eta \right) \mathbf{o}_k}\\
			\sum_{k=1}^K{\hat{g}_{\phi ,k}\left( \eta \right) \mathbf{o}_{k}^{*}}\\
		\end{array} \right] \left[ \begin{array}{c}
			\sum_{k=1}^K{\hat{g}_{\phi ,k}\left( \eta \right) \mathbf{o}_k}\\
			\sum_{k=1}^K{\hat{g}_{\phi ,k}\left( \eta \right) \mathbf{o}_{k}^{*}}\\
		\end{array} \right] ^{\mathrm{H}} \right) \notag
		\\
		&\overset{\left( a2 \right)}{=}-\sum_{k=1}^K{\hat{g}_{\phi ,k}\left( \eta \right) \left( \lambda _{\max}\left( \mathbf{\tilde{C}}_{\phi ,k} \right) \!+2\zeta \mathbf{o}_{k}^{\mathrm{H}}\mathbf{o}_k \right)}\overset{\left( a3 \right)}{\geqslant}-\max_k \left\{ \lambda _{\max}\left( \mathbf{\tilde{C}}_{\phi ,k} \right) \right\} -2\zeta \max_k \left\{ \left\| \mathbf{o}_k \right\|_2 ^2 \right\} .\tag{134}
	\end{align} 
\end{figure*}

\begin{figure*}[hb]
	\hrulefill
	\begin{align} \label{o_MM}
		\left\| \mathbf{o}_k \right\| _{2}^{2}&=\left\| \boldsymbol{\tilde{b}}_{\phi ,k}-\mathbf{\tilde{C}}_{\phi ,k}^{\mathrm{H}}\left( \boldsymbol{\phi }^{(n)}+\eta \left( \boldsymbol{\phi }^{(m)}-\boldsymbol{\phi }^{(n)} \right) \right) \right\| _{2}^{2} \notag
		\\
		&=\left\| \boldsymbol{\tilde{b}}_{\phi ,k} \right\| _{2}^{2}+\left\| \mathbf{\tilde{C}}_{\phi ,k}^{\mathrm{H}}\left( \boldsymbol{\phi }^{(n)}+\eta \left( \boldsymbol{\phi }^{(m)}-\boldsymbol{\phi }^{(n)} \right) \right) \right\| _{2}^{2}-2\mathrm{Re}\left\{ \boldsymbol{\tilde{b}}_{\phi ,k}^{\mathrm{H}}\mathbf{\tilde{C}}_{\phi ,k}^{\mathrm{H}}\left( \boldsymbol{\phi }^{(n)}+\eta \left( \boldsymbol{\phi }^{(m)}-\boldsymbol{\phi }^{(n)} \right) \right) \right\} \notag
		\\
		&\overset{\left( a4 \right)}{\leqslant} \lambda _{\max}\left( \mathbf{\tilde{C}}_{\phi ,k}\mathbf{\tilde{C}}_{\phi ,k}^{\mathrm{H}} \right) \left\| \boldsymbol{\phi }^{(n)}+\eta \left( \boldsymbol{\phi }^{(m)}-\boldsymbol{\phi }^{(n)} \right) \right\| _{2}^{2}+\left\| \boldsymbol{\tilde{b}}_{\phi ,k} \right\| _{2}^{2}-2\mathrm{Re}\left\{ \boldsymbol{\tilde{b}}_{\phi ,k}^{\mathrm{H}}\mathbf{\tilde{C}}_{\phi ,k}^{\mathrm{H}}\left( \boldsymbol{\phi }^{(n)}+\eta \left( \boldsymbol{\phi }^{(m)}-\boldsymbol{\phi }^{(n)} \right) \right) \right\} \notag
		\\
		&\overset{\left( a6 \right)}{\leqslant} M\lambda _{\max}\left( \mathbf{\tilde{C}}_{\phi ,k}\mathbf{\tilde{C}}_{\phi ,k}^{\mathrm{H}} \right) +\left\| \boldsymbol{\tilde{b}}_{\phi ,k} \right\| _{2}^{2}+2\left\| \mathbf{\tilde{C}}_{\phi ,k}\boldsymbol{\tilde{b}}_{\phi ,k} \right\| _1. \tag{136}
	\end{align}
\end{figure*}

\subsection{Impact of the Maximum Transmit Power}
Figure \ref{PowerFigure} illustrates the impact of the maximum transmit power on the WMSR.
In this context, it is worth recalling that the distortion noise at the transceiver is assumed to be proportional to the signal power.
Hence, increasing the signal power improves the SNR, but it increases the performance loss caused by the presence of HIs as well.
We see that the security performance gap between the \textbf{Non-Robust} and the \textbf{BCD-MM} algorithms gradually increases as the transmit power increases.
This is because the \textbf{Non-Robust} algorithm does not account for the HIs by design, and its performance degradation is more prominent. 
Additionally, we see that the WMSR of  the \textbf{Non-Robust} algorithm gradually decreases when the transmit power is greater than 32 dBm, which further strengthen the necessity of designing robust algorithms in the high transmit power regime.
Furthermore, we see that the security performance of the \textbf{BCD-MM} algorithm is always better than the \textbf{BCD-MM-SDR} algorithm, which substantiates the superiority of the proposed MM algorithm for solving the subproblem of the reflection coefficient vector $\boldsymbol{\phi}$ over the SDR method.

\subsection{Impact of the Number of RIS Elements}
Figure \ref{MFigure} illustrates the WMSR as a function of the number of RIS elements.
As expected, increasing the number of RIS elements improves the secrecy rate.
We can also see a diminishing return law as a function of the RIS elements. 
Fortunately, this decreasing trend is not very significant to our \textbf{BCD-MM} algorithm for the range of $M$ from 8 to 64.
However, the WMSR of the \textbf{Non-Robust} and \textbf{BCD-MM-Rand} algorithms is significantly lower than that of the \textbf{BCD-MM-2bit} algorithm, which further corroborates the advantages of the proposed robust design against the HIs.
Furthermore, the WMSR of the \textbf{BCD-MM-No-RIS} algorithm is much lower than that of the \textbf{BCD-MM-2bit} algorithm, which demonstrates the potential benefits of deploying an RIS for enhancing the secrecy rate.

\section{Conclusion} \label{conclusion}
In this paper, we studied the secrecy rate of an RIS-aided multi-user wireless network in the presence of hardware impairments.
We demonstrated that the deployment of an RIS can effectively increase the secrecy rate of the legitimate users through appropriate adjustment of the RIS phase shifts and the precoding matrix of the BS.
We introduced a BCD framework for jointly optimizing the precoding at the BS and the phase shifts of the RIS.
Specifically, we decoupled the original problem into two tractable subproblems and  proposed an SOCP-based algorithm to alternately optimize them.
To reduce the computational complexity, we  proposed an MM algorithm based on surrogate functions that are formulated in a closed-form expression.
Simulation results demonstrated the advantages of the proposed robust transmission design that accounts for the hardware impairments by design, as well as the computational efficiency of the proposed solutions in terms of number of iterations and CPU time.

% 开始附录
\begin{appendices}

% 附录1
\section{Proof of Lemma 3}\label{appendixA}
By introducing the auxiliary variables $p_{\mathrm{w}}$ and
$\mathbf{q}_{\mathrm{w}}$, $f_{3}\left( \mathbf{\tilde{w}} \right)$ in \eqref{f_3} can be written as 
{\small
\begin{align}
&f_3\left( \mathbf{\tilde{w}} \right) =\log \left( 1+\frac{\mathrm{Tr}\left[ \mathbf{\bar{H}}_{\mathrm{E}}^{\mathrm{H}}\kappa _{\mathrm{t}}\mathrm{diag}\left( \mathbf{WW}^{\mathrm{H}} \right) \mathbf{\bar{H}}_{\mathrm{E}} \right]}{\delta _{\mathrm{E}}^{2}} \right) \notag
\\
&=\log \left( 1+\frac{\mathrm{Tr}\left[ \mathbf{W}^{\mathrm{H}}\kappa _{\mathrm{t}}\mathrm{diag}\left( \mathbf{\bar{H}}_{\mathrm{E}}\mathbf{\bar{H}}_{\mathrm{E}}^{\mathrm{H}} \right) \mathbf{W} \right]}{\delta _{\mathrm{E}}^{2}} \right) \notag
\\
&=\log \left( 1+\frac{\kappa _{\mathrm{t}}\mathbf{\tilde{w}}^{\mathrm{H}}\left( \mathbf{I}_K\otimes \mathrm{diag}\left( \mathbf{\bar{H}}_{\mathrm{E}}\mathbf{\bar{H}}_{\mathrm{E}}^{\mathrm{H}} \right) \right) \mathbf{\tilde{w}}}{\delta _{\mathrm{E}}^{2}} \right) \notag
\\
&\overset{\left( b1 \right)}{=}\log \left( 1+\frac{\mathbf{\tilde{w}}^{\mathrm{H}}\mathbf{LL}^{\mathrm{T}}\mathbf{\tilde{w}}}{\delta _{\mathrm{E}}^{2}} \right) \notag
\\
&=-\log \left( \frac{\delta _{\mathrm{E}}^{2}}{\left\| \mathbf{L}^{\mathrm{T}}\mathbf{\tilde{w}} \right\|_2^2+\delta _{\mathrm{E}}^{2}} \right) \notag
\\
&\overset{\left( a1 \right)}{\geqslant}-p_{\mathrm{w}}\left( \frac{\delta _{\mathrm{E}}^{2}}{\left\| \mathbf{L}^{\mathrm{T}}\mathbf{\tilde{w}} \right\|_2^{2}+\delta _{\mathrm{E}}^{2}} \right) +\log p_{\mathrm{w}}+1 \notag
\\
&\overset{\left( a2 \right)}{\geqslant}-p_{\mathrm{w}}\left( \left( \left\| \mathbf{L}^{\mathrm{T}}\mathbf{\tilde{w}} \right\|_2 ^{2}+\delta _{\mathrm{E}}^{2} \right) \left\| \mathbf{q}_{\mathrm{w}} \right\|_2^2-2\mathrm{Re}\left\{ \mathbf{q}_{\mathrm{w}}^{\mathrm{H}}\mathbf{L}^{\mathrm{T}}\mathbf{\tilde{w}} \right\} +1 \right) \notag
\\
&\quad+\log p_{\mathrm{w}}+1 \notag
\\
&=-\mathbf{\tilde{w}}^{\mathrm{H}}\mathbf{\tilde{C}}_{3,\mathrm{w}}\mathbf{\tilde{w}}+2\mathrm{Re}\left\{ \mathbf{\tilde{b}}_{3,\mathrm{w}}^{\mathrm{H}}\mathbf{\tilde{w}} \right\} +\tilde{c}_{3,\mathrm{w}}=\tilde{f}_{3,\mathbf{\tilde{w}}}\left( \mathbf{\tilde{w}} \right),
\end{align} }
where $\mathbf{\tilde{C}}_{3,\mathrm{w}}$, $\mathbf{\tilde{b}}_{3,\mathrm{w}}$ and $\tilde{c}_{3,\mathrm{w}}$ are given in \eqref{f_3_w_paramter}, and the Cholesky decomposition is applied in step $(b1)$. 
Also, Lemma 1 and Lemma 2 are applied in steps $(a1)$ and $(a2)$, respectively.
In step $(a1)$, we set $\bar{x} = \frac{\delta _{\mathrm{E}}^{2}}{\left\| \mathbf{L}^{\mathrm{T}}\mathbf{\tilde{w}} \right\|_2 ^2+\delta _{\mathrm{E}}^{2}}$ and $\bar{y} = p_{\mathrm{w}}$, and the optimal solution for $p_{\mathrm{w}}$ is given in \eqref{optimal_p_w}.
In step $(a2)$, we set $\mathbf{\bar{x}} = \mathbf{L}^{\mathrm{T}}\mathbf{\tilde{w}}$ and $\mathbf{\bar{y}} = \mathbf{q}_{\mathrm{w}}$, and the optimal solution for $\mathbf{q}_{\mathrm{w}}$ is given in \eqref{optimal_g_w}.

Hence, the proof is completed.

% 附录2
\section{Proof of Lemma 4}\label{appendixB}

Similar to Lemma 3, we introduce the auxiliary variables $p_{\phi}$ and $\mathbf{q}_{\phi}$. 
Then, $f_{3}\left( \boldsymbol{\phi } \right)$ in \eqref{f_3} can be formulated as
{\small 
		\begin{align} \label{lemma4_1}
			&f_3\left( \boldsymbol{\phi } \right) \overset{\left( b2 \right)}{=}\log \left( 1+\frac{\mathrm{Tr}\left[ \mathbf{\bar{H}}_{\mathrm{E}}^{\mathrm{H}}\mathbf{JJ}^{\mathrm{T}}\mathbf{\bar{H}}_{\mathrm{E}} \right]}{\delta _{\mathrm{E}}^{2}} \right) \notag
			\\
			&=-\log \left( \frac{\delta _{\mathrm{E}}^{2}}{\left\| \mathbf{J}^{\mathrm{T}}\mathbf{\bar{H}}_{\mathrm{E}} \right\| _{F}^{2}+\delta _{\mathrm{E}}^{2}} \right) \notag
			\\
			&\overset{\left( a3 \right)}{\geqslant}-p_{\phi}\left( \frac{\delta _{\mathrm{E}}^{2}}{\left\| \mathbf{J}^{\mathrm{T}}\mathbf{\bar{H}}_{\mathrm{E}} \right\| _{F}^{2}+\delta _{\mathrm{E}}^{2}} \right) +\log p_{\phi}+1 \notag
			\\
			&\overset{\left( a4 \right)}{\geqslant}-p_{\phi}\left( \left( \left\| \mathbf{J}^{\mathrm{T}}\mathbf{\bar{H}}_{\mathrm{E}} \right\| _{F}^{2}+\delta _{\mathrm{E}}^{2} \right) \left\| \mathbf{Q}_{\phi} \right\| _{F}^{2}\!-\!2\mathrm{Re}\left\{ \mathrm{Tr}\left[\! \mathbf{\bar{H}}_{\mathrm{E}}^{\mathrm{H}}\mathbf{JQ}_{\phi}\! \right]\! \right\} \!+\!1 \!\right) \notag
			\\
			&\quad +\log p_{\phi}+1,
		\end{align}}
where the Cholesky decomposition is applied in step $(b2)$. 
Also, Lemma 2 and Lemma 3 are applied in steps $(a3)$ and $(a4)$, respectively.
In step $(a3)$, we set $\bar{x}=\frac{\delta _{\mathrm{E}}^{2}}{\left\| \mathbf{J}^{\mathrm{T}}\mathbf{\bar{H}}_{\mathrm{E}} \right\| _{F}^{2}+\delta _{\mathrm{E}}^{2}}
$ and $\bar{y} = p_{\phi}$, and the optimal solution for $p_{\phi}$ is given in \eqref{optimal_p_phi}.
In step $(a4)$, we set $\mathbf{\bar{X}}=\mathbf{J}^{\mathrm{T}}\mathbf{\bar{H}}_{\mathrm{E}}$ and $\mathbf{\bar{y}}=\mathbf{Q}_{\phi}$, and the optimal solution for $\mathbf{Q}_{\phi}^{\mathrm{opt}}$ is given in \eqref{optimal_g_phi}.

Next, we rewrite $\left\| \mathbf{J}^{\mathrm{T}}\mathbf{\bar{H}}_{\mathrm{E}} \right\| _{F}^{2}$ and $\mathrm{Tr}\left[ \mathbf{\bar{H}}_{\mathrm{E}}^{\mathrm{H}}\mathbf{JQ}_{\phi} \right]$ to further simplify \eqref{lemma4_1}.
$\left\| \mathbf{J}^{\mathrm{T}}\mathbf{\bar{H}}_{\mathrm{E}} \right\| _{F}^{2}$ can be rewritten as
{\small\begin{align}
&\left\| \mathbf{J}^{\mathrm{T}}\mathbf{\bar{H}}_{\mathrm{E}} \right\| _{F}^{2}=\mathbf{JJ}^{\mathrm{T}}\mathbf{\hat{h}}_{\mathrm{E}}\mathbf{\hat{h}}_{\mathrm{E}}^{\mathrm{H}}+\mathbf{JJ}^{\mathrm{T}}\mathbf{\hat{H}}_{\mathrm{E}}\mathbf{\hat{H}}_{\mathrm{E}}^{\mathrm{H}} \notag
\\
&=\frac{4}{\pi ^2}\mathrm{Tr}\left[ \mathbf{\Phi }^{\mathrm{H}}\mathbf{h}_{\mathrm{RE}}\mathbf{h}_{\mathrm{RE}}^{\mathrm{H}}\mathbf{\Phi H}_{\mathrm{BR}}\mathbf{JJ}^{\mathrm{T}}\mathbf{H}_{\mathrm{BR}}^{\mathrm{H}} \right]  +\mathbf{JJ}^{\mathrm{T}}\mathbf{h}_{\mathrm{BE}}\mathbf{h}_{\mathrm{BE}}^{\mathrm{H}} \notag
\\
&\quad+2\mathrm{Re}\left\{ \frac{2}{\pi}\mathbf{JJ}^{\mathrm{T}}\mathbf{h}_{\mathrm{BE}}\mathbf{h}_{\mathrm{RE}}^{\mathrm{H}}\mathbf{\Phi H}_{\mathrm{BR}} \right\} \notag
\\
&\quad+\mathrm{Tr}\left[ \mathbf{\Phi }^{\mathrm{H}}\mathrm{diag}\left( \mathbf{h}_{\mathrm{RE}} \right) \mathbf{TT}^{\mathrm{T}}\mathrm{diag}\left( \mathbf{h}_{\mathrm{RE}}^{\mathrm{H}} \right) \mathbf{\Phi H}_{\mathrm{BR}}\mathbf{JJ}^{\mathrm{T}}\mathbf{H}_{\mathrm{BR}}^{\mathrm{H}} \right] \notag
\\
&=\boldsymbol{\phi }^{\mathrm{H}}\mathbf{C}_{3,\phi}\boldsymbol{\phi }+2\mathrm{Re}\left\{ \mathbf{b}_{3,\phi}^{\mathrm{H}}\boldsymbol{\phi } \right\} +\mathbf{h}_{\mathrm{BE}}^{\mathrm{H}}\mathbf{JJ}^{\mathrm{T}}\mathbf{h}_{\mathrm{BE}},
\end{align}}
where $\mathbf{C}_{3,\phi}$, $\mathbf{{b}}_{3,\phi}$ are given in \eqref{f_3_phi_paramter_2}.
$\mathrm{Tr}\left[ \mathbf{\bar{H}}_{\mathrm{E}}^{\mathrm{H}}\mathbf{JQ}_{\phi} \right] $ can be rewritten as
{\small\begin{align}
&\mathrm{Tr}\left[ \mathbf{\bar{H}}_{\mathrm{E}}^{\mathrm{H}}\mathbf{JQ}_{\phi} \right] =\mathrm{Tr}\left[ \mathbf{JQ}_{\phi}\mathbf{\bar{H}}_{\mathrm{E}}^{\mathrm{H}} \right] \notag
\\
&=\mathrm{Tr}\left[ \mathbf{J}\left[ \begin{matrix}
	\mathbf{\hat{q}}_{\phi}&		\mathbf{\hat{Q}}_{\phi}\\
\end{matrix} \right] \left[ \begin{array}{c}
	\mathbf{\hat{h}}_{\mathrm{E}}^{\mathrm{H}}\\
	\mathbf{\hat{H}}_{\mathrm{E}}^{\mathrm{H}}\\
\end{array} \right] \right] =\mathrm{Tr}\left[ \mathbf{J\hat{q}}_{\phi}\mathbf{\hat{h}}_{\mathrm{E}}^{\mathrm{H}}+\mathbf{J\hat{Q}}_{\phi}\mathbf{\hat{H}}_{\mathrm{E}}^{\mathrm{H}} \right] \notag
\\
&=\mathrm{Tr}\left[ \left( \frac{2}{\pi}\mathbf{H}_{\mathrm{BR}}\mathbf{J\hat{q}}_{\phi}\mathbf{h}_{\mathrm{RE}}^{\mathrm{H}}+\mathbf{H}_{\mathrm{BR}}\mathbf{J\hat{Q}}_{\phi}\mathbf{T}^{\mathrm{T}}\mathrm{diag}\left( \mathbf{h}_{\mathrm{RE}}^{\mathrm{H}} \right) \right) \mathbf{\Phi } \right] \notag 
\\
&\quad+\mathrm{Tr}\left[ \mathbf{J\hat{q}}_{\phi}\mathbf{h}_{\mathrm{BE}}^{\mathrm{H}} \right] \notag
\\
&=\mathbf{a}_{3,\phi}^{\mathrm{H}}\boldsymbol{\phi }+\mathrm{Tr}\left[ \mathbf{J\hat{q}}_{\phi}\mathbf{h}_{\mathrm{BE}}^{\mathrm{H}} \right] 
\end{align}}
where $\mathbf{a}_{3,\phi}$ is given in \eqref{f_3_phi_paramter_2}.
Then, \eqref{lemma4_1} can be further written as
{\small\begin{align}
f_3\left( \boldsymbol{\phi } \right) \geqslant -\boldsymbol{\phi }^{\mathrm{H}}\mathbf{\tilde{C}}_{3,\phi}\boldsymbol{\phi }+2\mathrm{Re}\left\{ \mathbf{\tilde{b}}_{3,\phi}^{\mathrm{H}}\boldsymbol{\phi } \right\} +\tilde{c}_{3,\phi}\triangleq \tilde{f}_3\left( \boldsymbol{\phi } \right).
\end{align}},
where $\mathbf{\tilde{C}}_{3,\phi}$, $\mathbf{\tilde{b}}_{3,\phi}$ and $\tilde{c}_{3,\phi}$ are given in \eqref{f_3_phi_paramter_1}.

Hence, the proof is completed.

% 附录3
\section{Proof of Lemma 5}\label{appendixC}
Considering that the objective function $\tilde{r}_k\left( \mathbf{\tilde{w}} \right)$ of the optimization problem in \eqref{WMMSE_Problem_W} is a quadratic function, we assume that there exists a minorizing function $\bar{f}\left( \mathbf{\tilde{w}} \right)$ satisfying the following quadratic form

\begin{align}\label{tildefF_origin}
\bar{f}\left( \mathbf{\tilde{w}}|\mathbf{\tilde{w}}^{(n)} \right) =f\left( \mathbf{\tilde{w}}^{(n)} \right)+2\mathop {\mathrm{Re}}_{}\left\{ \mathbf{g}_{\mathrm{w}}^{\mathrm{H}}\left( \mathbf{\tilde{w}}-\mathbf{\tilde{w}}^{(n)} \right) \right\} \notag
\\
+\left( \mathbf{\tilde{w}}-\mathbf{\tilde{w}}^{(n)} \right) ^{\mathrm{H}}\mathbf{M}_{\mathrm{w}}\left( \mathbf{\tilde{w}}-\mathbf{\tilde{w}}^{(n)} \right),
\end{align}
where ${\mathbf{g}_{\mathrm{w}}}\in {\mathbb{C}}^{{NK} \times 1}$ and ${\mathbf{M}_{\mathrm{w}}}\in {\mathbb{C}}^{{NK} \times {NK}}$ are parameters to be determined.
Denote $\tilde{\mathcal{S}}_{\mathrm{w}}\triangleq \left\{ \mathbf{\tilde{w}}|\mathbf{\tilde{w}}^{\mathrm{H}}\mathbf{\tilde{w}}\leqslant P \right\} 
$.
For $\bar{f}\left( \mathbf{\tilde{w}} \right)$ to be a minorizing function, it needs to fulfill the  following conditions:

\noindent(C1) $\bar{f}( \mathbf{\tilde{w}}|\mathbf{\tilde{w}}^{(n)} )$
is continuous in $\mathbf{\tilde{w}}$ and $\mathbf{\tilde{w}}^{(n)}$;

\noindent(C2) $\bar{f}( \mathbf{\tilde{w}}^{(n)}|\mathbf{\tilde{w}}^{(n)} ) =f( \mathbf{\tilde{w}}^{(n)}) ,\forall \mathbf{\tilde{w}}^{(n)}\in \mathcal{\tilde{S}}_{\mathrm{w}}$;

\noindent(C3) $\bar{f}( \mathbf{\tilde{w}}|\mathbf{\tilde{w}}^{(n)} ) \leqslant f( \mathbf{\tilde{w}} ) ,\forall \mathbf{\tilde{w}},\mathbf{\tilde{w}}^{(n)}\in \mathcal{\tilde{S}}_{\mathrm{w}}$;

\noindent(C4) $ \bar{f}^{\prime}( \mathbf{\tilde{w}}^{(n)}|\mathbf{\tilde{w}}^{(n)};\eta ) |_{\mathbf{\tilde{w}}=\mathbf{\tilde{w}}^{(n)}}=f^{\prime}( \mathbf{\tilde{w}}^{(n)};\eta ) ,\forall \eta \,$with$\,\mathbf{\tilde{w}}^{(n)}+\eta \in \mathcal{\tilde{S}}_{\mathrm{w}}$.

To this end, we derive ${\mathbf{g}_{\mathrm{w}}}$ and ${\mathbf{M}_{\mathrm{w}}}$ that satisfy the conditions (C1) - (C4). 
Since $\bar{f}\left( {\mathbf{\tilde{w}}} \right)$ is a quadratic function, in addition, the condition (C1) is satisfied.
By substituting $\mathbf{\tilde{w}}^{(n)}$ into $\bar{f}\left( {\mathbf{\tilde{w}}|{\mathbf{\tilde{w}}^{(n)}}} \right)$, it can be verified that $\bar{f}\left( {\mathbf{\tilde{w}}} \right)$ satisfies condition (C2).
Hence,  $\mathbf{g}_{\mathrm{w}}$ and $\mathbf{M}_{\mathrm{w}}$ need to be determined in order to fulfill the conditions (C3) and (C4).

First, we derive an expression for $\mathbf{g}_{\mathrm{w}}$ that fulfills (C4), which requires that the first-order derivatives of $f\left( {{{\mathbf{\tilde{w}}}^{(n)}}} \right)$ and $\bar{f}\left( {\mathbf{\tilde{w}}|{\mathbf{\tilde{w}}^{(n)}}} \right)$ are equal in any direction.
Let $\mathbf{\tilde{w}}^{m}$ belongs to ${{\cal S}_{\mathrm{w}}}$. 
The directional derivative of $f\left( {\mathbf{\tilde{w}}} \right)$ in the direction ${\mathbf{\tilde{w}}^{m}}-{\mathbf{\tilde{w}}^{(n)}}$ is given by
{\small\begin{equation}\label{direc_deriv_fF}
2\mathrm{Re}\left\{ \sum_{k=1}^K{h_{\mathrm{w},k}\left( \mathbf{\tilde{w}}^{(n)} \right)\! \left(\! \boldsymbol{\tilde{b}}_{\mathrm{w},k}^{\mathrm{H}}\!-\!\left( \mathbf{\tilde{w}}^{(n)} \right) ^{\mathrm{H}}\mathbf{\tilde{C}}_{\mathrm{w},k}\! \right) \!\left(\! \mathbf{\tilde{w}}^{(m)}\!-\!\mathbf{\tilde{w}}^{(n)} \!\right)}\! \right\}\!,\!
\end{equation}}
where ${h_{\mathrm{w},k}}\left( {{{\mathbf{\tilde{w}}}^{(n)}}} \right)$ is defined in \eqref{lemma4_h}.
Moreover, the directional derivative of $\tilde f\left( {\mathbf{\tilde{w}}|{\mathbf{\tilde{w}}^{(n)}}} \right)$ in \eqref{tildefF_origin} evaluated at $\mathbf{\tilde{w}}^{(n)}$ in the same direction is given by
\begin{equation}\label{direc_deriv_tildefF}
2\mathop {\mathrm{Re}}_{}\left\{ \mathbf{g}_{\mathrm{w}}^{\mathrm{H}}\left( \mathbf{\tilde{w}}^{(m)}-\mathbf{\tilde{w}}^{(n)} \right) \right\}.
\end{equation}
Therefore, the vector $\mathbf{g}_{\mathrm{w}}$ is derived as
\begin{equation}
\mathbf{g}_{\mathrm{w}} = \sum_{k=1}^K{h_{\mathrm{w},k}\left( \mathbf{\tilde{w}}^{(n)} \right) \left( \boldsymbol{\tilde{b}}_{\mathrm{w},k}-\mathbf{\tilde{C}}_{\mathrm{w},k}^{\mathrm{H}}\mathbf{\tilde{w}}^{(n)} \right)}.
\end{equation}

Then, we derive an expression for $\mathbf{M}_{\mathrm{w}}$ that fulfills (C3). 
This requires that $\tilde f\left( {\mathbf{\tilde{w}}|{\mathbf{\tilde{w}}^{(n)}}} \right)$ is a lower bound of $f\left( \mathbf{\tilde{w}} \right)$ for each linear cut in any direction.
Therefore, for any auxiliary variable $\eta \in \left[0,1\right]$ and $\mathbf{\tilde{w}}^{(m)} \in {\cal S}_{\mathrm{w}}$, $\mathbf{M}_{\mathrm{w}}$ needs to be chosen so that the following expression is fulfilled
{\small\begin{align}\label{sufficient_A2}
&f( \mathbf{\tilde{w}}^{(n)}+\eta ( \mathbf{\tilde{w}}^{(m)}-\mathbf{\tilde{w}}^{(n)} ) ) \geqslant f( \mathbf{\tilde{w}}^{(n)} ) +2\eta \mathrm{Re}\left\{ \mathbf{g}_{\mathrm{w}}^{\mathrm{H}}( \mathbf{\tilde{w}}-\mathbf{\tilde{w}}^{(n)} ) \right\} \notag
\\
&\qquad\qquad+\eta ^2( \mathbf{\tilde{w}}-\mathbf{\tilde{w}}^{(n)} ) ^{\mathrm{H}}\mathbf{M}_{\mathrm{w}}( \mathbf{\tilde{w}}-\mathbf{\tilde{w}}^{(n)} ).
\end{align}}

\vspace{-0.5cm}
Let us define $\bar{m}_{\mathrm{w}}\left( \eta \right) \triangleq f\left( \mathbf{\tilde{w}}^{(n)}+\eta \left( \mathbf{\tilde{w}}^{(m)}-\mathbf{\tilde{w}}^{(n)} \right) \right) $ and $\bar{n}_{\mathrm{w}}\left( \eta \right) \triangleq f\left( \mathbf{\tilde{w}}^{(n)} \right) +2\eta \mathrm{Re}\left\{ \mathbf{g}_{\mathrm{w}}^{\mathrm{H}}\left( \mathbf{\tilde{w}}^{(m)}-\mathbf{\tilde{w}}^{(n)} \right) \right\} +\eta ^2\left( \mathbf{\tilde{w}}^{(m)}-\mathbf{\tilde{w}}^{(n)} \right) ^{\mathrm{H}}\mathbf{M}_{\mathrm{w}}\left( \mathbf{\tilde{w}}^{(m)}-\mathbf{\tilde{w}}^{(n)} \right) $.
By direct inspection, we note that $\bar{m}_{\mathrm{w}}\left( 0 \right)$ is equal to $\bar{n}_{\mathrm{w}}\left( 0 \right)$.
Also, the first-order derivative of $\bar{m}_{\mathrm{w}}\left( \eta \right)$ is given by
\begin{equation}\label{first_deriv_j}
	\nabla _{\eta}\bar{m}_{\mathrm{w}}\left( \eta \right) =\sum_{k=1}^K{\hat{g}_{\mathrm{w},k}\left( \eta \right) \nabla _{\eta}\hat{h}_{\mathrm{w},k}\left( \eta \right)},
\end{equation}
where
{\small\begin{subequations}
	\begin{align}
		&\hat{h}_{\mathrm{w},k}\left( \eta \right) \triangleq \tilde{r}_{\mathrm{w},k}\left( \mathbf{\tilde{w}}^{(n)}+\eta \left( \mathbf{\tilde{w}}^{(m)}-\mathbf{\tilde{w}}^{(n)} \right) \right),
		\\
		&\hat{g}_{\mathrm{w},k}\left( \eta \right) \triangleq \frac{\exp \left\{ -\zeta \hat{h}_{\mathrm{w},k}\left( \eta \right) \right\}}{\sum_{i=1}^K{\exp \left\{ -\zeta \hat{h}_{\mathrm{w},i}\left( \eta \right) \right\}}},
		\\
		&\nabla _{\eta}\hat{h}_{\mathrm{w},k}\left( \eta \right) =-2\eta \left( \mathbf{\tilde{w}}^{(m)}-\mathbf{\tilde{w}}^{(n)} \right) ^{\mathrm{H}}\mathbf{\tilde{C}}_{\mathrm{w},k}\left( \mathbf{\tilde{w}}^{(m)}-\mathbf{\tilde{w}}^{(n)} \right) \notag
		\\
		&\label{deriv_hat_hDk} +2\mathrm{Re}\!\left\{\! \boldsymbol{\tilde{b}}_{\mathrm{w},k}^{\mathrm{H}}\left( \mathbf{\tilde{w}}^{(m)}\!-\!\mathbf{\tilde{w}}^{(n)} \right) \!-\!\left( \mathbf{\tilde{w}}^{(n)} \right) ^{\mathrm{H}}\!\mathbf{\tilde{C}}_{\mathrm{w},k}\left( \mathbf{\tilde{w}}^{(m)}-\mathbf{\tilde{w}}^{(n)} \right) \!\right\}\!. 
	\end{align}
\end{subequations}}
It can be verified that $\nabla _{\eta}\bar{m}_{\mathrm{w}}\left( 0 \right)$ is equal to $\nabla _{\eta}\bar{n}_{\mathrm{w}}\left( 0 \right)$.
Hence, we obtain the sufficient condition for \eqref{sufficient_A2} as follows
\begin{equation}\label{2nd_deriv_m_geq_n_F}
\nabla _{\eta}^{2}\bar{m}_{\mathrm{w}}\left( \eta \right) \ge \nabla _{\eta}^{2}\bar{n}_{\mathrm{w}}\left( \eta \right) ,\forall \eta \in \left[ 0,1 \right].
\end{equation}

Next, we further manipulate \eqref{2nd_deriv_m_geq_n_F} to solve for $\mathbf{M}_{\mathrm{w}}$.
To this end, we define
\begin{align}
\mathbf{e}_{k}&\triangleq \boldsymbol{\tilde{b}}_{\mathrm{w},k}-\mathbf{\tilde{C}}_{\mathrm{w},k}^{\mathrm{H}}\left( \mathbf{\tilde{w}}^{(n)}+\eta \left( \mathbf{\tilde{w}}^{(m)}-\mathbf{\tilde{w}}^{(n)} \right) \right), 
\\
\mathbf{\bar{w}}&\triangleq \mathbf{\tilde{w}}^{(m)}-\mathbf{\tilde{w}}^{(n)},
\end{align}
so that $\nabla _\eta ^2 \bar{n}_{\mathrm{w}}\left( \eta \right)$ can be derived as
{\small\begin{align}\label{second_n_w}
\nabla _{\eta}^{2}\bar{n}_{\mathrm{w}}\left( \eta \right) &=2\left( \mathbf{\tilde{w}}^{(m)}-\mathbf{\tilde{w}}^{(n)} \right) ^{\mathrm{H}}\mathbf{M}_{\mathrm{w}}\left( \mathbf{\tilde{w}}^{(m)}-\mathbf{\tilde{w}}^{(n)} \right) \notag
\\
&=\left[ \begin{matrix}
	\mathbf{\bar{w}}^{\mathrm{H}}&		\mathbf{\bar{w}}^{\mathrm{T}}\\
\end{matrix} \right] \left[ \begin{matrix}
	\mathbf{M}_{\mathrm{w}}&		0\\
	0&		\mathbf{M}_{\mathrm{w}}^{\mathrm{T}}\\
\end{matrix} \right] \left[ \begin{array}{c}
	\mathbf{\bar{w}}\\
	\mathbf{\bar{w}}^*\\
\end{array} \right], 
\end{align}}
where we used $\mathrm{Tr}\left( \!\mathbf{ABC}\! \right) \!=\!\left( \!\mathrm{vec}\left( \mathbf{A}^{\mathrm{T}} \right) \!\right) ^{\mathrm{T}}\left( \mathbf{I}\! \otimes\! \mathbf{B} \right) \mathrm{vec}\left( \mathbf{C} \right)$\cite{Zhang2017Matrix}.

Similarly, $\nabla _{\eta}^{2}\bar{m}_{\mathrm{w}}\left( \eta \right)$ is given by
\begin{align}\label{second_m_w}
&\nabla _{\eta}^{2}\bar{m}_{\mathrm{w}}\left( \eta \right) \notag
\\
&=\sum_{k\in \mathcal{K}}{\left( \hat{g}_{\mathrm{w},k}\left( \eta \right) \nabla _{\eta}^{2}\hat{h}_{\mathrm{w},k}\left( \eta \right) -\zeta \hat{g}_{\mathrm{w},k}\left( \eta \right) \left( \nabla _{\eta}\hat{h}_{\mathrm{w},k}\left( \eta \right) \right) ^2 \right)} \notag
\\
&\quad +\zeta \left( \sum_{k\in \mathcal{K}}{\hat{g}_{\mathrm{w},k}\left( \eta \right) \nabla _{\eta}\hat{h}_{\mathrm{w},k}\left( \eta \right)} \right) ^2 \notag
\\
&=\left[ \begin{matrix}
	\mathbf{\bar{w}}^{\mathrm{H}}&		\mathbf{\bar{w}}^{\mathrm{T}}\\
\end{matrix} \right] \mathbf{\bar{N}}_{\mathrm{w}}\left[ \begin{array}{c}
	\mathbf{\bar{w}}\\
	\mathbf{\bar{w}}^*\\
\end{array} \right],
\end{align}
where 
\begin{align}
\nabla _{\eta}\hat{h}_{\mathrm{w},k}\left( \eta \right) =2\mathrm{Re}\left\{ \mathbf{e}_{k}^{\mathrm{H}}\mathbf{\bar{w}} \right\},
\\
\nabla _{\eta}^{2}\hat{h}_{\mathrm{w},k}\left( \eta \right) =-2\mathbf{\bar{w}}^{\mathrm{H}}\mathbf{\tilde{C}}_{\mathrm{w},k}\mathbf{\bar{w}},
\end{align}
and $\mathbf{\bar{N}}_{\mathrm{w}}$ is given in \eqref{N_w_wave} at the bottom of the previous page. 
As a result, we have
\setcounter{equation}{117}
\begin{align}
\mathbf{\bar{N}}_{\mathrm{w}}\succeq \left[ \begin{matrix}
	\mathbf{M}_{\mathrm{w}}&		0\\
	0&		\mathbf{M}_{\mathrm{w}}^{\mathrm{T}}\\
\end{matrix} \right] . 
\end{align}
Choosing $\mathbf{M}_{\mathrm{w}}=\bar{\alpha}\mathbf{I}=\lambda _{\min}\left( \mathbf{\bar{N}}_{\mathrm{w}} \right) \mathbf{I}
$, \eqref{tildefF_origin} can be rewritten as
\begin{align}
\bar{f}\left( \mathbf{\tilde{w}}|\mathbf{\tilde{w}}^{(n)} \right) =\bar{c}_{\mathrm{w}}+2\mathrm{Re}\left\{ \mathbf{\bar{v}}_{\mathrm{w}}^{\mathrm{H}}\mathbf{\tilde{w}} \right\} +\bar{\alpha} \mathbf{\tilde{w}}^{\mathrm{H}}\mathbf{\tilde{w}},
\end{align}
where $\mathbf{\bar{v}}_{\mathrm{w}}$ and $\bar{c}_{\mathrm{w}}$ are given in \eqref{v_MM_w} and \eqref{cons_MM_w}, respectively.

However, the complexity of computing $\bar{\alpha}$ cannot be ignored.
We introduce the following lemmas to reduce the complexity:

\begin{enumerate}[({\textit{a}}1)]
\setlength{\itemsep}{0.7ex}
\item ${\lambda _{\min }}\left( {\bf{A}} \right) + {\lambda _{\min }}\left( {\bf{B}} \right) \le {\lambda _{\min }}\left( {{\bf{A}} + {\bf{B}}} \right)$, if ${\bf{A}}$ and ${\bf{B}}$ are Hermitian matrices \cite{lutkepohl1996handbook};
	
\item ${\lambda _{\max }}\left( {\bf{A}} \right) = {\mathop{\rm Tr}\nolimits} \left( {\bf{A}} \right)$ and ${\lambda _{\min }}\left( {\bf{A}} \right) = 0$, if ${\bf{A}}$ is a rank one matrix \cite{lutkepohl1996handbook};
	
\item $\sum\nolimits_{m = 1}^{(m)} {{a_m}{b_m}}  \le \max _{m = 1}^{(m)}\left\{ {{b_m}} \right\}$, if ${a_m},{b_m}\ge 0$ and $\sum\nolimits_{m = 1}^{(m)} {{a_m}}=1$ \cite[Theorem 30]{Matrix-differential};

\item ${\mathop{\rm Tr}\nolimits} \left( {{\bf{AB}}} \right) \le {\lambda _{\max }}\left( {\bf{A}} \right){\mathop{\rm Tr}\nolimits} \left( {\bf{B}} \right)$, if $\bf A$ and $\bf B$ are positive semidefinite matrices \cite{lutkepohl1996handbook}.
\end{enumerate}

Additionally, it can be readily verified that

\begin{enumerate}[({\textit{a}}5)]
\setlength{\itemsep}{0.7ex}
\item $ - \sqrt {{P}} {\left\| {{{\bf{B}}}{{\bf{c}}}} \right\|_2}$ is the solution to the following problem:
\begin{subequations}\label{problem_in_a5}
	\begin{alignat}{2}
		\min_{\bf{x}} \quad & {\mathop{\rm Re}\nolimits} \left\{  {{{\bf{c}}^{\rm H}}{{\bf{B}}^{\rm H}}{\bf{x}}}  \right\} \\
		\mbox{s.t.}\quad
		& {{{\bf{x}}^{\rm H}}{\bf{x}}} \le {P}.
	\end{alignat}
\end{subequations}
	
\end{enumerate}

Using the inequalities (\textit{a}1) and (\textit{a}3) and the equalities (\textit{a}2), a lower bound for $\bar{\alpha}$ can be derived as given in \eqref{a_MM} at the bottom of the previous page.

Recall that $\mathbf{\tilde{w}}=\mathbf{\tilde{w}}^{(n)}+\eta \left( \mathbf{\tilde{w}}^{(m)}-\mathbf{\tilde{w}}^{(n)} \right)$, $\forall \eta \in \left[ 0,1 \right]$,
thus $\left\| {\mathbf{\tilde{w}}=\mathbf{\tilde{w}}^{(n)}+\eta \left( \mathbf{\tilde{w}}^{(m)}-\mathbf{\tilde{w}}^{(n)} \right)} \right\|_2 \le \sqrt{P}$.
Furthermore, using (\textit{a}4) and (\textit{a}5), an upper bound for $\left\| {{{\bf{e}}_k}} \right\|_2^2$ can be derived as given in \eqref{e_MM} at the bottom of this page.

Finally, by combining \eqref{a_MM} with \eqref{e_MM}, we  obtain the simple lower bound for $\bar{\alpha}$ in \eqref{bar_alpha}.

Hence, the proof is completed.

% 附录3
\section{Proof of Lemma 6}\label{appendixD}
Similar to the proof of Lemma 5, we first assume that exists a quadratic function $\bar{f}\left( \boldsymbol{\phi }|\boldsymbol{\phi }^{(n)} \right)$ to minorize $f\left( \boldsymbol{\phi } \right)$ as follows 
\setcounter{equation}{122}
\begin{align}\label{f_phi_MM}
\bar{f}( \boldsymbol{\phi }|\boldsymbol{\phi }^{(n)} ) &\!=f( \boldsymbol{\phi }^{(n)}) +\!2\mathrm{Re}\{ \mathbf{g}_{\phi}^{\mathrm{H}}( \boldsymbol{\phi }-\boldsymbol{\phi }^{(n)} ) \} \!\! \notag
\\
&+( \boldsymbol{\phi }-\boldsymbol{\phi }^{(n)} ) ^{\mathrm{H}}\mathbf{M}_{\phi}( \boldsymbol{\phi }-\boldsymbol{\phi }^{(n)} ), 
\end{align}
where the parameters $\mathbf{M}_{\phi} \in \mathbb{C}^{M\times M}
$ and $\mathbf{g}_{\phi} \in \mathbb{C}^{M\times 1}$
are undetermined parameters.
The hypothesis holds if the following four conditions are satisfied:

\noindent(D1) $\bar{f}( \boldsymbol{\phi }|\boldsymbol{\phi }^{(n)} )$
is continuous in $\boldsymbol{\phi }$ and $\boldsymbol{\phi }^{(n)}$;

\noindent(D2) $\bar{f}( \boldsymbol{\phi }^{(n)}|\boldsymbol{\phi }^{(n)} ) =f( \boldsymbol{\phi }^{(n)}) ,\forall \boldsymbol{\phi }^{(n)}\in \mathcal{\tilde{S}}$;

\noindent(D3) $\bar{f}( \boldsymbol{\phi }|\boldsymbol{\phi }^{(n)} ) \leqslant f( \boldsymbol{\phi } ) ,\forall \boldsymbol{\phi },\boldsymbol{\phi }^{(n)}\in \mathcal{\tilde{S}}$;

\noindent(D4) $ \bar{f}^{\prime}( \boldsymbol{\phi }^{(n)}|\boldsymbol{\phi }^{(n)};\eta ) |_{\boldsymbol{\phi }=\boldsymbol{\phi }^{(n)}}=f^{\prime}( \boldsymbol{\phi }^{(n)};\eta ) ,\forall \eta \,$with$\,\boldsymbol{\phi }^{(n)}+\eta \in \mathcal{\tilde{S}}$.

\noindent Then, (D1) and (D2) is always satisfied,
Hence, the expression of $\mathbf{M}_{\phi}$ and $\mathbf{g}_{\phi}$ are determined by (D3) and (D4).

To satisfy the condition (D4), the directional derivatives of the left and right hand sides of \eqref{f_phi_MM} need to be equal in any direction, which yields
\begin{align}
\mathbf{g}_{\phi}=\sum_{k=1}^K{h_{\phi ,k}( \boldsymbol{\phi }^{(n)} ) \left( \mathbf{\tilde{C}}_{\phi ,k}^{\mathrm{H}}\boldsymbol{\phi }^{(n)} -\boldsymbol{\tilde{b}}_{\phi ,k} \right)},
\end{align}
where $h_{\phi ,k}( \boldsymbol{\phi }^{(n)}  )$ is defined in \eqref{h_phi}.

Then, we try to derive an expression for $\mathbf{M}_{\phi}$ that fulfills (D3).
If $\bar{f}( \boldsymbol{\phi }|\boldsymbol{\phi }^{(n)} )$ is the lower bound of $f( \boldsymbol{\phi } )$ for each linear cut in any direction, condition (D3) would hold.
Let $\boldsymbol{\phi }^{(m)}$ belongs to $\mathcal{\tilde{S}}$. 
By defining $\boldsymbol{\phi }=\boldsymbol{\phi }^{(n)} +\eta ( \boldsymbol{\phi }^{(m)} -\boldsymbol{\phi }^{(n)})$, $\forall \eta \in \left[0,1\right]$, (D3) can be reformulated as
\begin{align}\label{sufficient_A2_phi}
&f( \boldsymbol{\phi }^{(n)} +\eta ( \boldsymbol{\phi }^{(m)} -\boldsymbol{\phi }^{(n)})) \geqslant  2\eta \mathrm{Re}\{ \mathbf{g}_{\phi}^{\mathrm{H}}( \boldsymbol{\phi }^{(m)}-\boldsymbol{\phi }^{(n)} ) \} \notag
\\
&+f( \boldsymbol{\phi }^{(n)} ) + \eta ^2( \boldsymbol{\phi }^{(m)}-\boldsymbol{\phi }^{(n)} ) ^{\mathrm{H}}\mathbf{M}_{\phi}( \boldsymbol{\phi }^{(m)}-\boldsymbol{\phi }^{(n)} ). 
\end{align} 
Similar to Appendix C, \eqref{2nd_deriv_j_geq_J_F} is a sufficient condition for \eqref{sufficient_A2_phi} to hold, i.e.,
\begin{equation}\label{2nd_deriv_j_geq_J_F}
\nabla _\eta ^2 {\bar{m}_{\phi}}\left( \eta  \right) \geqslant \nabla _\eta ^2 {\bar{n}_{\phi}}\left( \eta  \right), \forall \eta \in \left[0,1\right],
\end{equation}
where $\bar{m}_{\boldsymbol{\phi }}\left( \eta \right) $ and $\bar{n}_{\boldsymbol{\phi }}\left( \eta \right) $ represent the left and right hand sides of \eqref{sufficient_A2_phi}, respectively.

Then, to obtain the expression for $\mathbf{M}_{\phi}$,  we calculate the second-order derivatives of $\bar{m}_{\boldsymbol{\phi }}\left( \eta \right) $ and $\bar{n}_{\boldsymbol{\phi }}\left( \eta \right) $.
We first define
\begin{align}
\mathbf{o}_k&\triangleq \mathbf{\tilde{C}}_{\phi ,k}^{\mathrm{H}}( \boldsymbol{\phi }^{(n)} +\eta ( \boldsymbol{\phi }^{(m)} -\boldsymbol{\phi }^{(n)}) ) -\boldsymbol{\tilde{b}}_{\phi ,k}, 
\\
\boldsymbol{\bar{\phi}} &\triangleq \boldsymbol{\phi }^{(m)}-\boldsymbol{\phi }^{(n)}, 
\end{align}
so that $\nabla _\eta ^2 {\bar{m}_{\phi}}\left( \eta  \right)$ can be written as
\begin{align}\label{l_2_phi}
	\nabla _\eta ^2 {\bar{m}_{\phi}}\left( \eta  \right)=\left[ \begin{matrix}
		\boldsymbol{\bar{\phi}}^{\mathrm{H}}&		\boldsymbol{\bar{\phi}}^{\mathrm{T}}\\
	\end{matrix} \right] \mathbf{\bar{N}}_{\phi}\left[ \begin{array}{c}
		\boldsymbol{\bar{\phi}}\\
		\boldsymbol{\bar{\phi}}*\\
	\end{array} \right], 
\end{align}
where
\begin{subequations}
\begin{align}
\hat{h}_{\phi ,k}\left( \eta \right) &\triangleq -\tilde{r}_{\phi ,k}( \boldsymbol{\phi }^{(n)} +\eta ( \boldsymbol{\phi }^{(m)} -\boldsymbol{\phi }^{(n)}) ),
\\
\hat{g}_{\phi ,k}\left( \eta \right) &\triangleq \frac{\exp \left\{ -\zeta\hat{h}_{\phi ,k}\left( \eta \right) \right\}}{\sum_{i=1}^K{\exp \left\{ -\zeta\hat{h}_{\phi ,i}\left( \eta \right) \right\}}},
\end{align}
and
\begin{align}\label{h_phi_one}
\nabla _{\eta}\hat{h}_{\phi ,k}\left( \eta \right) &\!=\!2\mathrm{Re}\left\{ \boldsymbol{\phi }^{\left( n \right) ,\mathrm{H}}\mathbf{\tilde{C}}_{\phi ,k}\boldsymbol{\bar{\phi}}-\boldsymbol{\tilde{b}}_{\phi ,k}\boldsymbol{\bar{\phi}} \right\} \!+\!2 \eta \boldsymbol{\bar{\phi}}^{\mathrm{H}}\mathbf{\tilde{C}}_{\phi ,k}\boldsymbol{\bar{\phi}}, 
\\
\nabla _{\eta}^{2}\hat{h}_{\phi ,k}\left( \eta \right) &\!=\!2\boldsymbol{\bar{\phi}}^{\mathrm{H}}\mathbf{\tilde{C}}_{\phi ,k}\boldsymbol{\bar{\phi}} \notag
\\
&\!=\!\left[ \begin{matrix}
		\boldsymbol{\bar{\phi}}^{\mathrm{H}}&		\boldsymbol{\bar{\phi}}^{\mathrm{T}}\\
	\end{matrix} \right] \left[ \begin{matrix}
		 \mathbf{\tilde{C}}_{\phi ,k}&		0\\
		0&		 \mathbf{\tilde{C}}_{\phi ,k}^{\mathrm{T}}\\
	\end{matrix} \right] \left[ \begin{array}{c}
		\boldsymbol{\bar{\phi}}\\
		\boldsymbol{\bar{\phi}}*\\
	\end{array} \right],
\end{align}
and $\mathbf{\bar{N}}_{\phi}$ is defined in \eqref{N_phi} at the bottom of the previous page.

\end{subequations}

In addition, $\nabla _\eta ^2 {\bar{n}_{\phi}}\left( \eta  \right)$ can be further rewritten as
\setcounter{equation}{131}
\begin{align}\label{r_2_phi}
\nabla _\eta ^2 {\bar{n}_{\phi}}\left( \eta  \right)
&=2\boldsymbol{\bar{\phi}}^{\mathrm{H}} \mathbf{M}_{\phi}  \boldsymbol{\bar{\phi}} \notag
\\
&=\left[ \begin{matrix}
	\boldsymbol{\bar{\phi}}^{\mathrm{H}}&		\boldsymbol{\bar{\phi}}^{\mathrm{T}}\\
\end{matrix} \right] \left[ \begin{matrix}
	 \mathbf{M}_{\phi}&		0\\
	0&		 \mathbf{M}_{\phi}^{\mathrm{T}}\\
\end{matrix} \right] \left[ \begin{array}{c}
	\boldsymbol{\bar{\phi}}\\
	\boldsymbol{\bar{\phi}}*\\
\end{array} \right].
\end{align}

As a result, we have
\begin{equation}
\mathbf{\bar{N}}_{\phi}\succeq 
 \left[ \begin{matrix}
	 \mathbf{M}_{\phi}&		0\\
	0&		 \mathbf{M}_{\phi}^{\mathrm{T}}\\
\end{matrix} \right]. 
\end{equation}
For simplicity, we choose $\mathbf{M}_{\phi}=\bar{\beta}\mathbf{I}=\lambda _{\min}\left( \mathbf{\bar{N}}_{\phi} \right)\mathbf{I}$.
By using the properties (\textit{a}1)-(\textit{a}3) in Appendix C, we replace $\bar{\beta}$ with its upper
bound, as shown in \eqref{b_MM} at the bottom of this page.

Then, we introduce the following result to deal with $\left\| \mathbf{o}_k \right\| _{2}^{2}$:
\begin{enumerate}[({\textit{a}}6)]
\setlength{\itemsep}{0.7ex}
\item $ - {\left\| {{{\bf{B}}}{{\bf{c}}}} \right\|_1}$ is the solution to the following problem for $\mathbf{x}=\left[ x_1,...,x_M \right] ^{\mathrm{T}}$:
\setcounter{equation}{134}
\begin{subequations}\label{problem_in_a6}
	\begin{alignat}{2}
		\min_{\bf{x}} \quad & {\mathop{\rm Re}\nolimits} \left\{  {{{\bf{c}}^{\rm H}}{{\bf{B}}^{\rm H}}{\bf{x}}}  \right\} \\
		\mbox{s.t.}\quad
		& \left| x_m \right|\leqslant 1,1\le m\le M.
	\end{alignat}
\end{subequations}
\end{enumerate}
 
By using (\textit{a}4) in Appendix C and (\textit{a}6), we obtain the upper bound for $\left\| \mathbf{o}_k \right\| _{2}^{2}$ given in \eqref{o_MM} at the bottom of this page.
%The upper bound of $\bar{\beta}$ is independent of $\eta$.
%Hence, the conclusion based on the condition (A2')  satisfies the condition (A2) as well.

Finally, \eqref{f_phi_MM} can be rewritten as
\setcounter{equation}{136}
\begin{align}
\bar{f}\left( \boldsymbol{\phi }|\boldsymbol{\phi }^{(n)} \right) =\!\bar{c}_{\phi}+2\mathrm{Re}\left\{ \mathbf{\bar{v}}_{\phi}^{\mathrm{H}}\boldsymbol{\phi } \right\} +\bar{\beta}\boldsymbol{\phi }^{\mathrm{H}}\boldsymbol{\phi },
\end{align}
where $\mathbf{\bar{v}}_{\phi}$ and $\bar{c}_{\phi}$ are given in \eqref{v_MM_phi} and \eqref{const_MM_phi}, respectively.

Hence, the proof is completed.

\end{appendices}

\bibliographystyle{IEEEtran}
\bibliography{IEEEabrv,Refer}

\end{document}